\documentclass{article}
\usepackage[utf8]{inputenc}
\usepackage{amsmath}
\usepackage{amsbsy}
\usepackage{xcolor}
\newtheorem{theorem}{Theorem}
\newtheorem{mydef}{Definition}
\newtheorem{corollary}{Corollary}
\usepackage{graphicx}
\usepackage{caption}
\usepackage{subcaption}
\usepackage{mathtools}
\usepackage{float}
\usepackage{natbib}
\usepackage{adjustbox}
\usepackage{amsfonts}

\title{Semiparametric Estimation of Individual Coefficients in a Dyadic Link Formation Model Lacking Observable Characteristics\thanks{I thank Koen Jochmans for his supervision and mentoring and, in particular, for providing me with the opportunity to develop as a researcher during my postdoctoral research. I further thank Pantelis Karapanagiotis for his substantial advice and our discussions that helped me tackle down and understand the numerical problems that arose in the estimation.}}

\author{L.\ Sanna Stephan\footnote{University of Groningen,  Faculty of Economics and Business, Zernike Campus, Duisenberg Building, Nettelbosje 2, 9747 AE Groningen, Netherlands} }
\date{}

\begin{document}

\maketitle

\begin{abstract}
Dyadic network formation models have wide applicability in economic research, yet are difficult to estimate in the presence of individual specific effects and in the absence of distributional assumptions regarding the model noise component. The availability of (continuously distributed) individual or link characteristics generally facilitates estimation. Yet, while data on social networks has recently become more abundant, the characteristics of the entities involved in the link may not be measured. 
Adapting the procedure of \citet{KS}, I propose to use network data alone in a semiparametric estimation of the individual fixed effect coefficients, which carry the interpretation of the individual relative popularity. This entails the possibility to anticipate how a new-coming individual will connect in a pre-existing group. The estimator, needed for its fast convergence, fails to implement the monotonicity assumption regarding the model noise component, thereby potentially reversing the order if the fixed effect coefficients. This and other numerical issues can be conveniently tackled by my novel, data-driven way of normalising the fixed effects, which proves to outperform a conventional standardisation in many cases. I demonstrate that the normalised coefficients converge both at the same rate and to the same limiting distribution as if the true error distribution was known. The cost of semiparametric estimation is thus purely computational, while the potential benefits are large whenever the errors have a strongly convex or strongly concave distribution. 
\end{abstract}

\section{Introduction}
To date, economists still lack a precise understanding of how households, firms or other economic entities form the relationships among themselves. 
Furthermore, while data on social and economic networks has become more abundant, relevant individual characteristics may not be observable. 
This speaks against fully parametric network formation models that include covariates (such as \citet{graham}, \citet{mele}) and potentially a game-theoretic microeconomic foundation (such as \citet{sheng}, \citet{ShengRidder1}, \citet{ShengRidder2}).  
Given the diversity of socioeconomic settings in which networks matter (such as trade, finance and household decision making), estimation of a parsimonious, flexible network formation model is important.
In this paper, I use a dyadic network formation model\footnote{Network formation models are dyadic when the probability of linking between $i$ and $j$ given the model parameters depends only depends on the characteristics of these two individuals, while the characteristics of the remaining $N-2$ agents in the network have no impact on it.} with individual fixed effects for capturing all individuals characteristics.
I assume a monotone cumulative distribution function (CDF) of the noise component and that the dependent variable suffices an index restriction, the index being the sum of the two individual effects of the agents involved in the link. 

The parsimonious model at hand is nested in the model considered by \citet{GWY}, who limits his attention primarily to identification of the model parameters\footnote{A tentative sketch of a possible estimation procedure is given in the appendix. However, the author does not concretize the procedure or investigate its feasibility and the resulting estimator's properties. }. I estimate linking probabilities using the procedure of \citet{KS}, which does not require knowledge of the error distribution. Linking probabilities feed into individual moment conditions and thus a Generalized Method of Moments (GMM) objective function. 
I provide a concise proof showing that if the estimated linking probabilities coincide with those of the true data generating process (DGP), so do the normalised fixed effect coefficients. Because the estimates of the linking probabilities are consistent, the normalised fixed effect coefficients can be estimated consistently, achieve the parametric convergence rate and are asymptotically normally distributed. Consequently, the cost of semiparametric estimation is purely computational, given a large enough sample. An erroneously specified error distribution, on the other hand, can have tremendous consequence, for example in presence of a strongly convex or concave error distribution.    

The (asymptotic) identification result and the asymptotic properties derived both apply to the normalised fixed effect coefficients. It could be presumed that the choice of normalisation is inconsequential for the estimation and only alters the interpretation of the estimated coefficients. 
I demonstrate however that numerical issues that arise with the chosen estimator can conveniently be solved by a novel normalisation method that is based on the minimal and maximal observed degree (i.e.\ number of friends), thereafter referred to as the ``degree-based minimax (DBMM) normalisation". The numerical issues arise because first, in order for the estimator to converge at the parametric rate, a kernel density estimation needs to be employed, but in absence of observable regressors, the kernel density estimator fails to implement the aforementioned monotonicity assumption, thereby risking to reverse the order of the fixed effect coefficients. Second, whenever the fixed effects are very concentrated, the matrix of derivatives of the moment conditions can be noninvertible due to collinearity while when they are very scattered (such as in the presence of an outlier), the moment condition of the respective individual will be largely uninformative, making it hard to situate the outlying fixed effect coefficient(s) relative to the others. Both of these issues are tackled by the DBMM normalisation, which guarantees that the normalised fixed effect coeffcients are (approximately) in the unit interval.    

The numerical issues are non negligible as they can plausibly arise in a variety of socio-economic dataset. 
Applications of outlier clusters include online dating sites on which typically a small fraction of males is perceived highly desirable in comparison to all others, trend imitation from celebrities that are much more talented than average individuals or cross linking of programming libraries where some are so well-designed that they critically stand out compared to all others.
Little variation in individual characteristics can arise for example when we consider networks among small scale businesses or junior academics. 
This shows that this novel way of normalising the parameters has distinct advantages in many real world settings. 

Equivalently, the features that imply huge cost of an erroneous parametric assumption (that is, a strongly convex or concave error distribution resulting in a skewed degree distribution) are also plausible to occur in socio economic datasets. Applications include follower networks among social media users, funding networks among startup firms, trend propagation networks, application networks among cryptocurrency blockchains or user networks among online retailers. Given that more cumbersome computations have become (and are becoming) increasingly cheap, the inconveniences of a semiparametric estimation as opposed to a logistic regression are becoming less important while the advantages prevail.
 
My paper contributes to a relatively novel strand of research on semiparametric estimation of dyadic link formation models. \citet{Koen} and \citet{graham} assume a logistic, \citet{dzemski} a normally distributed noise components, while \citet{candelaria} and \citet{toth} abstain from such an assumption. 
The emphasis of previous work has been on the estimation of the (common) coefficients of the observable independent variables, while the fixed effects have mostly been treated as nuisance parameters to be eliminated by a suitable transformation. 
These transformations such as pairing individuals with (close to) identical fixed effect (i.e.\ \citet{zeleneev}), pairwise differencing (i.e.\ \citet{candelaria}) or using variation over subgraph configurations (i.e.\ \citet{graham}) do require that, respectively, for every individual, a sufficient number of others with similar fixed effect coefficients exist, there being enough within individual and between individual variation in links or the number of identifying subgraphs being sufficient (and growing with the order of the network) and there being a sufficient variation over the subgraph configurations. Further,  
this proceeding is warranted when the emphasis is primarily on inference regarding the observed characteristics. However, even if the dyadic link formation model depends on characteristics that are, in principle, observable, oftentimes only network data is available.   
In this case individual specific fixed effects can be estimated and interpreted as the individual's relative popularity, with minimal requirements on the network data. 
While the semiparametric estimator I apply does not require knowledge of the error distribution, the latter can also be estimated following the estimation of the fixed effects parameters. 
This entails the possibility to predict the new links that could form once an individual is added to a preexisting group\footnote{Note that this is only possible if there are at least two such ''newcomers".}. 
There are various settings in which one would want to predict beneficial links, for example in order to optimise recruitment decisions: universities benefit from academic collaboration, management teams from a cooperative atmosphere and peer effects among students require a network to unfold. In other setting, policy makers may wish to predict links in order prevent them from arising, such as when dealing with criminal networks. This implies that predicting new links is highly relevant and important. 
Finally, since previous work has largely focused on common slope parameters, the asymptotic properties of the semiparameteric fixed effects estimator are not yet well understood. I contribute to this by proving consistency and asymptotic normality of the fixed effects coefficients that converge at the parametric rate. 


The remainder is organised as follows. Section two outlines the model, the necessary assumptions and some notations. Section three proofs asymptotic identification up to location and scale, given a consistent estimator of the linking probabilities. This estimator is introduced in section four and numerical issues are being analyzed. Section five introduces a novel normalisation that solves the numerical issues as well as a popular alternative, which is, standardisation. Section six outlines consistency and asymptotic normality (to be proved in more detail in the Appendix) and provides simulation evidence supporting the derivations. Section seven uses Monte Carlo result to compare the novel normalisation to the conventional standardisation and highlights cases in which the former outperforms the latter. Section eight demonstrates that in the presence of strongly convex or concave error distributions, an erroneously applied logit estimator can hardly generate any knowledge, while the semiparametric estimation performs well. This, together with section six demonstrates that the semiparametric estimator has potentially large advantages, while, as it converges at the parametric rate to the parametric asymptotic variance, it does not have substantial drawbacks. Section nine concludes and formal proofs are outlined in more detail in the appendix.

\section{Model and Assumptions}

Consider a dyadic network formation model.
\newline

{\bf Assumption 1: The Sample and the Population}
\begin{itemize}
    \item[1.1] $N$ individuals, indexed $i=1,...,N$ are randomly sampled from a population.
    \item[1.2] All characteristics of individual $i$ are unobserved and modelled using a fixed effect coefficient $\eta_i$.  
    \item[1.3] Individual $i's$ true characteristic is $\eta_{i,0}$. In the population, $\eta_{i,0}$ is distributed according to a density function $f_{\eta,0}$, which is continuous and has bounded support. 
\end{itemize}

 I use $\boldsymbol{\eta}_0
 $ denote the true fixed effect coefficient vector for the concrete sample at hand and $\mathbb{N}_0$ to denote the (infinite) set of all possible fixed effect coefficient vectors of size $N$ that can be drawn from the population.
\newline

{\bf Notation: Pairs and Links}\\

There exist $L=N(N-1)/2$ unique pairs that the $N$ agents can form. I use $\{i,j\}$ to denote the pair formed by agent $i$ and $j$,  $\mathbb{P}$ to denote the set of all pairs and $\mathbb{P}\backslash\{i,j\}$ to denote the set of all pairs {\it except} the one composed by $i$ and $j$. 
I use $w_{ij}$ to denote a vector of indicators featuring ones at position $i$ and $j$, all other elements being zero\footnote{There are always exactly two entries of value one, what varies is their position.}. This implies that $\eta_{i,0}+\eta_{j,0}=w_{ij}'\boldsymbol{\eta}_0 
\overset{def}{=} v_{ij}
$. I use $W$ of dimension $L \times N$ to denote the matrix comprising all vectors $w_{ij}$.
 \newline

 {\bf Assumption 2: The Network}
 \begin{itemize}
     \item[2.1] The network adjacency matrix, denoted $G$, is symmetric, featuring binary elements $g_{ij} \in [0;1]$ at position row $i$, column $j$\footnote{As $G$ is symmetric and $g_{ij}$ is binary, the network is undirected and unweighted.} and zeros on the main diagonal\footnote{That is, there are no self-links.}.
     \item[2.2] The linking outcome is observed for each pair $\{i,j\} \in \mathbb{P}$.
 \end{itemize}

{\bf Assumption 3: The Noise Component}

\begin{itemize}
    \item[3.1] The stochastic element of link formation is modeled using a random noise term $u_{ij}$. 
    \item[3.2] In the population, $u_{ij,0}$ is distributed according to a continuous density function $f_{u,0}$ and a continuously differentiable, monotone CDF $F_{u,0}$\footnote{The requirement for the error CDF to be differentiable and for the error density to be continuous can in principle be relaxed, albeit this requires an adaptation of the estimation procedure.}.
    \item[3.3] Conditional on the fixed effects, linking is independent across pairs, i.e.\ $u_{ij,0} \perp u_{km,0} \forall \{k,m\} \in \mathbb{P}\backslash \{i,j\}$.
\end{itemize}

{\bf Assumption 4: Index Sufficiency}
\begin{itemize}
    \item[4.1] The dependent variable $g_{ij}$ suffices an index restriction, such that 
\[P(g_{ij}=1|\boldsymbol{\eta}_0)
=F_{u,0}(w_{ij}'\boldsymbol{\eta}_0)=
F_{u,0}(v_{ij,0})=
F_{u,0}(\eta_{i,0}+\eta_{j,0})
\hspace{.2cm} \forall 
\boldsymbol{\eta}_0 \in \mathbb{N}_0.
\]
\end{itemize}

The model admits a latent variable representation. Define $g^*_{ij}$ to be a continuous latent random variable, the value of which is determined by the individual coefficients of $i$ and $j$ and a noise component $\varepsilon_{ij,0}=-u_{ij,0}$. Then we can also state the model as
\begin{align*}
g^*_{ij}= \eta_{i,0}+\eta_{j,0}+\varepsilon_{ij,0}
\hspace{.5cm} \varepsilon_{ij,0} \stackrel{i.i.d.}{\sim} F_{\varepsilon_0}
\\
g_{ij}=1 \hspace{.5cm} if \hspace{.5cm} g^*_{ij} \geq 0.
\end{align*}

{\bf Assumption 5: Probability Bounds}
\begin{itemize}
    \item[5.1] $0<F_{u,0}(\eta_{i,0}+\eta_{j,0})<1 
    \hspace{.25cm}\forall i,j \in N, i\neq j$\footnote{A sufficient condition for this is that $p_{1,ij,0}>\underline{p}, p_{0,ij,0}>\underline{p}$, for $p_{1,ij,0},p_{0,ij,0}$ as introduced below.}. This assumptions requires that the unconditional linking probability ($P(g_{ij}=1)$) is not zero or one (in which case links would never (respectively always) be formed for all parameter vectors). It further requires that for each possible realisation of the pair specific sum of fixed effects $v_{ij}$, there is a positive probability to observe this realisation ($v_{ij}$) in both the set of linked and unlinked pairs\footnote{That is $f_{v|g_{ij}=1}(v_{ij})\neq 0, f_{v|g_{ij}=0}(v_{ij})\neq 0$.}.
\end{itemize}

\section{Identification}

In the model at hand, applying a positive affine transformation to all model parameters does not affect any linking probability, since
 (using $\iota_N$ to denote a vector of ones of size $N$ and since $w_{ij}'b\iota_N=2b$)
\begin{equation}\label{problem}
    P(u_{ij} \leq w_{ij} ' \boldsymbol{\eta_0} )=
P(a u_{ij} + 2b \leq w_{ij} ' (a \boldsymbol{\eta_0}+b
\iota_N
) ) \hspace{.25cm} a>0; b \in \mathbb{R}; \{i,j\} \in \mathbb{P}.
\end{equation}

Presume that we carry out the normalisation  $\Tilde{\eta}_0(\eta_{i,0})=\eta_{i,0}-\eta_{k,0} \forall i$, such that the fixed effects coefficient of an arbitrarily selected individual $k$ is normalised to zero. Let $\Tilde{u}_0(u_{ij,0})=u_{ij,0}- 2\eta_{k,0}$ be the normalised error, then,
\[ P\left(\Tilde{u}_0(u_{ij,0}) \leq 
\Tilde{\eta}_0(\eta_{i,0})+
\Tilde{\eta}_0(\eta_{j,0})\right)=
P\left(a\Tilde{u}_0(u_{ij,0}) \leq 
a\Tilde{\eta}_0(\eta_{i,0})+
a\Tilde{\eta}_0(\eta_{j,0})\right) \]
\[ \forall a>0;
\{i,j\}\in \mathbb{P}.
\] 
  
Similarly, if we normalise  $\Tilde{\eta}_0(\eta_{i,0})=\eta_{i,0}/\eta_{k,0}$, resulting in $k's$ fixed effects coefficient being normalised to one, then letting $\Tilde{u}_0(u_{ij,0})=u_{ij,0}/\eta_{k,0}$ be the transformed error, then
\[ \forall a>0; \{i,j\} \in \mathbb{P}; i,j \neq k
\hspace{.55cm}
P\left(\Tilde{u}_0(u_{ij,0}) \leq 
\Tilde{\eta}_0(\eta_{i,0})+
\Tilde{\eta}_0(\eta_{j,0})\right)=\]
\[
 P\left(a\Tilde{u}_0(u_{ij,0}) +2(1-a) \leq 
a (\Tilde{\eta}_0(\eta_{i,0})+
\Tilde{\eta}_0(\eta_{j,0}))+2(1-a) \right)
\hspace{.25cm}
\mbox{and}
\] 
\[ \forall a>0; i \in N; i \neq k \hspace{.55cm}
P\left(\Tilde{u}_0(u_{ik,0}) \leq 
\Tilde{\eta}_0(\eta_{i,0})+1\right)
=  \]
\[P\left( a \Tilde{u}_0(u_{ik,0}) +2(1-a) 
\leq a \Tilde{\eta}_0(\eta_{i,0})+(1-a)+1\right)\footnote{This corresponds to an affine transformation in which $b=1-a$.}.
\] 
This shows that unless I carry out a normalisation that fixes location {\it and} scale of the (normalised) fixed effect coefficients, there is no hope to identify the (normalised) true coefficients. 

I now show that if an error CDF together with a normalised fixed effect coefficient vector is observationally equivalent to the true DGP, then the normalised coefficient vector must be the result of applying a positive affine transformation to the true coefficient vector. 
This has the implication that 
an error CDF that creates identical linking probabilities for all  possible pairs of agents point identifies the (scale and location) normalised fixed effect coefficients. 

\begin{mydef}
Given the random network adjacency matrix $G$, featuring zeros on the main diagonal, define
\[ P(G|\boldsymbol{\eta},F_u)\]
as the matrix collecting all linking probabilities given the fixed effects of the randomly sampled individuals and the error distribution, i.e.\ 
\[ P(G|\boldsymbol{\eta},F_u)_{ij}=
P(G|\boldsymbol{\eta},F_u)_{ji}=
P(g_{ij}=1|w_{ij}'\boldsymbol{\eta},F_u)=
F_u(\eta_i+\eta_j).
\]
\end{mydef}
\begin{mydef}
If for any given sample of $N$ agents, 
    \[ P(G|\boldsymbol{\eta}_1,F_{u_1})
    =P(G|\boldsymbol{\eta}_2,F_{u_2})
    \]
    then we say that the model parameters $\boldsymbol{\eta}_1,F_{u_1}$ are {\bf observationally equivalent} to the model parameters $\boldsymbol{\eta}_2,F_{u_2}$.
\end{mydef}
Before exploring the implications of the symmetry of the index function chosen and outlining how the fixed effects coefficient vector can be identified up to a twofold normalisation, it is useful to highlight a general feature of this network formation model that remains valid regardless which index function is chosen:
\begin{theorem}
If any two coefficient vectors are observationally equivalent and $N\geq4$, then any particular entry in the first coefficient vector can be expressed as a function of only three entries of the respective other coefficient vector, one being the particular individuals' coefficient in this other vector. 
\end{theorem}
\begin{corollary}
If a coefficient vector is observationally equivalent to the true coefficient vector, and there are at least four individuals, then the $i'th$ entry of that vector (i.e. the coefficient of individual $i$) can be expressed as a function of three entries of the true coefficient vector, one being it's $i'th$ entry. 
\end{corollary}
 This feature will be demonstrated here with the particular index function chosen, though an analogue argument can be made for any choice of a dyadic index function. 

Assume that the not further specified model parameters $\boldsymbol{\eta}_1, F_{u_1}$ are observationally equivalent to the model parameters of the true DGP, i.e.\
\noindent 
Due to the observational equivalence, we have
\begin{equation}
 F_{u_0}(\eta_{i,0}+\eta_{j,0}) =
 F_{u_1}(\eta_{i,1}+\eta_{j,1}) 
\hspace{.25cm}
\forall i \neq j.
\label{start}
\end{equation} 
With at least the four individuals $i,j,k$ and $m$ (and potentially more individuals), we have 
\[\eta_{i,1}=F_{u_1}^{-1}\Big(F_{
u_0}\big( \eta_{i,0}+\eta_{k,0}\big)\Big)-\eta_{k,1}
\]
\[\eta_{k,1}=F_{u_1}^{-1}\Big(F_{
u_0}\big( \eta_{k,0}+\eta_{m,0}\big)\Big)-\eta_{m,1}
\]
\[\eta_{m,1}=F_{u_1}^{-1}\Big(F_{
u_0}\big( \eta_{m,0}+\eta_{i,0}\big)\Big)-\eta_{i,1}
\]
which can be combined to lead 
\[\eta_{i,1}=F_{u_1}^{-1}\Big(F_{
u_0}\big( \eta_{i,0}+\eta_{k,0}\big)\Big)-
F_{u_1}^{-1}\Big(F_{
u_0}\big( \eta_{k,0}+\eta_{m,0}\big)\Big)+
F_{u_1}^{-1}\Big(F_{
u_0}\big( \eta_{m,0}+\eta_{i,0}\big)\Big)-\eta_{i,1}
\]
\[ 
\eta_{i,1}=\frac{1}{2} 
\Bigg(
F_{u_1}^{-1}\Big(F_{
u_0}\big( \eta_{i,0}+\eta_{k,0}\big)\Big)-
F_{u_1}^{-1}\Big(F_{
u_0}\big( \eta_{k,0}+\eta_{m,0}\big)\Big)+
F_{u_1}^{-1}\Big(F_{
u_0}\big( \eta_{m,0}+\eta_{i,0}\big)\Big) \Bigg),
\]
implying that there is a functional relationship 
\begin{equation}
    \eta_{i,1}=g\big( \eta_{i,0},\eta_{k,0},\eta_{m,0} \big)
    \label{g}
\end{equation} that is constant across individuals. That is, we can express the coefficient of individual $i$ as a function of three true coefficients, independent of the coefficient of individual $j$. 
Note that for any individual $i$, any choice of $k,m \neq i$ will lead to the same solution for $i's$ coefficient. Note also that this functional relationship is the same for all individuals. 
I replace $\eta_{i,1}$ and $\eta_{j,1}$ in \eqref{start} using \eqref{g} and take the derivative with respect to $\eta_{i,0}$ and $\eta_{j,0}$, respectively.  
\[
F_{u_1}\big(\eta_{i,1}+\eta_{j,1}\big)
=
F_{u_1}\Big(
g\big(\eta_{i,0},\eta_{k,0},\eta_{m,0}\big)+
g \big(\eta_{j,0},\eta_{k,0},\eta_{m,0}\big)\Big)
\]
\[  
\frac{\partial  F_{u_0}\big(
\eta_{i,0}+\eta_{j,0}\big) }{\partial 
\eta_{i,0}}
=
\frac{ \partial F_{u_1}\Big(
g \big(\eta_{i,0},\eta_{k,0},\eta_{m,0}\big)+
g \big(\eta_{j,0},\eta_{k,0},\eta_{m,0}\big) \Big) }{\partial 
\eta_{i,0}}\]
\[=
\frac{ \partial F_{u_1}\Big(
g\big(\eta_{i,0},\eta_{k,0},\eta_{m,0}\big)+
g \big(\eta_{j,0},\eta_{k,0},\eta_{m,0} \big) \Big) }{
\partial 
g \big(\eta_{i,0},\eta_{k,0},\eta_{m,0}\big)
}
\frac{\partial g \big(\eta_{i,0},\eta_{k,0},\eta_{m,0}\big)}{
\partial \eta_{i,0}}+\]
\[
\frac{ \partial F_{u_1}\Big(
g\big(\eta_{j,0},\eta_{k,0},\eta_{m,0}\big)+
g \big(\eta_{j,0},\eta_{k,0},\eta_{m,0}\big) \Big) }{
\partial 
g \big(\eta_{j,0},\eta_{k,0},\eta_{m,0} \big)
}\underbrace{
\frac{\partial g \big(\eta_{j,0},\eta_{k,0},\eta_{m,0}\big)}{
\partial \eta_{i,0}}}_{=0}
\] 
and
\[  
\frac{\partial  F_{u_0}\big(
\eta_{i,0}+\eta_{j,0}\big) }{\partial 
\eta_{j,0}}
=
\frac{ \partial F_{u_1}\Big(
g \big(\eta_{i,0},\eta_{k,0},\eta_{m,0}\big)+
g \big(\eta_{j,0},\eta_{k,0},\eta_{m,0}\big) \Big) }{\partial 
\eta_{j,0}}\]
\[=
\frac{ \partial F_{u_1}\Big(
g\big(\eta_{i,0},\eta_{k,0},\eta_{m,0}\big)+
g \big(\eta_{j,0},\eta_{k,0},\eta_{m,0} \big) \Big) }{
\partial 
g \big(\eta_{i,0},\eta_{k,0},\eta_{m,0}\big)
}
\underbrace{
\frac{\partial g \big(\eta_{i,0},\eta_{k,0},\eta_{m,0}\big)}{
\partial \eta_{j,0}}}_{=0}+\]
\[
\frac{ \partial F_{u_1}\Big(
g\big(\eta_{j,0},\eta_{k,0},\eta_{m,0}\big)+
g \big(\eta_{j,0},\eta_{k,0},\eta_{m,0}\big) \Big) }{
\partial 
g \big(\eta_{j,0},\eta_{k,0},\eta_{m,0}\big)
}
\frac{\partial g \big(\eta_{j,0},\eta_{k,0},\eta_{m,0}\big)}{
\partial \eta_{j,0}}
\]
and the symmetry of the index implies that both 
\[
\frac{ \partial F_{u_1}\Big(
g\big(\eta_{i,0},\eta_{k,0},\eta_{m,0}\big)+
g \big(\eta_{j,0},\eta_{k,0},\eta_{m,0}\big) \Big) }{
\partial 
g \big(\eta_{i,0},\eta_{k,0},\eta_{m,0}\big)
}
=
\frac{ \partial F_{u_1}\Big(
g\big(\eta_{i,0},\eta_{k,0},\eta_{m,0}\big)+
g \big(\eta_{j,0},\eta_{k,0},\eta_{m,0}\big) \Big) }{
\partial 
g \big(\eta_{j,0},\eta_{k,0},\eta_{m,0}\big)
}
\] 
and 
\[
\frac{\partial  F_{u_0}\big(
\eta_{i,0}+\eta_{j,0}\big) }{\partial 
\eta_{i,0}}=\frac{\partial  F_{u_0}\big(
\eta_{i,0}+\eta_{j,0}\big) }{\partial 
\eta_{j,0}}
\]
consequently 
\[
\frac{\partial  F_{u_0}\big(
\eta_{i,0}+\eta_{j,0}\big) }{\partial 
\eta_{i,0}}=
\frac{ \partial F_{u_1}\Big(
g\big(\eta_{i,0},\eta_{k,0},\eta_{m,0}\big)+
g \big(\eta_{j,0},\eta_{k,0},\eta_{m,0} \big) \Big) }{
\partial 
g \big(\eta_{i,0},\eta_{k,0},\eta_{m,0}\big)
}
\frac{\partial g \big(\eta_{i,0},\eta_{k,0},\eta_{m,0}\big)}{
\partial \eta_{i,0}}=\]
\[\frac{\partial  F_{u_0}\big(
\eta_{i,0}+\eta_{j,0}\big) }{\partial 
\eta_{j,0}}=
\frac{ \partial F_{u_1}\Big(
g\big(\eta_{i,0},\eta_{k,0},\eta_{m,0}\big)+
g \big(\eta_{j,0},\eta_{k,0},\eta_{m,0} \big) \Big) }{
\partial 
g \big(\eta_{j,0},\eta_{k,0},\eta_{m,0}\big)
}
\frac{\partial g \big(\eta_{j,0},\eta_{k,0},\eta_{m,0}\big)}{
\partial \eta_{j,0}}
\]
which implies that 
\[
\frac{\partial g \left(\eta_{i,0},\eta_{l,0},\eta_{k,0} \right)}{\partial \eta_{i,0}}=
\frac{\partial g \left(\eta_{j,0},\eta_{l,0},\eta_{k,0} \right)}{\partial \eta_{j,0}}.
\]
This shows that the derivative of the \eqref{g} with respect to the individual coefficient is the same for all individuals, in which case the derivative must be a constant. 
The above holds true irrespective of the sign of the derivative and as such, the sign of this constant is only identified by the assumption of a monotonous error CDF.  
Since the derivative is constant, this in turn implies that the function is an affine function and positive affine due to the monotonicity, i.e.\ that 
 \[ \eta_{i,1}= 
a \eta_{i,0}+b
\hspace{.25cm} \forall i.
\]
Taking the derivative of the linking probability of $i$ and $j$ with respect to $\eta_{i,0}$ amounts to asking the question how $j$'s linking probability changed if we switched her partner from being $i$ to being an individual with a slightly different (true) coefficient. If both error distributions and both parameter sets lead to the same linking probabilities for all possible samples that can be drawn from the population, then the change in $j$'s probability must be the same.  
Furthermore, if we consider $i$ and $j$ and what happens to $j$'s probability if her partner ($i$) changes, this is equivalent to considering what happens to $i$'s probability if her partner ($j$) changes. 

As a consequence, the symmetry of the index function has an important implication: 
\begin{theorem}
    In addition to Theorem 1, a symmetric index function implies that if two sets of coefficients lead to the same probability distribution for each link, then there is an affine relationship between individual $i's$ two coefficients. Under Assumption 2, this relationship is positive affine. 
\end{theorem}
\begin{corollary}
    With at least four individuals, a symmetric index and the monotonicity assumption, all model  coefficients are identified up to a location and a scale normalisation.
\end{corollary}
The intuition is that with only three individuals, the relationship between the entries of two observationally equivalent coefficient vectors need not be constant across individuals, even if the index is symmetric. The symmetry of the index, together with the existence of at least four individuals in turn guarantees this relationship that in particular is constant across individuals and due to the symmetry must be linear. 
\\ 
\noindent For the model at hand, this implies that 
\begin{equation}
    P(G| \boldsymbol{\eta}_0,F_{u_0})=P(G|\boldsymbol{\eta}_1,F_{u_1}) \hspace{.2cm} 
    \Longleftrightarrow
    \boldsymbol{\eta}_1=a\boldsymbol{\eta}_0+b,
    \hspace{.2cm}
    a>0,b \in \mathcal{R}
    \label{equalimplication1}.
\end{equation}

Using the estimator proposed by \citet{KS} I can consistently estimate the linking probabilities, which in turn implies consistent estimation of the location and scale normalised fixed effect coefficients. 

\section{Estimation}

\subsection{The Estimator }

Rank correlation estimators feature slow convergence. Kernel based estimators, on the other hand, can achieve the parametric rate. 
I estimate the linking probabilities using the procedure from \citet{KS}. I use $f_v(v), \hspace{.15cm} f_{v|g_{ij}=1}(v)$ and $f_{v|g_{ij}=0}(v)$ to denote the population density of $v_{ij}$, unconditionally, or, respectively, conditional on the link observed. 
 The probability to observe a link in the sample at hand is a function of the choice of model parameters $\boldsymbol{\eta},F_u$. Using Bayes rule and the Law of Total Probabilities, it can be rewritten as
 \begin{eqnarray}
P(g_{ij}=1|\boldsymbol{\eta})\big(\boldsymbol{\eta},F_u \big) =
\frac{P(g_{ij}=1 ) f_{v|g_{ij=1}}(v_{ij})}{f_v(v_{ij})}= 
\nonumber
\\
\frac{P(g_{ij}=1 ) f_{v|g_{ij=1}}(v_{ij})}{
P(g_{ij}=1 ) f_{v|g_{ij=1}}(v_{ij})+
P(g_{ij}=0 ) f_{v|g_{ij=0}}(v_{ij})
} 
\overset{def}{=}
\frac{p_{1,ij}}{p_{1,ij}+p_{0,ij}}. \label{truth}
 \end{eqnarray}
My aim is to choose $\boldsymbol{\eta}$ that maximises this expression for an objective function that depends on all links, subject to the normalisation restrictions. As shown above, this parameter vector will coincide with the normalised true coefficient vector if and only if the linking probabilities implied by $\hat{\boldsymbol{\eta}}, \hat{F}_{u}$ are observationally equivalent to those of the true DGP. 
In absence of knowledge of $F_{u,0}$, my strategy is to estimate the linking probabilities consistently. Observe that in \eqref{truth},  $P(g_{ij}=1)$ and $P(g_{ij}=0)$ both depend on $F_u$, but are observed. Further, for each pair $\{i,j\}$, given a choice of $\boldsymbol{\eta}$, I do observe whether a link is formed and I do know the value of $v_{ij}$. As a consequence, I can estimate $p_{1,ij}$ (respectively $p_{0,ij}$) using the set of links that are present (respectively absent) and running a kernel density estimator over $v_{ij}$. 
I use $K(z)$ to denote a Kernel and $K'(z)$ ($K''(z)$) to denote its first (respectively second) derivative. 
\newline 

{\bf Assumption 6: The Kernel}
As conventional, the kernel 
\begin{enumerate}
    \item is symmetric,
    \item integrates to one, 
    \item has an expected value of zero,
    \item has a variance of zero,
    \item is not skewed and 
    \item $|K(z)|<c, |K'(z)|<c,  |K''(z)|<c,$ 
    \item[] $
    \int |K(z)| dz <c, \int |K'(z)|dz<c,  \int |K''(z)|dz<c$
\end{enumerate}
I use $h$ to denote the chosen bandwidth, which in the estimation, I choose to $h=L^{-1/7}$. 
Then, applying the strategy from above leads to the estimates
\[  
\hat{p}_{1,ij} \big(\boldsymbol{\eta}\big)=
\frac{1}{h(L-1)}
\sum_{
\substack{
\{k,m\} \in \\
\mathbb{P}\setminus \{i,j\} }
}
I(g_{km}=1) K \left(\frac{v_{ij}-v_{km}}{h} \right)
\]
and 
\[  
\hat{p}_{0,ij} \big(\boldsymbol{\eta}\big)=
\frac{1}{h(L-1)}
\sum_{\substack{
\{k,m\} \in \\
\mathbb{P}\setminus \{i,j\} }}
I(g_{km}=0) K \left(\frac{v_{ij}-v_{km}}{h} \right)
\]
 Consequently, the estimated probabilities depend only on (the observed data and) the fixed effects coefficients (i.e.\ not on the (estimate of the) error distribution). 

A numerical problem can arise because the derivatives of the linking probabilities feature (an estimate of) the density of $v$ evaluated at $v_{l}$ in the denominator. Whenever the latter is close to zero, the objective function cannot be evaluated. 
The numerical problem gets amplified for higher order derivatives (because the denominator will be raised to higher powers). As such, when we aim for a Newton-Raphson algorithm for estimation, it would be preferable to not have to differentiate the linking probabilities more than once. This is why I opt for a GMM instead of Maximum Likeliood estimation. The $N$ moments are given by the (observed) $N$ individual degrees. I define the individual degree as 
\[
d_i =\frac{\sum_{j=1}^N g_{ij}}{N-1}.
\]
Then, for any choice of model parameters $\boldsymbol{\eta}, F_u$, the expected individual degree is   
\[E\big[d_i | \boldsymbol{\eta} \big] \big( \boldsymbol{\eta},F_u \big)
= P(g_{ij}=1| \boldsymbol{\eta}, \eta_i)\big( \boldsymbol{\eta},F_u \big)= \int F_{u}(\eta_i+\eta_j)f_{\eta}(\eta_j)d \eta_j \]
\[= E_{\eta_j} \left[F_u(\eta_i+\eta_j)|\eta_i\right]=
\int \left( \frac{p_{1,ij}}{p_{1,ij}+p_{0,ij}}\right) f_{\eta}(\eta_j)d \eta_j,
\]
which I can estimate by
\[
\hat{d}_i \big(\boldsymbol{\eta}\big)=
\frac{1}{N-1}
\sum_{j=1}^N 
\left(
\frac{\hat{p}_{1,ij}\big(\boldsymbol{\eta}\big)}{\hat{p}_{1,ij}\big(\boldsymbol{\eta}\big)+\hat{p}_{0,ij}\big(\boldsymbol{\eta}\big)}\right)=
\frac{1}{N-1}
\sum_{j=1}^N  \widehat{F_u}(\boldsymbol{\eta},v_{ij}).
\]
Note that I use $\widehat{F_u}(\boldsymbol{\eta},v_{ij})$ to denote the (kernel-density) estimated error distribution (which thus depends on all fixed effect coefficients) evaluated at $v_{ij}$. 
I define $N$ moment conditions 
\[ \hat{m}_i \big(\boldsymbol{\eta})=
d_i-\hat{d}_i \big(\boldsymbol{\eta}\big)=
\frac{1}{N-1}\sum_{j=1}^N 
\left(
g_{ij}-\left(
\frac{\hat{p}_{1,ij}\big(\boldsymbol{\eta}\big)}{\hat{p}_{1,ij}\big(\boldsymbol{\eta}\big)+\hat{p}_{0,ij}\big(\boldsymbol{\eta}\big)}\right) \right)
\]
which is the difference between the observed degree and its expected value. The optimisation problem is then
\[
min_{\boldsymbol{\eta}} 
\hspace{.1cm}
\hat{Q}(\boldsymbol{\eta})=
min_{\boldsymbol{\eta}}
\hspace{.1cm}
\hat{m}(\boldsymbol{\eta})'\hat{m}(\boldsymbol{\eta}).
\]
The derivatives of the moment conditions are estimated using 
\[
\frac{\partial \hat{p}_{1,ij}(\boldsymbol{\eta})}{\partial v_{j}}=
\frac{1}{h(L-1)}
\sum_{\substack{
\{k,m\} \in \\
\mathbb{P}\setminus \{i,j\} }}
I(g_{km}=1) K'  \left(\frac{v_{ij}-v_{km}}{h} \right)\frac{1}{h}\]
\[
\hspace{.1cm}
\overset{def}{=}\hspace{.1cm}
\hat{p}d_{1,ij}(\boldsymbol{\eta})
\]
and 
\[
\frac{\partial \hat{p}_{0,ij}(\boldsymbol{\eta})}{\partial v_{j}}=
\frac{1}{h(L-1)}
\sum_{\substack{
\{k,m\} \in \\
\mathbb{P}\setminus \{i,j\} }}
I(g_{km}=0) K'  \left(\frac{v_{ij}-v_{km}}{h} \right)\frac{1}{h}\]
\[
\hspace{.1cm}
\overset{def}{=}\hspace{.1cm}
\hat{p}d_{0,ij}(\boldsymbol{\eta})
\]
as
\[
\frac{\partial \hat{m}_i(\boldsymbol{\eta})}{\eta_i}=-
\frac{1}{N-1}\sum_{j=1}^N 
\frac{
\hat{p}_{1,ij}\big(\boldsymbol{\eta}\big)
\Big(\hat{p}d_{1,ij}\big(\boldsymbol{\eta})+
\hat{p}d_{0,ij}\big(\boldsymbol{\eta})\Big)-
\hat{p}d_{1,ij}\big(\boldsymbol{\eta}\big)
\Big(
\hat{p}_{1,ij}\big(\boldsymbol{\eta}\big)
+
\hat{p}_{0,ij}\big(\boldsymbol{\eta}\big)\Big)
}{\big(\hat{p}_{1,ij}\big(\boldsymbol{\eta}\big)+\hat{p}_{0,ij}\big(\boldsymbol{\eta}\big)\big)^2}
\]
\[
= -
\frac{1}{N-1}\sum_{j=1}^N 
\widehat{f_u}(\boldsymbol{\eta},v_{ij})
\]
and
\[
\frac{\partial \hat{m}_i(\boldsymbol{\eta})}{\eta_j}=-
\frac{1}{N-1} 
\frac{
\hat{p}_{1,ij}\big(\boldsymbol{\eta}\big)
\Big(\hat{p}d_{1,ij}\big(\boldsymbol{\eta})+
\hat{p}d_{0,ij}\big(\boldsymbol{\eta})\Big)-
\hat{p}d_{1,ij}\big(\boldsymbol{\eta}\big)
\Big(
\hat{p}_{1,ij}\big(\boldsymbol{\eta}\big)
+
\hat{p}_{0,ij}\big(\boldsymbol{\eta}\big)\Big)
}{\big(\hat{p}_{1,ij}\big(\boldsymbol{\eta}\big)+\hat{p}_{0,ij}\big(\boldsymbol{\eta}\big)\big)^2}
\]
\[
= - \frac{1}{N-1} \widehat{f_u}(\boldsymbol{\eta},v_{ij}).
\]
Note that here I am using $\widehat{f_u}(\boldsymbol{\eta},v_{ij})$ to denote the (kernel-density) estimated error density (which depends on all fixed effect coefficients), evaluated at $v_{ij}$.

\subsection{The Inability of Kernel Based Estimators To Correctly Identify the Order of the Fixed Effect Coefficients}

Because 
\[
P(g_{ij}=1|\boldsymbol{\eta})
\big(\boldsymbol{\eta},F_u\big)
=
F_{u}(\eta_{i}+\eta_{j})=
P(u_{ij} \leq \eta_{i}+\eta_{j})=
\]
\[
P(-u_{ij} > - \eta_{i}-\eta_{j})=
1- P(-u_{ij} \leq - \eta_{i}-\eta_{j})=
1-F_{-u}(-\eta_{i}-\eta_{j})=\]
\[
1-P(g_{ij}=1|\boldsymbol{\eta})
\big(-\boldsymbol{\eta},F_{-u}\big),
\]
consequently, the model identifies the order of the fixed effect coefficients.  
Intuitively, as 
\[P(g_{ij}=1|w_{ij}'\boldsymbol{\eta})=
1-
P(g_{ij}=1|-w_{ij}'\boldsymbol{\eta})
\]
we know that, in the model, multiplying all fixed effect coefficients by minus one (thereby reversing their order) will lead to complimentary probabilities and as such, the reverse expected degree distribution. 
In practice, however, we estimate the fixed effects given the network and its degree distribution, and thus we do not take into account that reversing the fixed effect coefficients would, according to the model, change all link observations into the respective complimentary event. 
In the following, we write $G(\boldsymbol{\eta}_0)$ and $g_l(\boldsymbol{\eta}_0)$ to denote more clearly that the estimator takes a given network (as generated by the true coefficients) and computes estimated probabilities for different parameter values. Consequently, in the estimation, $g_{ij}(\boldsymbol{\eta}_0)$ is unchanged for all links when the estimator uses $- \boldsymbol{\eta}$ instead of $\boldsymbol{\eta}$ as input. 
As I hold the observed adjacency matrix $G(\boldsymbol{\eta}_0)$ fixed, with any kernel that is symmetric around zero I obtain
\[  
\hat{p}_{1,ij} \big(-\boldsymbol{\eta}|
G(\boldsymbol{\eta}_0)\big)=
\frac{1}{h(L-1)}
\sum_{
\substack{
\{k,m\} \in \\
\mathbb{P}\setminus \{i,j\} }}
I(g_{km}(\boldsymbol{\eta}_0)=1) K \left(\frac{-(v_{ij}-v_{km})}{h} \right)
\]
\[  
\frac{1}{h(L-1)}
\sum_{\substack{
\{k,m\} \in \\
\mathbb{P}\setminus \{i,j\} }}
I(g_{km}(\boldsymbol{\eta}_0)=1) K \left(\frac{(v_{ij}-v_{km})}{h} \right)=
\hat{p}_{1,ij} \big(\boldsymbol{\eta}|G(\boldsymbol{\eta}_0)\big)
\]
despite the fact that 
\[f_{v|g_{ij}=1}(v) \neq f_{-v|g_{ij}=1}(-v)\] due to the monotonicity assumption. Similarly,  
\[  
\hat{p}_{0,ij} \big(-\boldsymbol{\eta}|G(\boldsymbol{\eta}_0)\big)=
\hat{p}_{0,ij} \big(\boldsymbol{\eta}|G(\boldsymbol{\eta}_0)\big)
\]
leading to 
\[
\frac{\hat{p}_{1,ij}\big(-\boldsymbol{\eta}|G(\boldsymbol{\eta}_0)\big)}{\hat{p}_{1,ij}\big(-\boldsymbol{\eta}|G(\boldsymbol{\eta}_0)\big)+\hat{p}_{0,ij}\big(-\boldsymbol{\eta}|G(\boldsymbol{\eta}_0)\big)}
=
\frac{\hat{p}_{1,ij}\big(\boldsymbol{\eta}|G(\boldsymbol{\eta}_0)\big)}{\hat{p}_{1,ij}\big(\boldsymbol{\eta}|G(\boldsymbol{\eta}_0)\big)+\hat{p}_{0,ij}\big(\boldsymbol{\eta}|G(\boldsymbol{\eta}_0)\big)}
\]
and consequently 
\[
\hat{Q}(\boldsymbol{\eta}|G(\boldsymbol{\eta}_0))=
\hat{Q}(-\boldsymbol{\eta}|G(\boldsymbol{\eta}_0)).
\]
Since the kernel based estimator fails to properly implement the monotonicity assumption, it can identify the order of the fixed effect coefficients only up to an affine transformation. In particular, two fixed effect coefficient vectors with opposite sign will lead to the same estimated probabilities, given the network. The consequence is that the order of the estimated fixed effect coefficients is either correct or completely reversed. 
Further, due to the symmetry of the Kernel 
\[
K'(z)=-K'(-z)
\] such that
\[
\frac{\partial \hat{p}_{1,ij}(\boldsymbol{\eta}|G(\boldsymbol{\eta}_0))}{\partial v_{ij}}=
\frac{1}{h(L-1)}
\sum_{
\substack{
\{k,m\} \in \\
\mathbb{P}\setminus \{i,j\} }
}
I(g_{km}(\boldsymbol{\eta}_0)=1) K'  \left(\frac{v_{ij}-v_{km}}{h} \right)\frac{1}{h}=
\]
\[
\frac{-1}{h(L-1)}
\sum_{\substack{
\{k,m\} \in \\
\mathbb{P}\setminus \{i,j\} }}
I(g_{km}(\boldsymbol{\eta}_0)=1) K' \left(\frac{-(v_{ij}-v_{km})}{h} \right)\frac{-1}{h}=
\frac{\partial \hat{p}_{1,ij}(-\boldsymbol{\eta}|G(\boldsymbol{\eta}_0))}{\partial v_{ij}},
\]
a Newton-Raphson algorithm, when given a starting value close to the wrong optimum, updates the coefficients into the wrong directions and converges into this optimum. 

\section{Normalisation or Standardisation}

From \eqref{equalimplication1} it follows that if we can consistently estimate linking probabilities, then the fixed effect coefficients are (asymptotically) identified up to location and scale. The way the normalisation or standardization is carried out alters the interpretation of the estimates. 
In the following, I use $\Tilde{\boldsymbol{\eta}}(\boldsymbol{\eta})$ and $\Tilde{\eta}(\eta_i)$ to denote, respectively, a vector-valued function and a scalar function that transform a given coefficient vector or respectively an element of it, the result being that the transformed coefficient vector exhibits two specific features, such as constant entries or moments. Note that the function $\Tilde{\boldsymbol\eta}$ that achieves the normalisation or standardisation is, in general, specific to the respective input vector, e.g.\  $\Tilde{\boldsymbol{\eta}}_0(.)$ is the transformation function for the true coefficient vector. 

 As evident from $\eqref{problem}$, as long as the transformation is positive affine, the transformed model parameters $\Tilde{\boldsymbol{\eta}}(\boldsymbol{\eta}),F_{\tilde{u}(u)}$ are observationally equivalent to the original ones.

\[
    P(g_{ij}=1|w_{ij}' \boldsymbol{\eta}_0,F_{u_0})=
    P(g_{ij}=1|w_{ij}'
    \boldsymbol{\Tilde{\eta}_0}(\boldsymbol{\eta}_0), F_{\Tilde{u}_0(u_0)}) \hspace{.25cm} \forall \{i,j\} \in \mathbb{P}
    \] with $F_{\Tilde{u}_0(u_0)}$ being a positive  affine transformation of $F_{u_0}$, and 
\[    
     P(g_{ij}=1|w_{ij}' \boldsymbol{\eta}_1,F_{u_1})=
     P(g_{ij}=1|w_{ij}'
    \boldsymbol{\Tilde{\eta}_1}(\boldsymbol{\eta}_1),
    F_{\Tilde{u}_1(u_1)})
    \hspace{.25cm} \forall \{i,j\} \in \mathbb{P}
\] with $F_{\Tilde{u}_1(u_1)}$ being a positive  affine transformation of $F_{u_1}$,
such that
\begin{eqnarray}
    P(g_{ij}=1|w_{ij}' \boldsymbol{\Tilde{\eta}_0}(\boldsymbol{\eta}_0), F_{\Tilde{u}_0(u_0)}
    )=P(g_{ij}|w_{ij}'\boldsymbol{\Tilde{\eta}_1}(\boldsymbol{\eta}_1), F_{\Tilde{u}_1(u_1)}
    ) \hspace{.25cm} \forall \{i,j\} \in \mathbb{P}
    \nonumber \\    \Longrightarrow
\boldsymbol{\Tilde{\eta}_1}(\boldsymbol{\eta}_1)
=a\boldsymbol{\Tilde{\eta}_0}(\boldsymbol{\eta}_0)+b .
    \label{equalimplication2}
\end{eqnarray}
In the following, I show that if the transformation sets two elements (normalisation) or moments (standardisation) of the fixed effect coefficient vector to constant values, then, in addition to \eqref{equalimplication2}, $a=1$ and $b=0$, which in turn (due to observational equivalence) implies that $F_{\Tilde{u}_0(u_0)}=F_{\Tilde{u}_1(u_1)}$. The proposed normalisation has the advantage that it guarantees the correct order and solves numerical problems. 

As a consequence, the estimator is defined as
\[
\widetilde{
\hat{
\boldsymbol{\eta}} }(\hat{\boldsymbol{\eta}})=
min_{\boldsymbol{\eta}} 
\hspace{.1cm}
\hat{Q}(
\Tilde{\boldsymbol{\eta}}
(\boldsymbol{\eta})
)=
min_{\boldsymbol{\eta}}
\hspace{.1cm}
\hat{m}(\Tilde{\boldsymbol{\eta}}
(\boldsymbol{\eta}))'\hat{m}(\Tilde{\boldsymbol{\eta}}
(\boldsymbol{\eta})).
\]

\subsection{Normalisation}

Let the normalisation function be defined by $\underline{i} \in N: \Tilde{\eta}(\eta_{\underline{i}})=0$ and $\overline{i} \in N:\Tilde{\eta}(\eta_{\overline{i} })=1$, that is, given any coefficient vector, the fixed effects coefficient of individual $\underline{i}$ ($\overline{i}$) is set to zero (one). The transformation function, specific to the coefficient vector $\boldsymbol{\eta}$ is then defined as 
\begin{equation}
    \Tilde{\eta}(\eta_i)=
\frac{\eta_{i}-\eta_{\underline{i}}}{\eta_{\overline{i}}-\eta_{\underline{i}}}
\label{norm}
\end{equation}
and achieves that the transformed fixed effects vector features a zero and a one at the same entry (respectively, $\underline{i}$ and $\overline{i}$).

\begin{theorem}
    If the true model parameters $\boldsymbol{\eta}_0, F_{u_0}$, normalised using \eqref{norm} into $\boldsymbol{\Tilde{\eta}_0}(\boldsymbol{\eta}_0), F_{\Tilde{u}_0(u_0)}$ are observationally equivalent with $\boldsymbol{\eta}_1, F_{u,1}$, normalised using \eqref{norm} into $\boldsymbol{\Tilde{\eta}_1}(\boldsymbol{\eta}_1), F_{\Tilde{u}_1(u_1)}$, then this implies that $\Tilde{\boldsymbol{\eta}}_0(\boldsymbol{\eta}_0)=\Tilde{\boldsymbol{\eta}}_1(\boldsymbol{\eta}_1)$.
\end{theorem}

{\bf If $ \mathbf{
\Tilde{\eta}_0(\eta_{\underline{i},0})=\Tilde{\eta}_1(\eta_{\underline{i},1})=0}$, then $\mathbf{b=0}$.} 
\\ \noindent 
Given the continuity of the population distribution of the individual effects, there exist three individuals $i,j,k \in N \backslash \underline{i}$ such that 
\[ \Tilde{\eta}_0(\eta_{i,0})+\Tilde{\eta}_0(\eta_{j,0})=\Tilde{\eta}_0(\eta_{k,0}) 
\hspace{.25cm} \Tilde{\eta}_0(\eta_{i,0}) \neq 0, \Tilde{\eta}_0(\eta_{j,0}) \neq 0, \Tilde{\eta}_0(\eta_{k,0}) \neq 0.
\]
Observational equivalence implies that 
\[F_{\Tilde{u}_0(u_0)}\big(\Tilde{\eta}_0(\eta_{i,0})+\Tilde{\eta}_0(\eta_{j,0})\big)=
F_{\Tilde{u}_1(u_1)}\big(\Tilde{\eta}_1(\eta_{i,1})+\Tilde{\eta}_1(\eta_{j,1})\big)=\]
\[
F_{\Tilde{u}_1(u_1)}\Big(a\big(\Tilde{\eta}_0(\eta_{i,0})+\Tilde{\eta}_0(\eta_{j,0})\big)+2b\Big).\]
But also (using the probability of individual $k$ to link with individual $\underline{i}$)
\[ 
F_{\Tilde{u}_0(u_0)}\big(\Tilde{\eta}_0(\eta_{i,0})+\Tilde{\eta}_0(\eta_{j,0})\big)=
F_{\Tilde{u}_0(u_0)}\big(
\Tilde{\eta}_0(\eta_{k,0})\big)=\]
\[F_{\Tilde{u}_1(u_1)}\big(
\Tilde{\eta}_1(\eta_{k,1})\big)=
F_{\Tilde{u}_1(u_1)}\Big(a \big( \Tilde{\eta}_0(\eta_{k,0})\big)+b\Big) = 
F_{\Tilde{u}_1(u_1)}\Big(a \big( \Tilde{\eta}_0(\eta_{i,0})+\Tilde{\eta}_0(\eta_{j,0})\big)+b\Big). 
\]
Therefore 
\[
F_{\Tilde{u}_1(u_1)}\Big(a\big(\Tilde{\eta}_0(\eta_{i,0})+\Tilde{\eta}_0(\eta_{j,0})\big)+2b\Big)
=
F_{\Tilde{u}_1(u_1)}\Big(a \big( \Tilde{\eta}_0(\eta_{i,0})+\Tilde{\eta}_0(\eta_{j,0})\big)+b\Big),\]
thus
\[ F_{\Tilde{u}_1(u_1)}^{-1}\Big( F_{\Tilde{u}_1(u_1)}\Big(a\big(\Tilde{\eta}_0(\eta_{i,0})+\Tilde{\eta}(\eta_{j,0})\big)+2b\Big)\Big)=F_{\Tilde{u}_1(u_1)}^{-1}\Big(F_{\Tilde{u}_1(u_1)}\Big(a\big(\Tilde{\eta}_0(\eta_{i,0})+\Tilde{\eta}_0(\eta_{j,0})\big)+b\Big)\Big)\]
\[\Rightarrow
a\big(\Tilde{\eta}_0(\eta_{i,0})+\Tilde{\eta}_0(\eta_{j,0})\big)+2b=
a\big(\Tilde{\eta}_0(\eta_{i,0})+\Tilde{\eta}_0(\eta_{j,0})\big)+b,
\]
which implies that $b=0$. \\
\\ \noindent
{\bf If $\mathbf{\Tilde{\eta}_0(\eta_{\overline{i},0})=\Tilde{\eta}_1(\eta_{\overline{i},1})=1}$, then $\mathbf{b=1-a}$.} \\
\noindent 
Given the continuity of the population distribution of the individual effects, there exist three individuals $i,j,k \in N \backslash \overline{i}$ such that 
\[\Tilde{\eta}_0( \eta_{i,0})+\Tilde{\eta}_0(\eta_{j,0})= \Tilde{\eta}_0(\eta_{k,0})+1, 
\hspace{.25cm} \Tilde{\eta}_0(\eta_{i,0}) \neq 1, \Tilde{\eta}_0(\eta_{j,0}) \neq 1,
\Tilde{\eta}_0(\eta_{k,0}) \neq 1.
\]
Then observational equivalence implies that   
\[ F_{\Tilde{u}_0(u_0)}
\big(\Tilde{\eta}_0(\eta_{i,0})+\Tilde{\eta}_0(\eta_{j,0})\big)=
F_{\Tilde{u}_1(u_1)}\Big(\Tilde{\eta}_1(\eta_{i,1})+\Tilde{\eta}_1(\eta_{j,1})\Big)=\]
\[F_{\Tilde{u}_1(u_1)}\Big(a\big(\Tilde{\eta}(\eta_{i,0})+\Tilde{\eta}(\eta_{j,0})\big)+2b\Big).\]
But also (using the probability of individual $k$ to link with individual $\overline{i}$)
\[
 F_{\Tilde{u}_0(u_0)}
\big(\Tilde{\eta}_0(\eta_{i,0})+\Tilde{\eta}_0(\eta_{j,0})\big)=
F_{\Tilde{u}_0(u_0)}(\Tilde{\eta}_0(\eta_{k,0})+1)=\]
\[
F_{\Tilde{u}_1(u_1)}\Big(\Tilde{\eta}_1(\eta_{k,1})+1\Big)=
F_{\Tilde{u}_1(u_1)}\Big(a\Tilde{\eta}_0(\eta_{k,0})+b+1\Big)=\]
\[
F_{\Tilde{u}_1(u_1)}\Big(a\big(\Tilde{\eta}_0(\eta_{i,0})+\Tilde{\eta}(\eta_{j,0})-1\big)+b+1\Big)=
F_{\Tilde{u}_1(u_1)}\Big(a\big(\Tilde{\eta}_0(\eta_{i,0})+\Tilde{\eta}(\eta_{j,0})\big)-a+b+1\Big),
\]
consequently 
\[
F_{\Tilde{u}_1(u_1)}^{-1}\Big(
F_{\Tilde{u}_1(u_1)}\Big(a\big(\Tilde{\eta}(\eta_{i,0})+\Tilde{\eta}(\eta_{j,0})\big)+2b\Big)\Big)=
F_{\Tilde{u}_1(u_1)}^{-1}\Big(
F_{\Tilde{u}_1(u_1)}\Big(a\big(\Tilde{\eta}_0(\eta_{i,0})+\Tilde{\eta}(\eta_{j,0})\big)-a+b+1\Big)\Big)
\]
\[
\Rightarrow
2b=-a+b+1 \Rightarrow b=1-a.
\]

As outlined above, the kernel-based estimator estimates the linking probabilities such that $\hat{Q}(\boldsymbol{\eta}|G(\boldsymbol{\eta}_0))=\hat{Q}(\boldsymbol{-\eta}|G(\boldsymbol{\eta}_0))$. One of these two coefficient vectors implies the wrong order (namely one that violates the monotonicity assumption). The incorrectly ordered vector can be identified post estimation (as it will have to be in the case of standardised fixed effect coefficients), but an advantage of the normalisation is that a data-based choice of $\underline{i}$ and $\overline{i}$ will guarantee that the ordering in maintained by the normalisation function whenever it is correct and reversed whenever it is incorrect. 

\subsubsection{	The Degree-Based Mini-Max (DBMM) Normalisation}

I propose to normalise the parameter vector such that the individual with the smallest (largest) number of friends has a fixed effects coefficient of zero (one). Let $\underline{i}$ and $\overline{i}$ be defined by
\[
\underline{i},j \in N; d_{\underline{i} } \leq d_j \forall j \in N
\]
\[
\overline{i},j \in N; d_{\overline{i}} \geq d_j \forall j \in N
\]

In expectation, hence asymptotically, this would result in all other coefficients being normalised into the interval from zero to one. In a finite sample, the individual with the largest true fixed effect coefficient need not be the individual with the largest degree. This however, usually only causes minor issues (unless the error has a very substantial variance) and entail that some of the normalised true coefficients are larger than one or smaller than zero (if the individual with the largest (respectively smallest) number of friends is almost, but not exactly the individual with the largest (respectively smallest) coefficient). 
 
The normalised coefficient of individual $i$ can be interpreted as the fraction of the sample who have a fixed effects coefficient that is smaller than $i's$, or $i's$ relative popularity.

\subsubsection{Implementation}

I maximize $Q(\boldsymbol{\eta})$ under the constraint that the most unpopular and popular individuals feature fixed effect coefficients of, respectively, zero and one and that the other coefficients are all within this range. As the indices $\underline{i}$ and $\overline{i}$ are known from the data, we could be tempted to fix these values in the input vector and subsequently maximise over $N-2$ parameters, restricting them to be located in the unit interval. This will not work because we would neglect the gradient with respect to $\eta_{\underline{i}}$ and $\eta_{\overline{i}}$. Despite the fact that the normalised coefficients of $\underline{i}$ and $\overline{i}$ are known, yet the (unnormalised) fixed effects of these two individuals affect all other normalised coefficients as they appear in the normalisation equation. Altering either of those (unnormalised) coefficients would alter all normalised coefficients. 
Putting it differently, optimising over normalised parameters will not work, because each of them can vary as a consequence of variation in any of the three nonnormalised coefficients involved in the normalisation equation. Fixing $\eta_{\underline{i}}$ and $\eta_{\overline{i}}$ would effectively not make them part of the search routine, the potential result being that they are distinct point situated off the cloud of all other fixed effect coefficients.

I implement the degree-based mini-max normalisation by optimising over unrestricted coefficients and carrying out a normalisation inside the routine. We know that at the optimum (or close to it), carrying out this normalisation using $\eta_{\underline{i}}$ and $\eta_{\overline{i}}$ is numerically unproblematic; however, we do know that the Newton-Raphson algorithm will, at time, wander far off the peak, thereby potentially creating problems as $ \eta_{\overline{i}}-\eta_{\underline{i}}$ gets very small and then the normalised coefficients get very large, leading to numerical problems. There are potentially numerous ways to circumvent these numerical problems, but the easiest apparently is to carry out a minimax normalisation using the maximal and minimal coefficient (as opposed to the individuals with the minimal and maximal degree) inside the routine and then in the very last step (when we are sure to be close to the optimum), to carry out the degree-based minimax normalisation. This will effectively switch the order of all coefficients in the last step in case the algorithm converged into the wrong optimum. 

 Since the fixed effects' order is guaranteed to be correct, the individual with the highest fixed effect will, in expectation and asymptotically, have the largest degree. If our sample is composed such that $ min (\boldsymbol{\eta}_0) \neq \eta_{\underline{i},0}$ and/or $ max(\boldsymbol{\eta}_0) \neq \eta_{\overline{i},0}$, this means that there were some random draws that are overproportionally large or small, compared to the average properties of the DGP, or, putting it differently, that this difference between $ min (\boldsymbol{\eta}_0) $ and $ \eta_{\underline{i}}$ (respectively $ max (\boldsymbol{\eta}_0) $ and $ \eta_{\overline{i}}$) is caused by noise. As a consequence, some of the normalised true coefficients are slightly outside the unit interval. For the normalised estimates, in turn, the most popular (unpopular) individual will always also be the one with the largest (respectively smallest) fixed effects by construction. Normalised estimated coefficients can only be outside of the unit interval if $\underline{i}$ and/or $\overline{i}$ are not unique, that is, if there are several individuals that are the ost (un)popular. In this case, the routine will select one of tem, which might then not be the one with the largest (respectively smallest) estimated coefficient.  Note that this issue vanishes as the sample size increases. 

\subsection{Standardisation}

When the transformation function $\boldsymbol{\Tilde{\eta}}(.)$ results in a vector with constant moments (as opposed to entries), we refer to it as a standardisation. 

A popular choice is 
\[ \tilde{\eta}(\eta_i)= \frac{\eta_i-\overline{\eta}}{s_{\eta}}\]
\[
\overline{\eta}=
\frac{1}{N}\sum_{i=1}^N
\eta_i
\] 
\[s_{\eta}=
\left(
\frac{1}{N-1}\sum_{i=1}^N (\eta_i-\overline{\eta})^2
\right)^{1/2}
\]
resulting in 
\[\overline{ \Tilde{\boldsymbol{\eta}}(\boldsymbol{\eta})}=
 \frac{1}{N} \sum_i \tilde{\eta}(\eta_i)=0\]
\[ s_{\Tilde{\boldsymbol{\eta}}(\boldsymbol{\eta})}=
\left(
\frac{1}{N-1} \sum_i \left(\tilde{\eta}(\eta_i)\right)^2
\right)^{1/2}
=1.\]

{\bf If $\mathbf{
\overline{ \Tilde{\boldsymbol{\eta}_0}(\boldsymbol{\eta}_0)}=
\overline{ \Tilde{\boldsymbol{\eta}_1}(\boldsymbol{\eta}_1)}=
0
}$, then $\mathbf{b=0}$}\\
\noindent
Since 
\[ N^{-1} \sum_{i=1}^N \Tilde{\eta}_0(\eta_{i,0})=0
=N^{-1} \sum_{i=1}^N \Tilde{\eta}_1(\eta_{i,1})=
 N^{-1} \sum_{i=1}^N \left( a \Tilde{\eta}_0(\eta_{i,0})+b\right)=\]
\[a N^{-1} \sum_i \tilde{\eta}_i(\eta_{i,0})+b=b=0,
\]
hence $b=0$. \\

{\bf If $\mathbf{
s_{ \Tilde{\boldsymbol{\eta}_0}(\boldsymbol{\eta}_0)}=
s_{ \Tilde{\boldsymbol{\eta}_1}(\boldsymbol{\eta}_1)}=
1
}$, then $\mathbf{|a|=1}$}\\
However, 
\[ N^{-1} \sum_{i=1}^N \left(\Tilde{\eta}_0(\eta_{i,0})\right)^2=1
=N^{-1} \sum_{i=1}^N \left(\Tilde{\eta}_1(\eta_{i,1})\right)^2=
a^2 N^{-1} \sum_i \left(\Tilde{\eta}_0(\eta_{i,0})\right)^2=a^2=1
\] only implies that $a^2=1$, such that the sign of the standardized coefficients is only identified by the monotonicity assumption. Because the estimator can not implement this assumption, the standardisation will not, in general identify the order of the fixed effect coefficients correctly. This problem can however be solved post estimation: if we observe that the order of the estimated fixed effects does not correspond to the order of the observed degrees, we multiply all estimates with minus one, thereby intentionally flipping their order.

 Note that an alternative would be to use an uneven moment instead of the standard deviation in the normalisation. This would identify the order, yet only be a valuable alternative if we know that in the true fixed effect distribution in the population, a particular third moment (for instance the skewness) is not (close to) zero. In absence of this knowledge, this is not a feasible alternative. Since there exists no linear transformation with which a skew can be achieved, it is also not possible for us to guarantee a skewed fixed effect coefficient distribution. 



\section{Consistency and Asymptotic Normality}

\subsection{Consistency}

Both the estimated error CDF and the estimated error density converge to the true CDF (respectively density) uniformly in $\boldsymbol{\eta}$ (for a detailed proof, kindly refer to the Appendix). Let $\mathbb{N}$ be a compact set $\in \mathbb{R}^N$ (chosen large enough so I am confident to cover $\mathbb{N}_0$) to which I restrict my search of candidates for $\boldsymbol{\eta}$, then 
\newline

\noindent
{\bf Lemma 1: }
 $\forall \boldsymbol{\eta} \in \mathbb{N}  \hspace{.25cm} \widehat{F}_{u}(\boldsymbol{\eta},v_{ij})-
F_{u_0}(v_{ij})=O(N^{-1/2}h^{-1} \underline{p}^{-1})$.
\newline
{\bf Lemma 2: }
 $\forall \boldsymbol{\eta} \in \mathbb{N}  \hspace{.25cm} \widehat{f}_{u}(\boldsymbol{\eta},v_{ij})-
f_{u_0}(v_{ij})=O(N^{-1/2}h^{-2} \underline{p}^{-1})$. 
\newline 

If I hold the coefficient vector $\boldsymbol{\eta}$ at any constant value, I can treat the $L$ pair specific linking probabilities $F_u(w_{ij}'\boldsymbol{\eta})$ as random variables (whose true values are $F_{u_0}(w_{ij}'\boldsymbol{\eta})$ and which are estimated by $\widehat{F_u}(w_{ij}'\boldsymbol{\eta})$). I can then carry out a Taylor expansion of the approximate objective function around the true objective function (both evaluated at $\boldsymbol{\eta})$). 
I use $F_{u_0}(\boldsymbol{\eta})$ to denote the vector stacking all $F_{u_0}(w_{ij}'\boldsymbol{\eta})$. 
This shows that for any coefficient vector, the difference between the approximate and the true objective functions vanishes asymptotically. For details, kindly refer to the Appendix. 

\[Q(\hat{F}_u
(\boldsymbol{\eta}))=
Q(F_{u_0}
(\boldsymbol{\eta}))+\]
\[\frac{2}{N-1}
\sum_{
\substack{\{i,j\} \\ \in \mathbb{P}}
} \Bigg(
-
\underbrace{\left(m_i(F_{u_0}(\boldsymbol{\eta}))+m_j(F_{u_0}(\boldsymbol{\eta}))\right)}_{
\substack{O(1)\\
sample \hspace{.1cm} averages \hspace{.1cm}  of \hspace{.1cm}  probabilities
}
}
\underbrace{\left(\widehat{F_u}(\boldsymbol{\eta},w_{ij}'\boldsymbol{\eta})-F_{u_0}(w_{ij}'\boldsymbol{\eta})\right)}_{
\substack{O(N(N-1)/2)^{-5/14})\\
by \hspace{.1cm} Lemma \hspace{.1cm}  1 
}}+\]
\[ 
\frac{1}{(N-1)}\Big(
\underbrace{
\widehat{F_u}(
\boldsymbol{\eta},
w_{ij}'\boldsymbol{\eta})+F_{u_0}(w_{ij}'\boldsymbol{\eta})
}_{
\substack{O(N(N-1)/2)^{-5/14})\\
by \hspace{.1cm} Lemma \hspace{.1cm}  1 
}
}\Big)^2\Bigg).
\]
Since a sum of asymptotically bounded terms is also asymptotically bounded (the bound being the largest among the summands), thus 
\[Q(\hat{F}_u
(\boldsymbol{\eta}))=
Q(F_{u_0}
(\boldsymbol{\eta}))
+\frac{2}{N-1} O(N(N-1)/2)^{-5/14})=
Q(F_{u_0}
(\boldsymbol{\eta}))+o((N-1)^{-1}),
\]
from which we can see that the approximate and the true objective function, evaluated at any constant coefficient vector $\boldsymbol{\eta}$, are asymptotically equivalent. Because the convergence results in Lemma 1 and 2 are uniform, consequently this asymptotical equivalence holds true not only at the true value, but also at the estimate for a given sample size (and all the values the estimate can take as it converges towards its expected value). In turn, the asyumptotical equivalence of the approximate and the true objective function, together with the fact that the latter is maximised at the true coefficient value, imply consistency of the estimated (normalised) fixed effects. 

\subsection{Asymptotic Normality}

For any parameter vector $\boldsymbol{\eta}$, I use 
\[ m(\boldsymbol{\eta})=
m(F_u(\boldsymbol{\eta}))
\] and 
\[M(\boldsymbol{\eta})=
\frac{\partial m(\boldsymbol{\eta})}{\partial \boldsymbol{\eta}}
\]
to denote the vector of moment conditions and the matrix of derivatives of the moment conditions with respect to the parameters for any given choice of $F_u$.  Therefore, I use
\[ \hat{m}(\boldsymbol{\eta})=
m(\hat{F}_u(\boldsymbol{\eta}))
\] and 
\[\hat{M}(\boldsymbol{\eta})=
\frac{\partial \hat{m}(\boldsymbol{\eta})}{\partial \boldsymbol{\eta}}
\] 
when using the kernel estimates of $F_u$ and $f_u$ 
and I use 
\[ m_0(\boldsymbol{\eta})=
m(F_{u_0}(\boldsymbol{\eta}))
\] and 
\[M_0(\boldsymbol{\eta})=
\frac{\partial m_0(\boldsymbol{\eta})}{\partial \boldsymbol{\eta}}
\] when using the error distribution of the true DGP, 
where $\hat{M}(\boldsymbol{\eta})$ and $M_0(\boldsymbol{\eta})$ are matrices of dimension $N\times N$.
In the following, I work with the normalised estimates $\widetilde{
\hat{
\boldsymbol{\eta}} }(\hat{\boldsymbol{\eta}})$ and the normalised true coefficients $\Tilde{\boldsymbol{\eta}}_0(
\boldsymbol{\eta}_0)$. 

With a second order Taylor expansion, I obtain 
\[
\widetilde{
\hat{
\boldsymbol{\eta}} }(\hat{\boldsymbol{\eta}})-
\Tilde{\boldsymbol{\eta}}_0(\boldsymbol{\eta}_0)
=
\left(\hat{M}(\widetilde{
\hat{
\boldsymbol{\eta}} }(\hat{\boldsymbol{\eta}}))'
\hat{M}(\Bar{\boldsymbol{\eta}})\right)^{-1} 
\hat{M}(\widetilde{
\hat{
\boldsymbol{\eta}} }(\hat{\boldsymbol{\eta}}))'
\hat{m}(
\Tilde{\boldsymbol{\eta}}_0(
\boldsymbol{\eta}_0)),\]
with $\Bar{\boldsymbol{\eta}}$ being a mean value between $\widetilde{
\hat{
\boldsymbol{\eta}} }(\hat{\boldsymbol{\eta}})$ and $\Tilde{\boldsymbol{\eta}}_0(
\boldsymbol{\eta}_0)$. 
\newline 

\noindent
{\bf Lemma 3: } $ \forall \boldsymbol{\eta} \in \mathbb{N }
\hspace{.2cm}
\sqrt{N-1}
\hat{m}(\boldsymbol{\eta}) =
\sqrt{N-1}
m_0(\boldsymbol{\eta})+o(1).$
\\
\noindent
{\bf Lemma 4: } $ \forall \boldsymbol{\eta} \in \mathbb{N }
\hspace{.2cm}
    \hat{M}(\boldsymbol{\eta})=
M_0(\boldsymbol{\eta})+o(1).
$
\newline 

Consistency in turn implies that 
\[ 
M_0(\widetilde{
\hat{
\boldsymbol{\eta}} }(\hat{\boldsymbol{\eta}}))
\overset{p}{\rightarrow}
M_0(\Tilde{
\boldsymbol{\eta}}_0 (\boldsymbol{\eta}_0))\]
and 
\[
M_0(\Bar{\boldsymbol{\eta}}) \overset{p}{\rightarrow}
M_0(\Tilde{
\boldsymbol{\eta}}_0 (\boldsymbol{\eta}_0)).
\]
Together, this 
implies that 
\[\sqrt{N-1}\left(
\widetilde{
\hat{
\boldsymbol{\eta}} }(\hat{\boldsymbol{\eta}})-
\Tilde{\boldsymbol{\eta}}_0(\boldsymbol{\eta}_0)\right)
\overset{p}{\rightarrow}\]
\[
\left(M_0(\Tilde{
\boldsymbol{\eta}}_0 (\boldsymbol{\eta}_0))'
M_0(\Tilde{
\boldsymbol{\eta}}_0 (\boldsymbol{\eta}_0))\right)^{-1} 
M_0(\Tilde{
\boldsymbol{\eta}}_0(\boldsymbol{\eta}_0))'
\sqrt{N-1}m_0(
\Tilde{\boldsymbol{\eta}}_0(
\boldsymbol{\eta}_0)),\]
such that the normalised estimates converge to the normalised true coefficients at the parametric rate and are asymptotically normally distributed by standard GMM asymptotic theory, that is 
\[
\sqrt{N-1}
\left(
\widetilde{
\hat{
\boldsymbol{\eta}} }(\hat{\boldsymbol{\eta}})-
\Tilde{\boldsymbol{\eta}}_0(\boldsymbol{\eta}_0)\right)
\overset{d}{\rightarrow}
N \left( 0, V_0\right)
\]
with \[
V_0=
\left(M_0(\Tilde{\boldsymbol{\eta}}_0(\boldsymbol{\eta}_0))'M_0(\Tilde{\boldsymbol{\eta}}_0(\boldsymbol{\eta}_0))\right)^{-1}
M_0(\Tilde{\boldsymbol{\eta}}_0(\boldsymbol{\eta}_0))' \Omega_0 
M_0(\Tilde{\boldsymbol{\eta}}_0(\boldsymbol{\eta}_0))
\left(M_0(\Tilde{\boldsymbol{\eta}}_0(\boldsymbol{\eta}_0))'M_0(\Tilde{\boldsymbol{\eta}}_0(\boldsymbol{\eta}_0))\right)^{-1}
\]
and 
\[
\Omega_0= E[m_0(\Tilde{\boldsymbol{\eta}}_0(\boldsymbol{\eta}_0))
m_0(\Tilde{\boldsymbol{\eta}}_0(\boldsymbol{\eta}_0))'].
\]
From this, I conclude that the cost of semiparametric estimation are purely computational as the estimator converges at the parametric rate and converges to the same asymptotic distribution as if the true error density was known.

I conduct a Monte Carlo experiment to verify the results stated above. Given a sample size of $100$,  
for $b=1,...,192$
I simulate a sample $\boldsymbol{\eta}_{0,b}$ (uniform draws from the interval $[-1,3]$ and use it to simulate the links $g_{b,ij}$ using standard logistic errors. I estimate the normalised coefficients $\widetilde{
\hat{
\boldsymbol{\eta}} }(\hat{\boldsymbol{\eta}})_b$ (the number of iterations being limited to 500) 
and compute the estimated variance 
\[
\hat{V}_b=
\left(\hat{M}(\widetilde{
\hat{
\boldsymbol{\eta}} }(\hat{\boldsymbol{\eta}})_b)'
\hat{M}(\widetilde{
\hat{
\boldsymbol{\eta}} }(\hat{\boldsymbol{\eta}})_b)\right)^{-1}
\hat{M}(\widetilde{
\hat{
\boldsymbol{\eta}} }(\hat{\boldsymbol{\eta}})_b)' \widehat{\Omega}_b
\hat{M}(\widetilde{
\hat{
\boldsymbol{\eta}} }(\hat{\boldsymbol{\eta}})_b)
\left(\hat{M}(\widetilde{
\hat{
\boldsymbol{\eta}} }(\hat{\boldsymbol{\eta}})_b)'\hat{M}(\widetilde{
\hat{
\boldsymbol{\eta}} }(\hat{\boldsymbol{\eta}})_b)\right)^{-1}
\]
with
\[
\widehat{\Omega}_{b,ij}=
\frac{1}{N-1}
\sum_{k=1}^N
\left( g_{b,ik}-
\widehat{F_u}(\widetilde{\hat{\boldsymbol{\eta}}}(\hat{\boldsymbol{\eta}})_b,w_{ik}'
\widetilde{\hat{\boldsymbol{\eta}}}(\hat{\boldsymbol{\eta}})_b\right)
\left( g_{b,jk}-
\widehat{F_u}(\widetilde{\hat{\boldsymbol{\eta}}}(\hat{\boldsymbol{\eta}})_b,w_{jk}'
\widetilde{\hat{\boldsymbol{\eta}}}(\hat{\boldsymbol{\eta}})_b\right).
\]
I select an arbitrary individual (here, the first) and compute
\[ \phi_b=
\sqrt{99} \frac{
\left( 
\widetilde{\hat{\boldsymbol{\eta}}}(\hat{\boldsymbol{\eta}})_{1,b}-
\tilde{\boldsymbol{\eta}}_{0}(\boldsymbol{\eta}_0)_{1,b}\right)
}{\sqrt{\hat{V}_{11},b}}
\]
which, according to the above, should asymptotically be standard normally distributed. 
In \ref{norm}, I plot the histogram of $\phi_b$. I choose to plot the probabilities (as opposed to the frequencies) on the vertical axis, that is, the interval $c$ is of height 
\[
\frac{\sum_b \mathbb{I}\left(\phi_b \in c \right)}{192 \mbox{width}_c},
\]
which is the relative frequency divided by the width of the interval. I use a kernel density estimator to estimate the density of $\phi_b$ and for comparison, I also plot the standard normal density. 
\begin{figure}[H]
    \centering
\caption{Histogram of $\rho_b$ for three bin sizes, estimated density and standard normal density}
 \begin{subfigure}[b]{\textwidth} \caption{width=1cm}
\includegraphics[width=\textwidth]{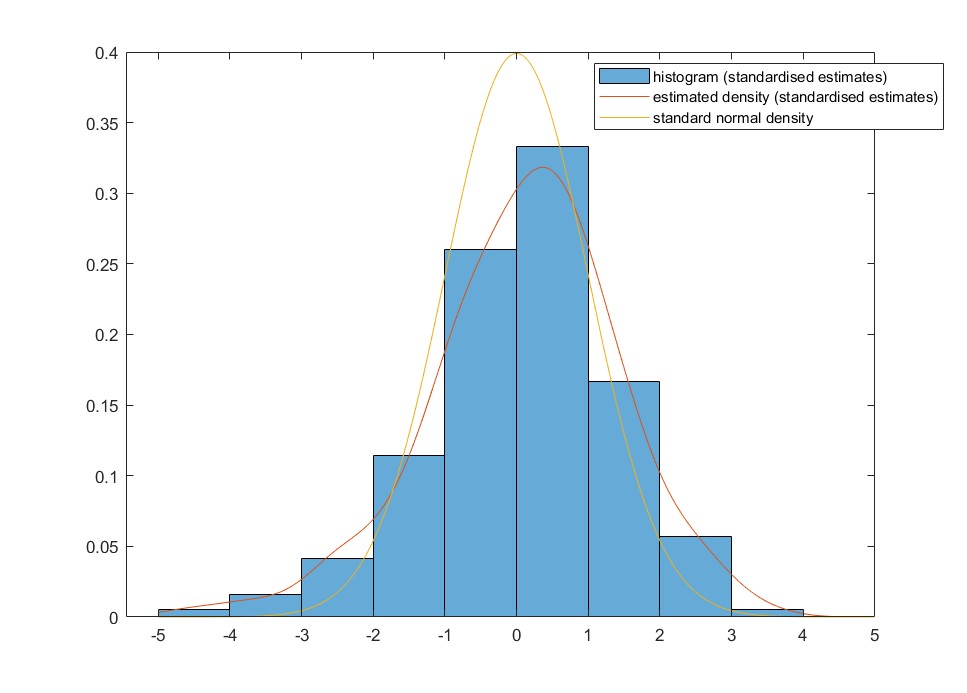}
\end{subfigure}
 \begin{subfigure}[b]{\textwidth}\caption{width=.2cm}
\includegraphics[width=\textwidth]{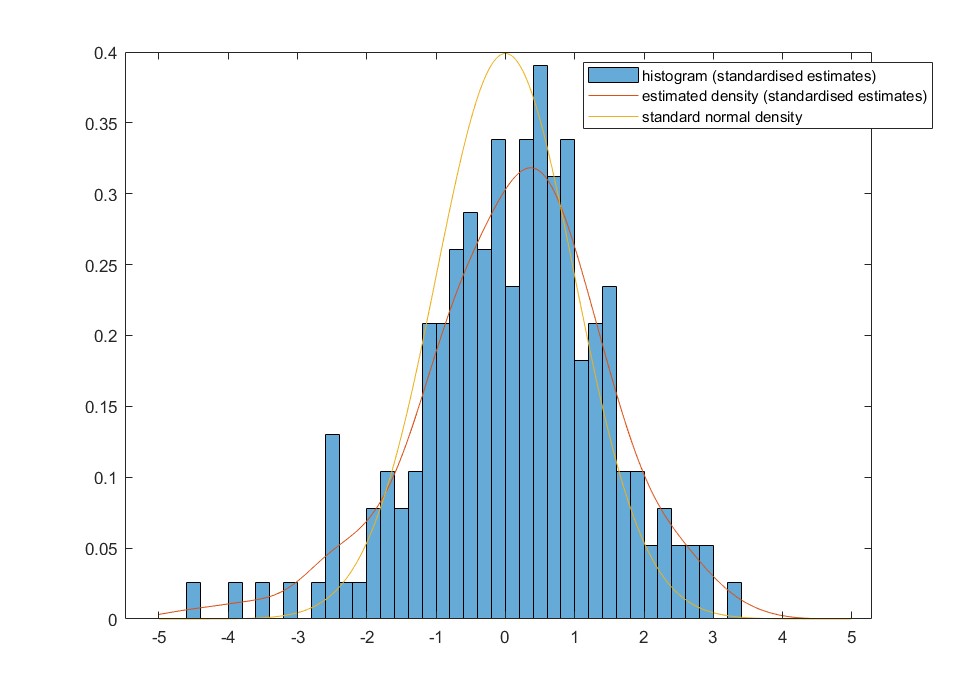}
\end{subfigure}
\end{figure}
\begin{figure}[H]\ContinuedFloat
    \centering
 \begin{subfigure}[b]{\textwidth}\caption{width=.1cm}
\includegraphics[width=\textwidth]{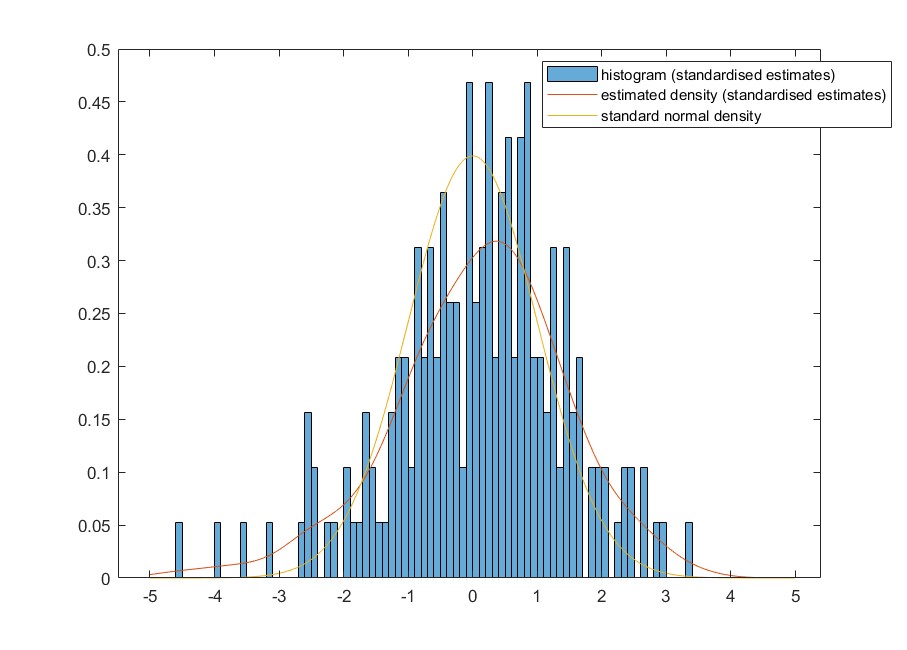}
\end{subfigure}\label{norm}
\end{figure}
Indeed, $\phi_b$ has an empirical distributed function that is bell-shaped, roughy symmetric around zero and resembles the standard normal distribution, despite a relaltively small sample size ($N=100$) and number of simulations ($B=192$). 

\begin{figure}[H]
    \centering
    \begin{subfigure}[b]{.45\textwidth} \caption{}
    \label{unadjusted}
\includegraphics[width=\textwidth]{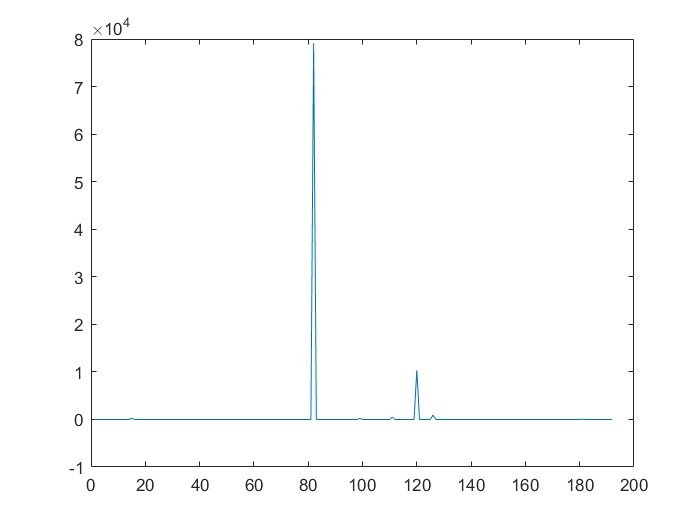}
    \end{subfigure}
     \begin{subfigure}[b]{.45\textwidth}
     \caption{} \label{adjusted_1}   
\includegraphics[width=\textwidth]{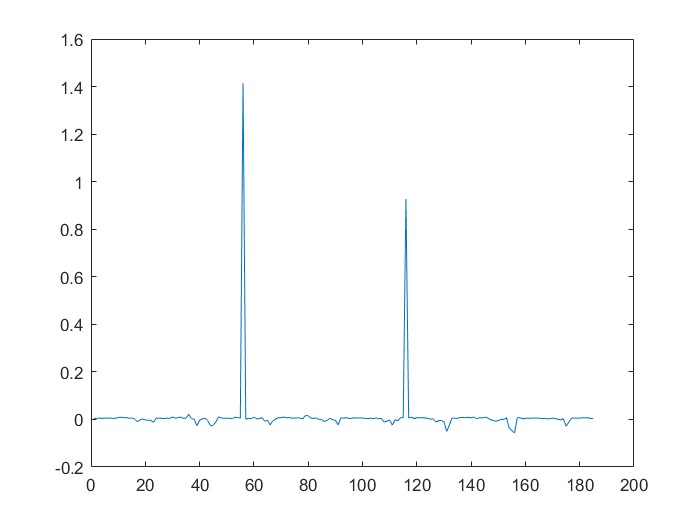}
    \end{subfigure}
     \begin{subfigure}[b]{.45\textwidth}
          \caption{} \label{adjusted_2}
\includegraphics[width=\textwidth]{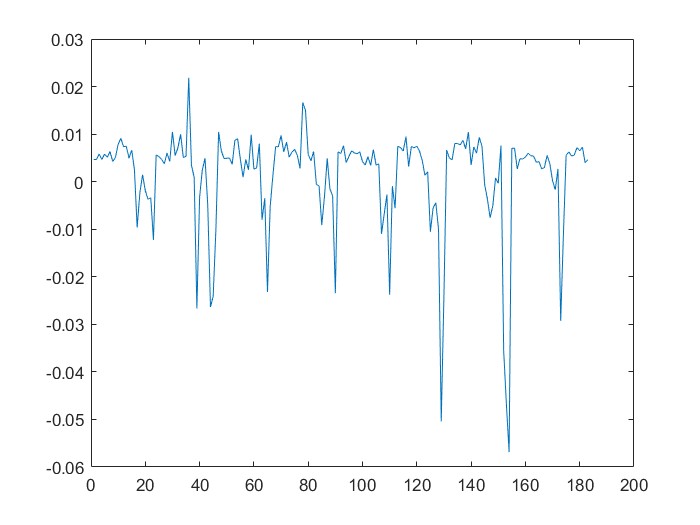}
    \end{subfigure}
\end{figure}

I use $D_b=V_{0,b}-\hat{V}_{b}$ to denote the matrix differences between the estimated and the asymptotic variance (the latter, being computed with the true normalised coefficients and the (accordingly transformed) logistic error distribution). 
Figure \ref{unadjusted} plots $D_{11,b} \forall b=1,...,B$. There are two potential reasons for extreme outliers, the first being that the normalised coefficients become very large as the difference in the coefficients between the most popular and the most unpopular individual is approximately zero. The estimated variance can also be inflated due to numerical issues when inverting the matrix of derivatives of the moment conditions. 
Both issues are likely to be resolved by an increase in the sample size and/or the number of iterations. 
that can be resolved by increasing the number of iterations.
Figure \ref{adjusted_1} excludes the seven cases for which the difference in variances exceeds two and Figure \ref{adjusted_2} excludes the two additional cases for which the difference is between 2 and 0.8. Excluding these extreme outliers, the estimated and asymptotic variance are indeed approximately equal. 

\section{Monte Carlo Results - DBMM-Normalisation Versus Standardisation}

In all Monte Carlo studies, the 
sample size is 100, the maximal number of iterations is 5000. 
For comparison of the two transformation functions, the true error has a logistic distribution. 
The estimates are compared to the true values (used to generate the network) after both were normalised or standardised as outlined above. 

\subsection{General Case}

\noindent
In general, it does not matter whether we standardise or normalise the coefficients, as long as we do (in case of the standardisation) correct for a potential erroneous ordering. The minimax normalisation features two distinct advantages namely 
\begin{itemize}
    \item the correct order is guaranteed and 
    \item no trimming is necessary.
\end{itemize}
Both issues are of non negligible importance as the standardised estimates do regularly converge into the wrong optimum (in which the order is reversed) and further are oftentimes subject to trimming. In the following, we use two cases to exemplify that the minimax normalisation can be advantageous. 

\subsection{Little Variation}

\noindent Networks with very little variation in the characteristics of the individuals involved, translating into little variation in the observed degree distribution are plausible to arise in socioeconomic settings. For example, it is typical especially in countries with a lot of industrial production that many micro or small scale businesses service pre-products to a larger production facility. These small businesses do not typically have an innovative business model or expert know-how, are highly informal and homogeneous. They are interrelated with one another through personal and business relationships that, given the informal setting, are often interwined and the resulting network is likely to feature little degree variation \\
\noindent Another examples could be co-authorship among junior academics. If we do not consider links to more senior researchers and focus exclusively on the links among researchers at the beginning of their career, the data is likely to exhibit a very homogeneous degree distribution caused by little variation in individual characteristics. Due to the lack of time of juniors to develop further any personal strengths that will, at a later stage of their career, distinguish them from others, juniors can be perceived as relatively homogeneous both in terms of their (inter-junior) network as well as their skills.  

In both settings, little variation in the degrees and the fixed effect coefficients are likely to be an issue. 

When there is little variation in the true fixed effect coefficients (translating into little variation in the degree distribution), the estimated fixed effect coefficients will deviate little from their mean. 
As the standard deviation becomes small, the standardised coefficients become large in absolute value. 
Since the estimates of the derivatives of the linking probabilities feature kernel estimates of the probability to observe a given sum of the (two link specific) fixed effect coefficients in the denominator (that is, an estimate of $f_v(v_{ij})$), whenever this estimate gets arbitrarily close to zero for even just one link the matrix $\hat{M}(\boldsymbol{\eta})'\hat{M}(\boldsymbol{\eta})$ can become close to being singular (or even feature a non-numerical entry). This in turn happens when a given $v_{ij}$ is at a very large distance from most $v_{km}, \{km\}\neq \{ij\}$. It is thus sufficient that a particular individual coefficient, by the aforementioned inflation through the division by a standard deviation close to zero, becomes very large relative to the others. 
Even in absence of the aforementioned invertibility problem, if an individual has a very large (or small) individual effect, then her moment condition will be very uninformative, as the error CDF will be insensitive to any further change in its input once the latter exceeds three (hence when index differences divided by the bandwidth exceed three for the majority of links the individual is involved in). This leads to a loss of information and, in case that more than two individuals are concerned, effectively under-identification. 

An attempt to resolve this would be to trim the fixed effects coefficients, e.g.\ to choose to restrict them to be, for instance, smaller than four in absolute value. Trimming however entails a substantial loss of information as the individuals beyond the borderline essentially only contribute little information (namely that their fixed effects coefficient is larger than the trimming threshold). It is unsurprising that this strategy performs poorly in the case of little variation in the fixed effects, as almost all individual are trimmed. The DBMM normalisation does not suffer from this problem because the difference between the coefficient of the most popular and the most unpopular individual does not become arbitrarily small when the degrees lack variation. As such, the estimated derivatives are well-behaved. Naturally, the estimation is rather poor, given the lack of variation in the data, yet, is is feasible and insightful.

 In the Monte Carlo study, the fixed effects coefficients to generate the network were drawn from a uniform distribution on the interval -.5 to .5. 
\begin{figure}[H]
    \centering
        \caption{Degree Distribution}
    \includegraphics{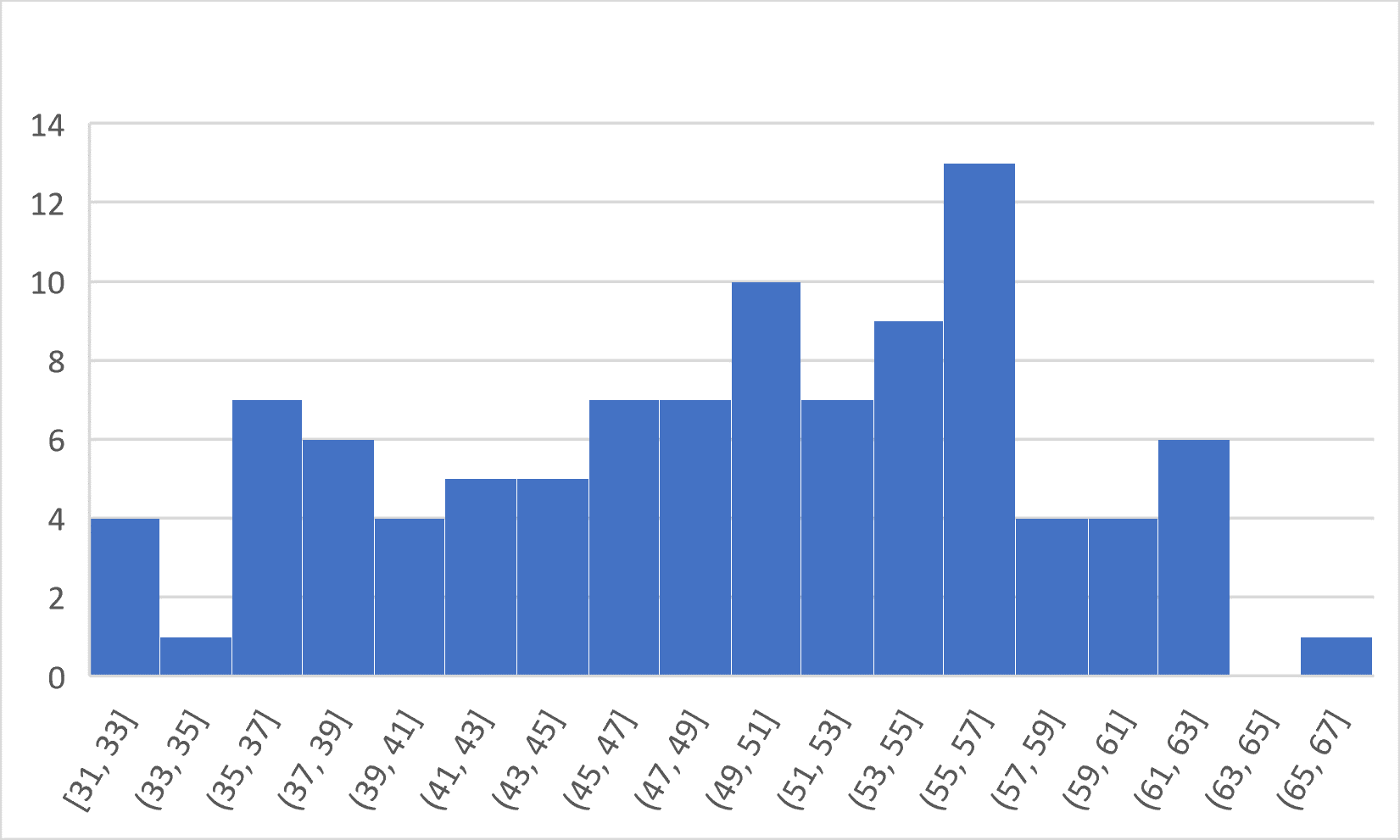}
    \label{fig:enter-label}
\end{figure}

\begin{figure}[H]
    \centering
        \caption{Normalised True Fixed Effect Coefficients (Horizontal Axis) Against Normalised Estimates (Vertical Axis), Degree-Based Minimax Normalisation }
    \includegraphics{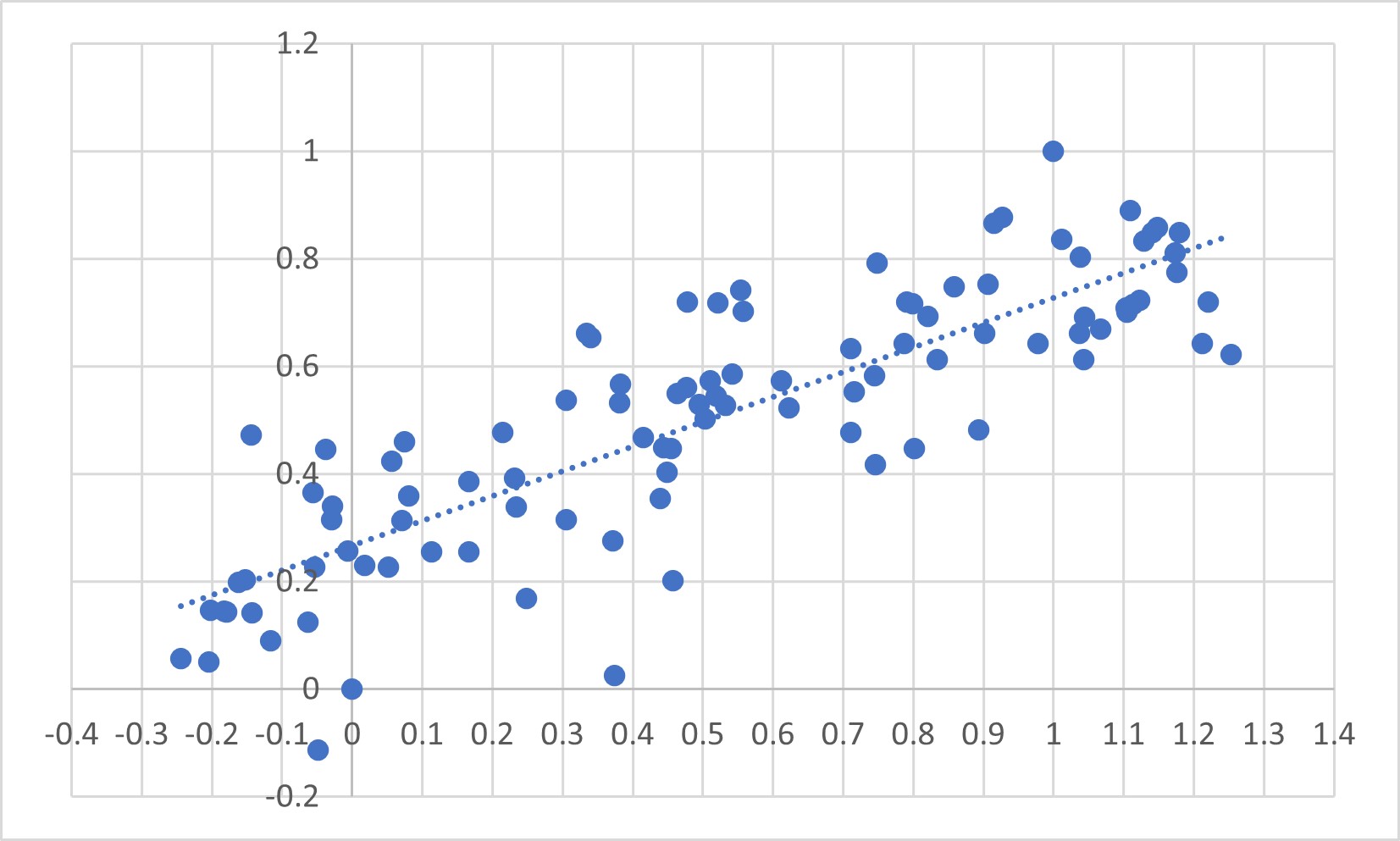}
    \label{fig:enter-label}
\end{figure}

Due to the lack of variation in the fixed effects coefficients, most individuals have a similar degree. We take notice that the most (un) popular individuals are by far not the ones with the smallest (respectively largest) true fixed effects coefficient. 
Accordingly, many of the true coefficients are outside of the unit interval after the normalisation\footnote{This fact would disappear with an increase in the sample size: by monotonicity, the individual with the largest (smallest) fixed effects coefficient will be the most popular (unpopular) in expectation, thus asymptotically, when coefficients are normalised with the DBMM normalisation, they all lie in the unit interval.}. The regression line illustrating the relationship between the two sets of coefficients is both shifted and too flat. Nonetheless, the performance of the estimator is remarkable, given the uninformative data.
We abstain from plotting the results for the estimation using standardisation because the latter either immediately runs into numerical issues (no trimming) or trims all individuals, the result being that the estimates carry no information.

\begin{figure}[H]
    \centering
    \caption{True Coefficients (Horizontal Axis) Against Logit Estimates (Little Variation In The Fixed Effects, Errors Are Truly Logistic) }
    \includegraphics[width=\textwidth]{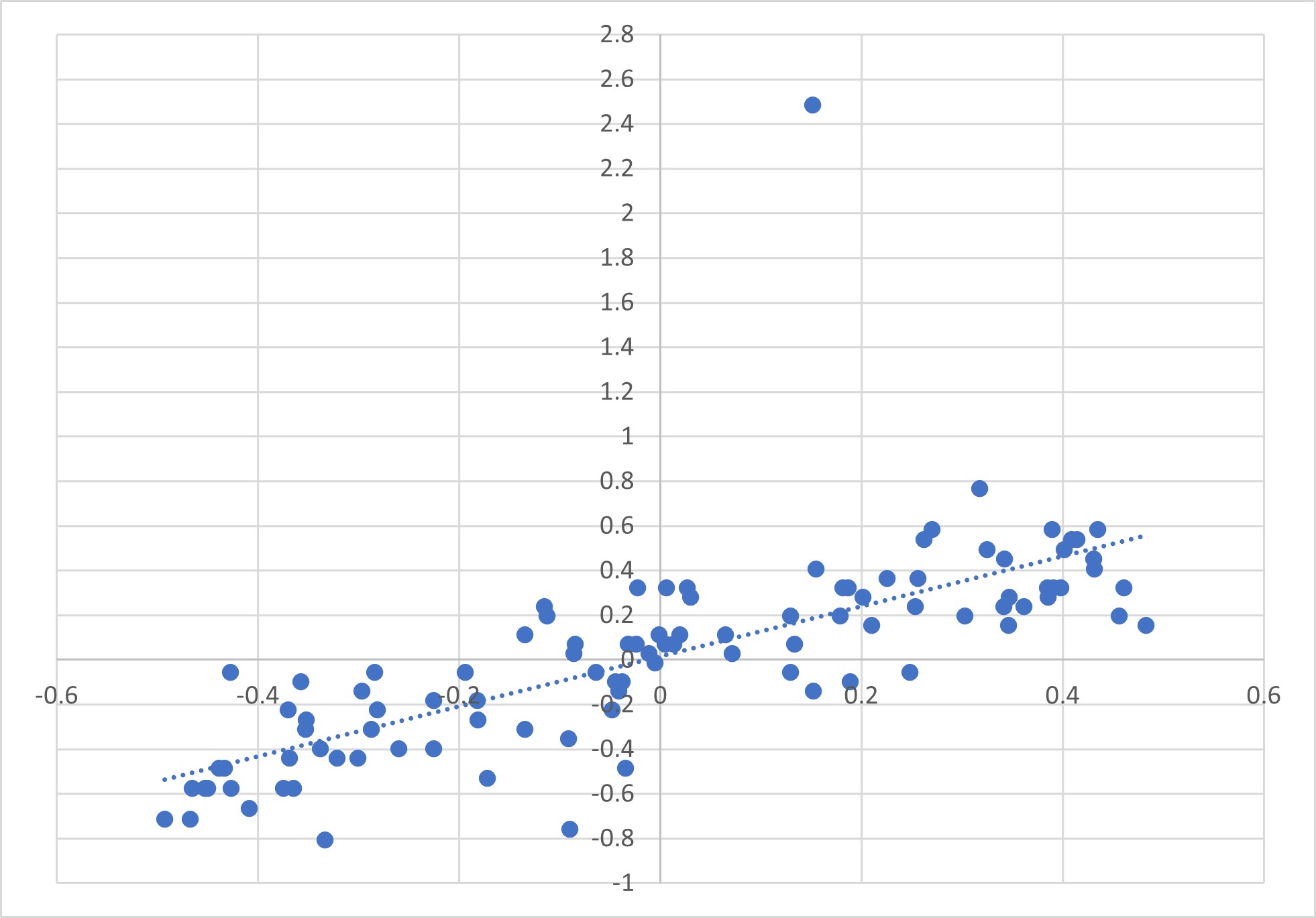}
\end{figure}

\noindent For comparison, we also plot the logit estimator. Knowing that the true errors here are indeed logistically distributed, any deviation from a perfect fit should thus be induced by noise or numerical issues. The latter arise for the logit estimator because the little variation in both degrees and fixed effect coefficients can also result in numerical problems when the Hessian is inverted.

\subsection{Clustering}

\noindent 
To investigate the impact of clustering, one cluster of 90 fixed effect coefficients was drawn from the uniform distribution on the interval $-1,1$, and the remaining 10 fixed effect coefficients were set at 2.5, 2.4, 2.3, 2.7, 2.6, 3, 3.5, 2.4, 3.1, 2.8. For the standardisation, trimming was implemented at -4 and 4, in order to prevent numerical issues. 
\begin{figure}[H]
    \centering
    \caption{Degree Distribution}  \includegraphics{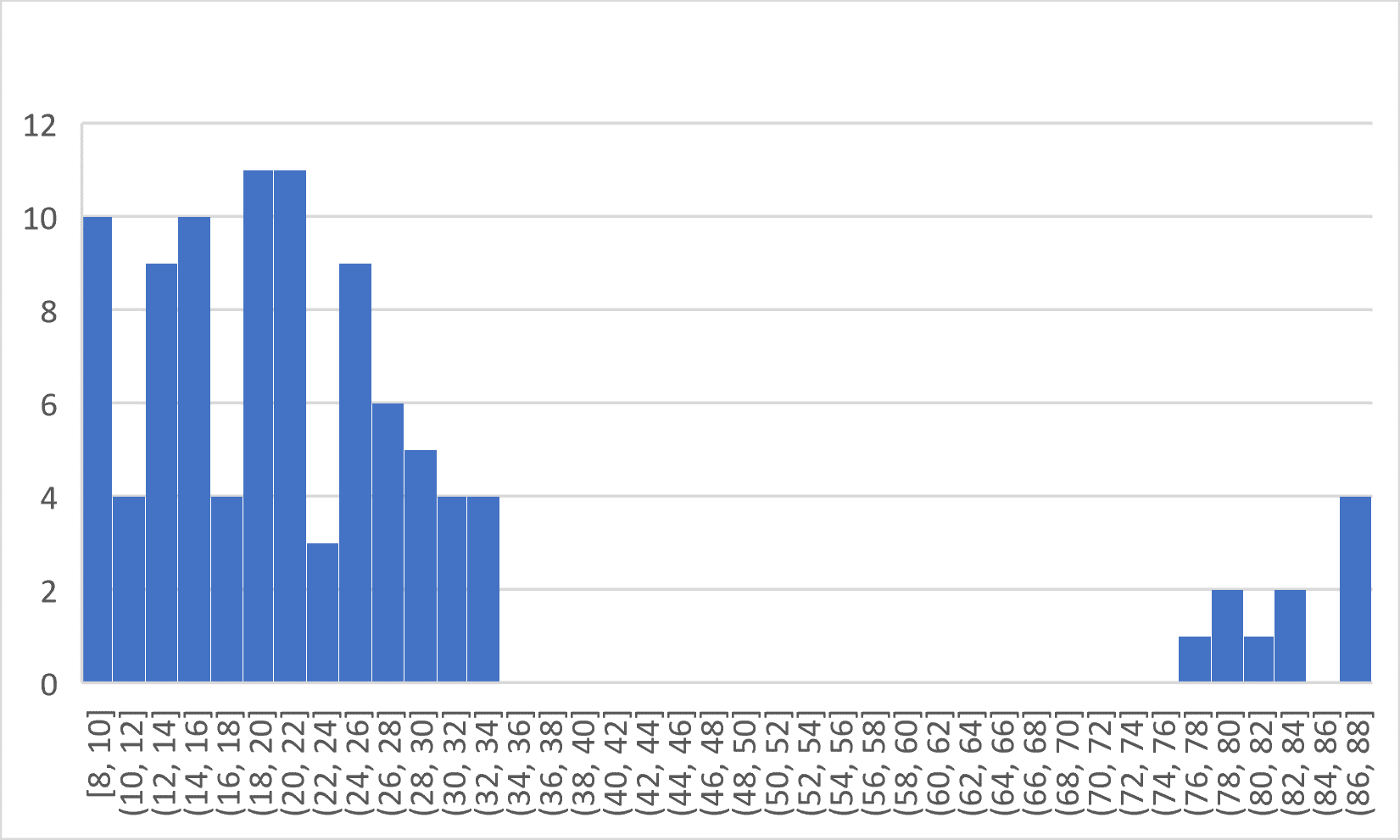}
    \label{fig:enter-label}
\end{figure}
\noindent
Unsurprisingly, we can also detect two clusters in the degree distribution. The estimation results show that in this case, the minimax transformation performs substantially better than the standardisation (the slope of the line connecting the transformed true and estimated coefficients is 0.8416 for the minimax normalisation and 0.5309 for the standardisation).
\begin{figure}[H]
    \centering
  \caption{Normalised True Fixed Effect Coefficients (Horizontal Axis) Against Normalised Estimates (Vertical Axis), Degree-Based Minimax Normalisation }  \includegraphics{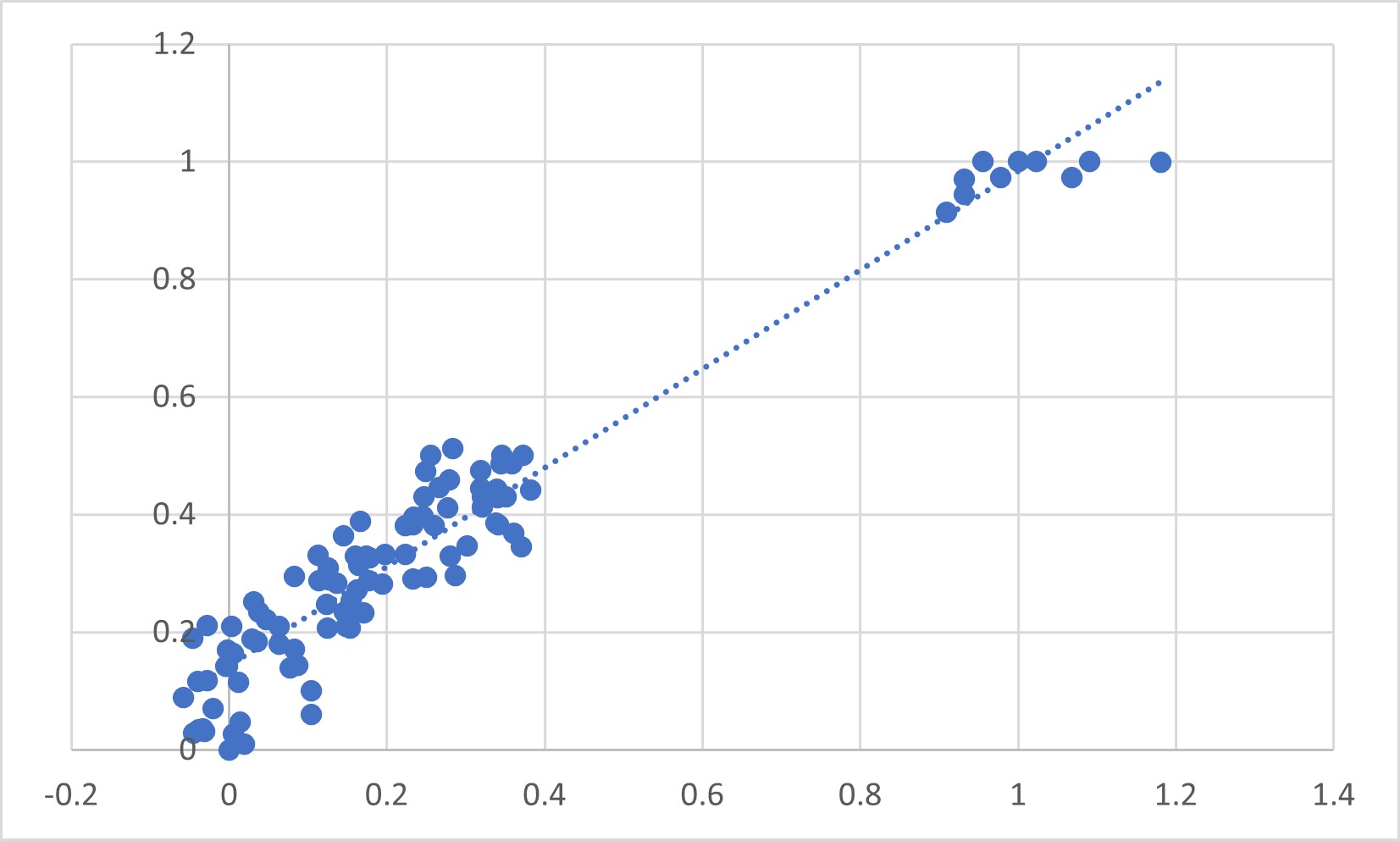}
    \label{fig:enter-label}
\end{figure}

\begin{figure}[H]
    \centering
  \caption{Standardised True Fixed Effect Coefficients (Horizontal Axis) Against Standardised Estimates (Vertical Axis) }  \includegraphics{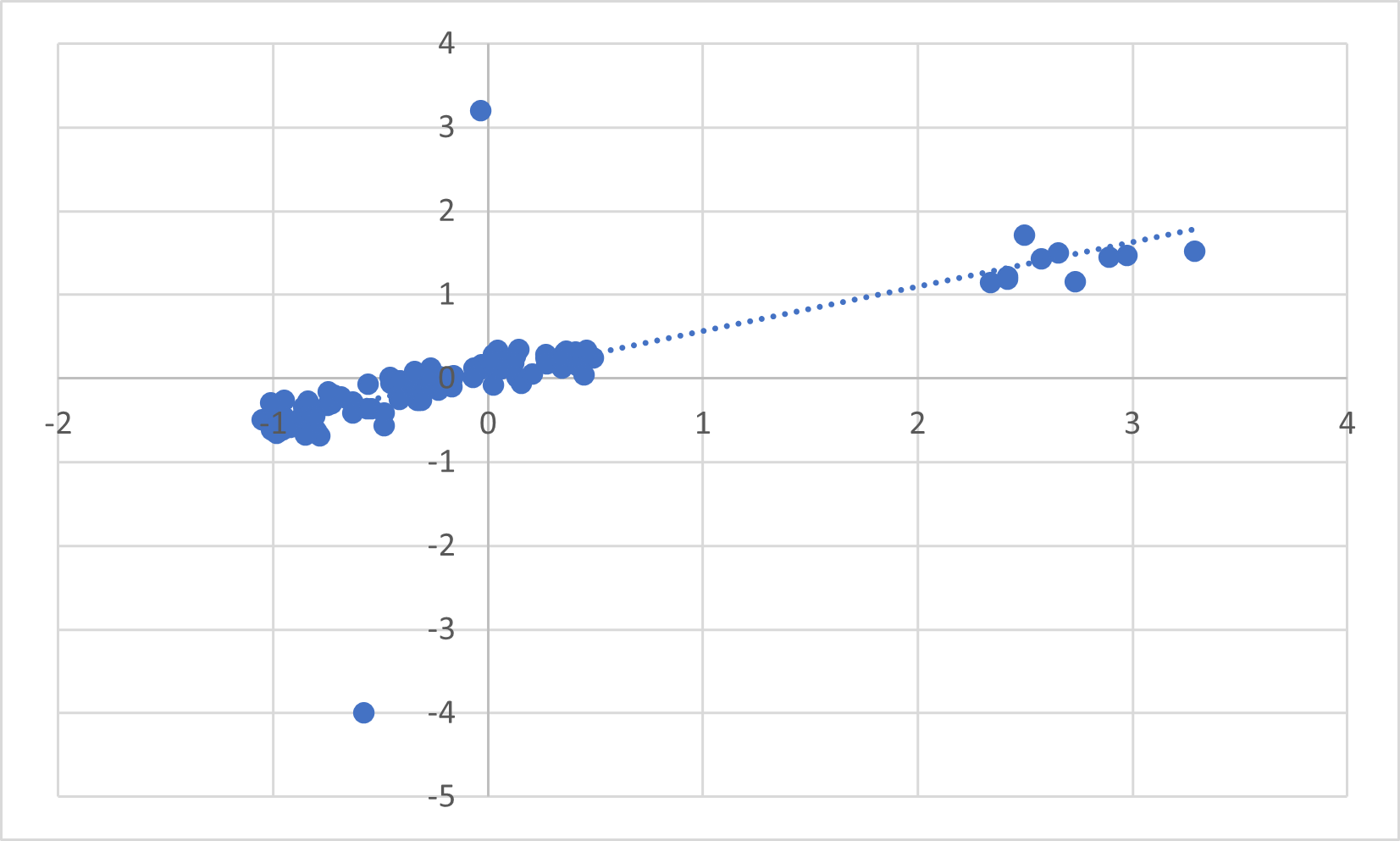}
    \label{fig:enter-label}
\end{figure}

Because the two clusters are very far apart from one another in the standardisation case, little can be learned for linking between clusters. This is why within each cluster, the estimation performs well, but due to the loss of information for linking between the clusters, the entire second cluster is estimated at values way too low. Some coefficients are estimated at values so high to cause numerical issue, making trimming necessary and thereby entailing further losses in information. 

The minimax normalisation does not suffer from this problem: despite the fact that both clusters are far away in relative terms, since all coefficients are bounded in (or close to) the zero to one interval, the absolute distance between the clusters is small enough to estimate inter-cluster linking probabilities. For the same reason, trimming is unnecessary, such that information can be maintained. 

For comparison again, we plot the logit estimator knowing that the errors in the simulation are in fact logistically distributed.

\begin{figure}[H]
    \centering
    \caption{True Coefficients (Horizontal Axis) Against Logit Estimates (Clustering In The Fixed Effects, Errors Are Truly Logistic) }
    \includegraphics{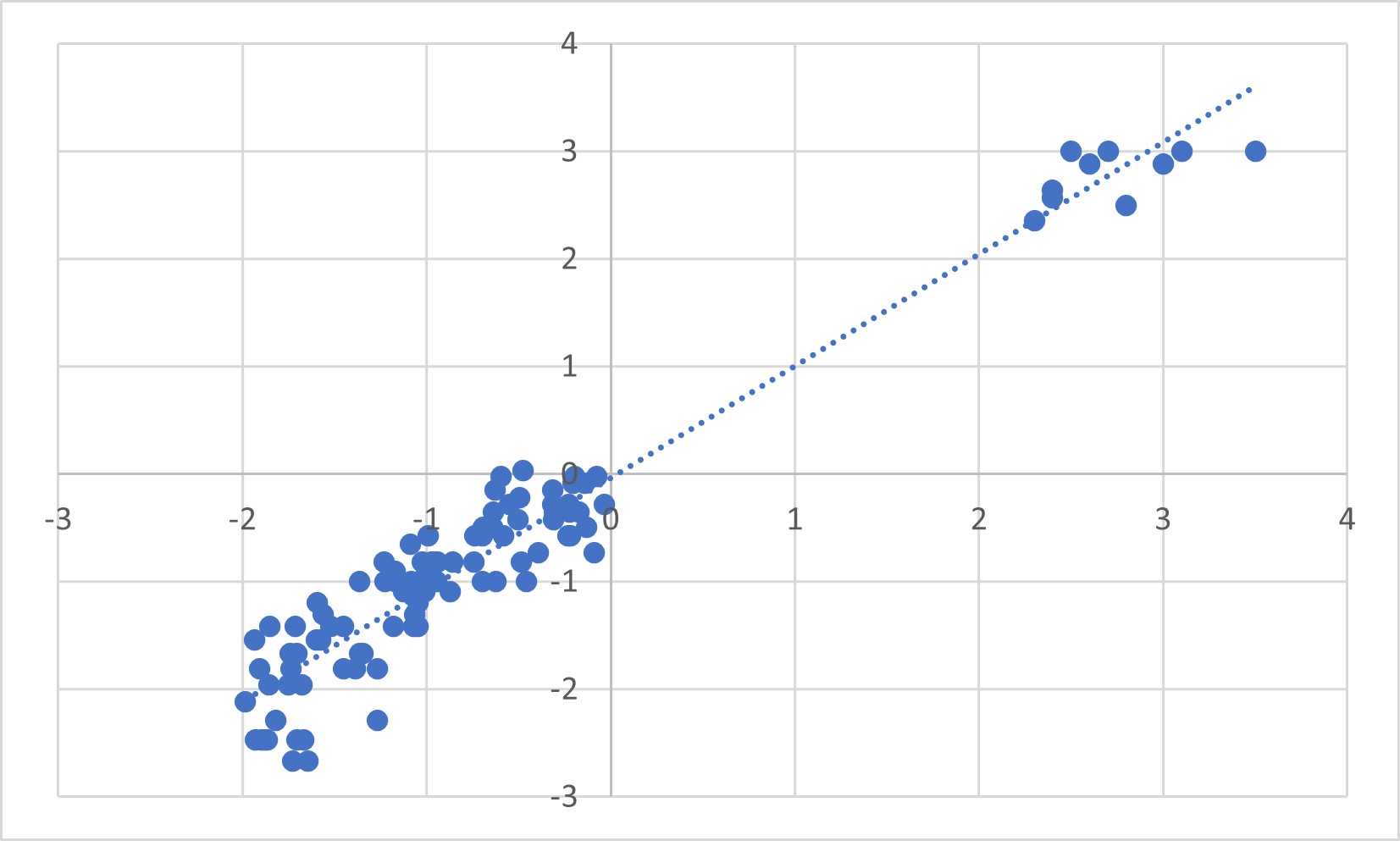}
\end{figure}

\section{Comparison to The Logit Estimator}

The Logit estimator is a popular choice for the dyadic link formation model. 
Admittedly, it is computationally easy and relatively unaffected when the true DGP features errors with a CDF that is not too far from the logistic distribution. That is, the parametric miss-specification of logistic errors is oftentimes inconsequential. However, if the true DGP error distribution exhibits certain characteristics, then the erroneously applied logit estimator will not produce any meaningful estimates. Below, we highlight these characteristics and provide examples of real world network formation processes in which they could occur.

\begin{itemize}
    \item {\bf Strong Convexity.} \\

For example, on modern social media (instagram, youtube), most active individuals have a negligible number of followers while at the same time, a substantial fraction of users (the ``influencers") features a stardom so large that most active users follow them. Presuming that it is rather randomness (as opposed to striking differences in individual characteristics) that drives these differences, we can model such a degree distribution assuming that small errors are very unlikely while large errors have a high probability to occur. A strongly convex error distribution can model the effect that small differences in individual characteristics lead to enormous differences in connections. \\
\noindent As an other example, consider 
startups competing for investor funding.  
They often have similar characteristics: given their lack of business history, there are few hard criteria to make an assessment. The outcome of the competition for funding is usually that some ideas attract practically all and the others almost no investors. If we think of a network in which links between startups signify the existence of at least one joint investor (assuming that all investors will invest in more than just one startup), then having many links in this networks indicates popularity among investors. If we investigate the probability of two startups sharing an investor and we observe that the latter is on average very high for a subset of startups and on average very low for the rest, a strongly convex error distribution seems more plausible than a tremendous difference in characteristics. 

\item {\bf Strong Concavity.}


A trend is popular if there are many users such as brands or designers incorporating it in their collections. Since all of these users will adopt several trends, we can say that two trends are ``connected" if they are adopted in two collections. Note that this notion of ``trend" is more close to a stylistic feature, item, color or novelty, such that it requires other ``trends" to become a marketable product. Most features will enjoy moderate popularity in the production industry, with some appearing in almost all collections. This could be modeled via a strongly concave error distribution, implying that small errors are very likely and nodes (here features) with individuals characteristics in the upper third of the sample are practically linked to everybody. 
\\
\noindent A similar argument can be made for blockchaines, which can be perceived as ``connected" if at least one app runs on both of them. Similarly to trends, chains to date distinguish themselves from one another, yet insufficiently so to warrant the tremendous adoption differences. In both cases, it appears plausible to model the latter by noise, as opposed to inherent advantages.  
\\
\noindent Finally, we can think of nodes as (online) shops or websites connected through joint users. Then, most active sites serve a moderate share of the user population while some will be used by almost everybody.
\end{itemize}

In the following, I generate links using a strongly convex or strongly concave error distribution. I then apply the semiparametric estimation routine (using the standardisation and the DBMM normalisation) and also the (erroneous) logit estimator to the data. Again, normalised true coefficients are plotted against normalised estimated. We can see that the logit estimation results are almost uninformative, while the semiparametric estimator performs almost perfectly regardless which transformation function is chosen.

\subsection{Beta Distribution With Parameters $\alpha=5, \beta=1$ (Strongly Convex)}

\begin{figure}[H]
    \centering
    \caption{Degree Distribution}
    \includegraphics{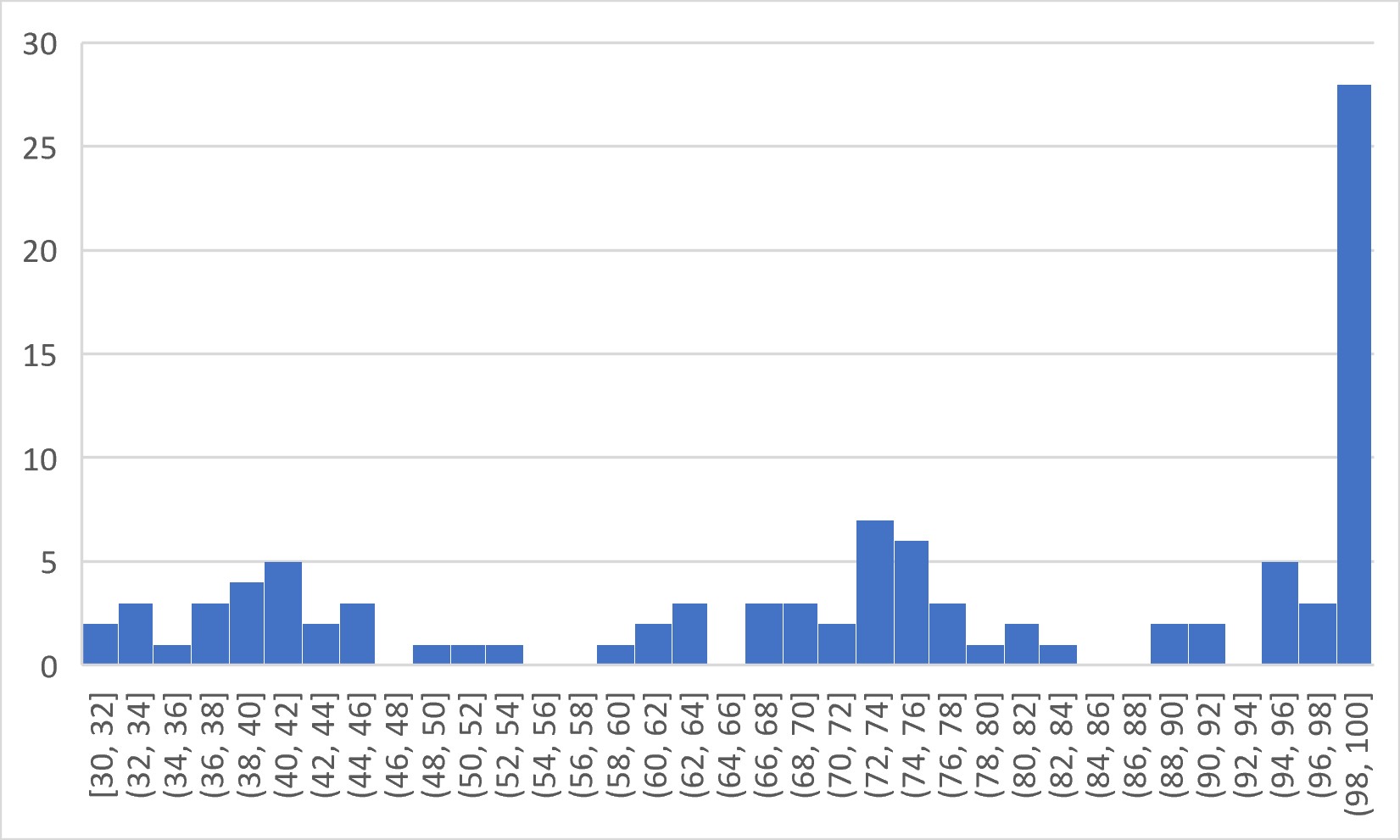}
\end{figure}

\begin{figure}[H]
    \centering
         \caption{True Fixed Effects Coefficients (Horizontal Axis) Against Estimates Assuming Standard Logistic Errors }
    \includegraphics{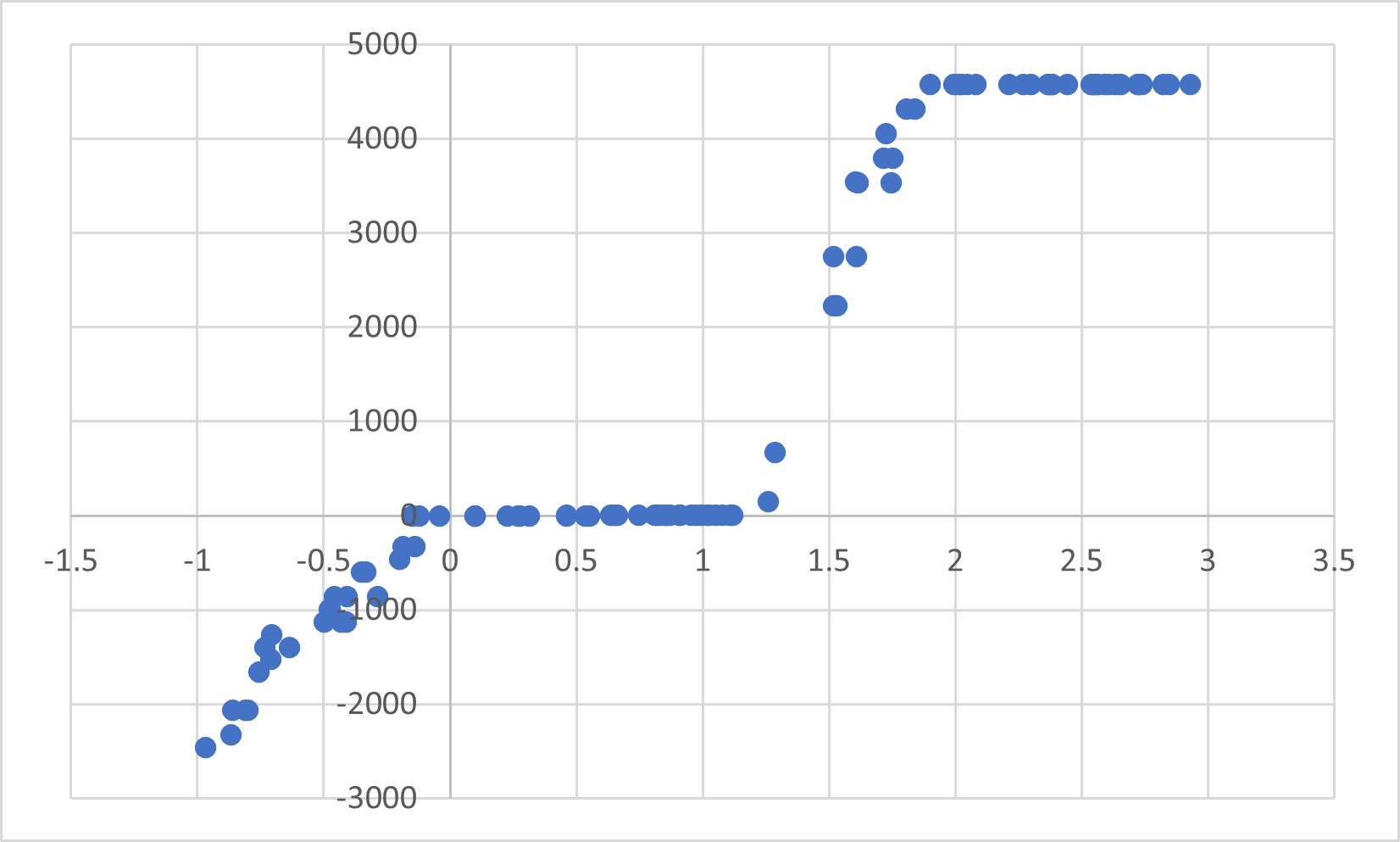}
\end{figure}

\begin{figure}[H]
    \centering
        \caption{Standardised True Fixed Effect Coefficients (Horizontal Axis) Against Standardised Estimates (Vertical Axis) } 
    \includegraphics{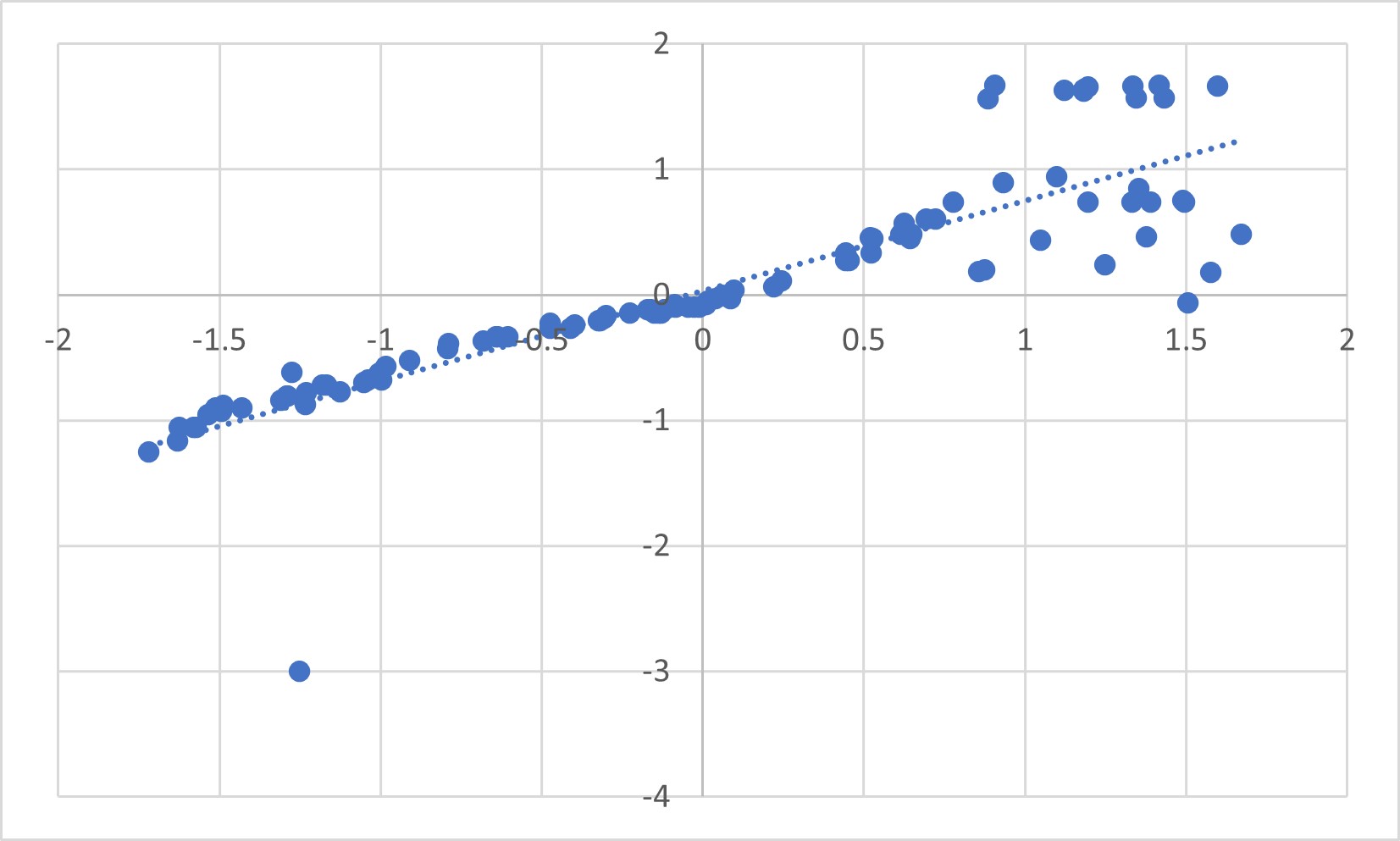}
\end{figure}

\begin{figure}
    \centering
       \caption{Normalised True Fixed Effect Coefficients (Horizontal Axis) Against Normalised Estimates (Vertical Axis), Degree-Based Minimax Normalisation }
    \includegraphics{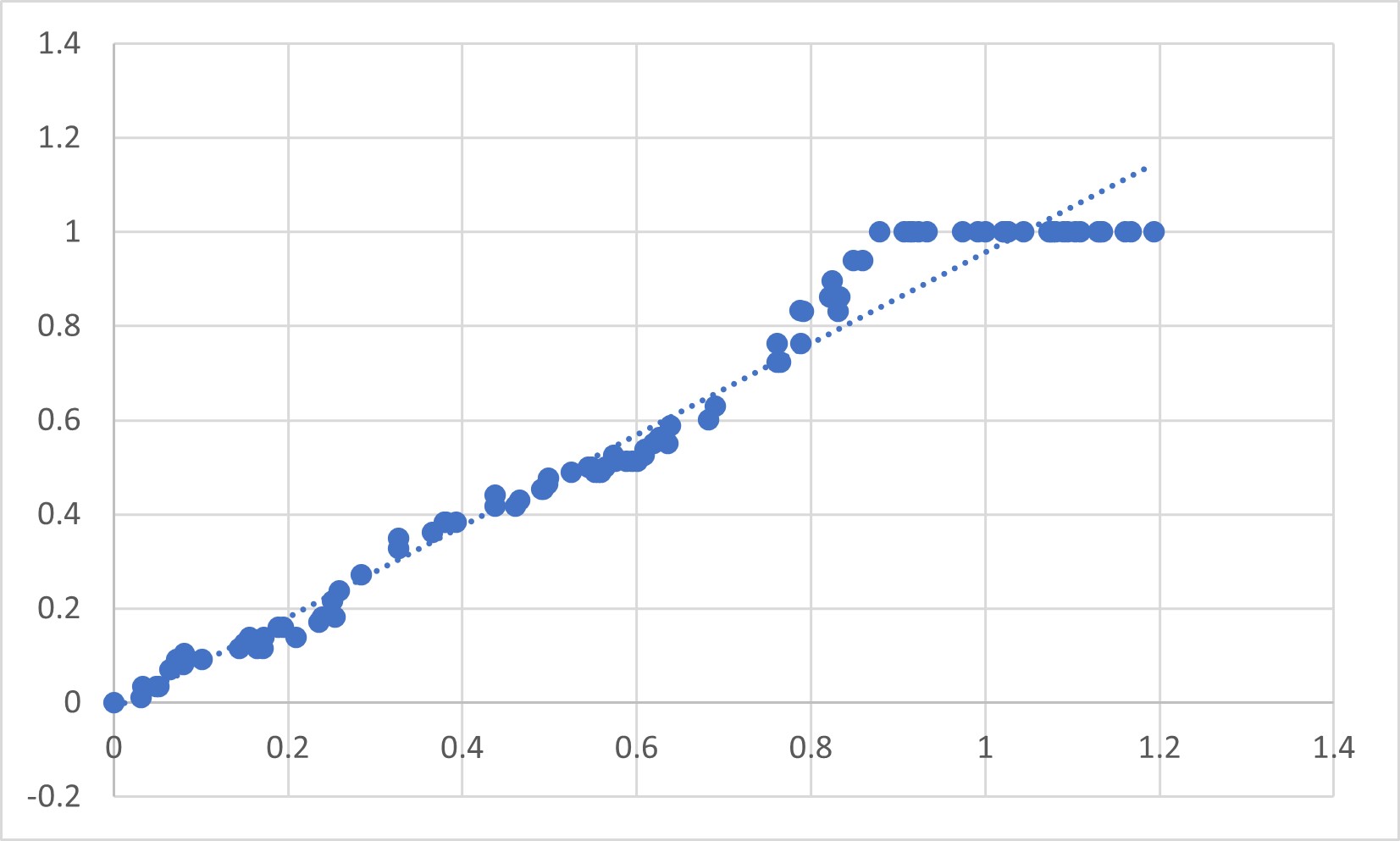}
\end{figure}

\subsection{Beta Distribution With Parameters $\alpha=2, \beta=5$ (Strongly Concave)}

\begin{figure}[H]
    \centering
        \caption{Degree Distribution}
    \includegraphics{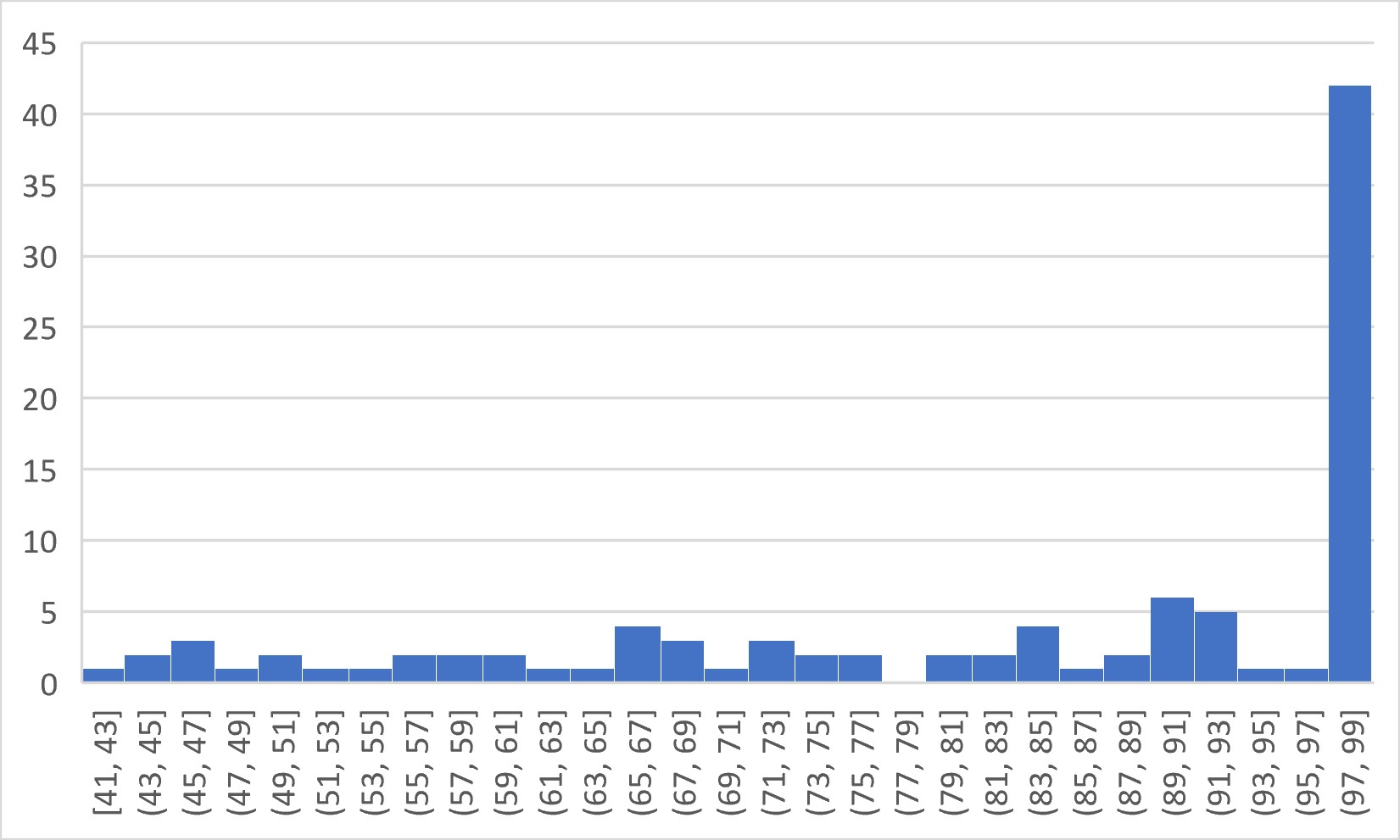}
\end{figure}

\begin{figure}[H]
    \centering
     \caption{True Fixed Effects Coefficients (Horizontal Axis) Against Estimates Assuming Standard Logistic Errors }
    \includegraphics{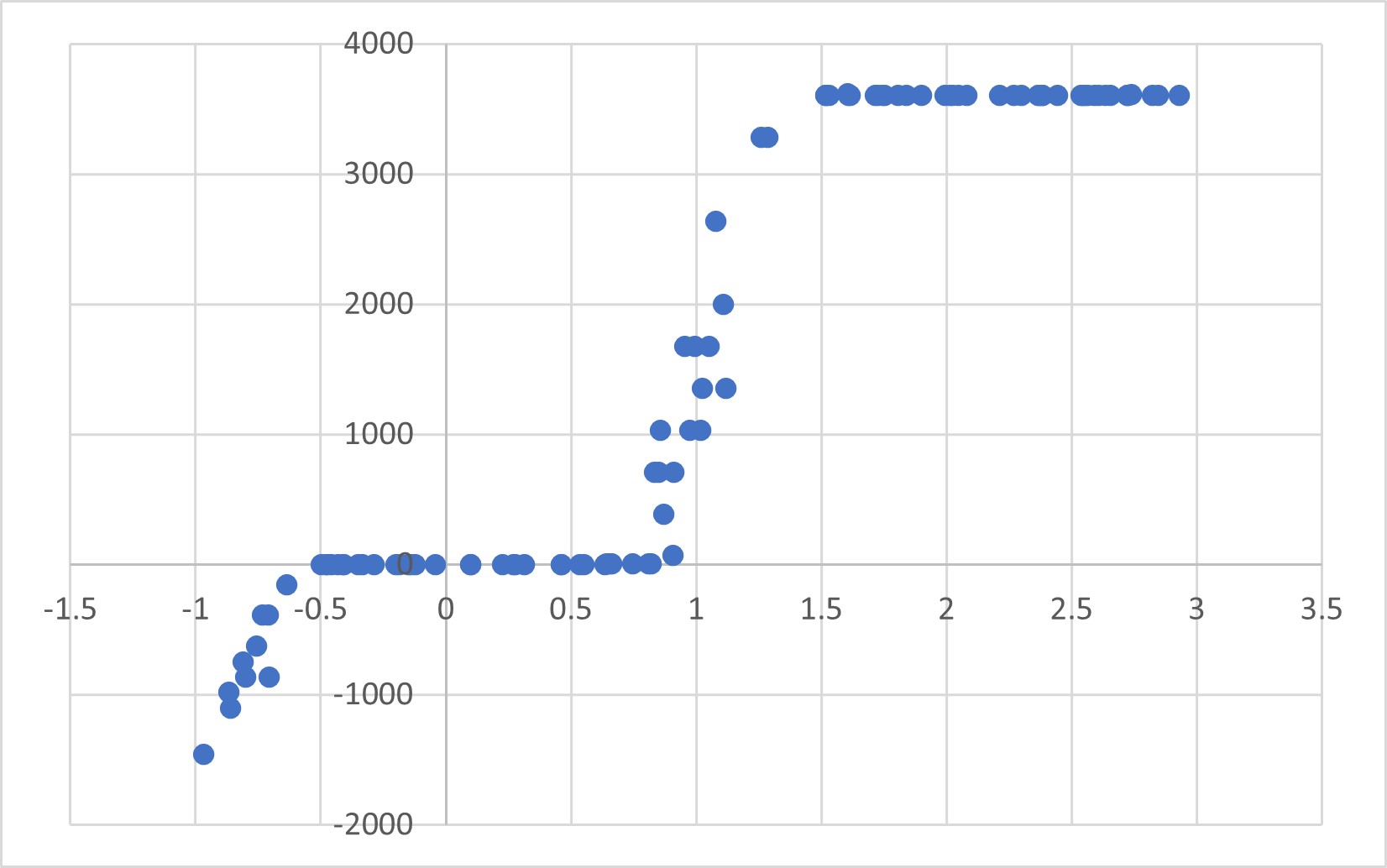}
\end{figure}

\begin{figure}[H]
    \centering
    \caption{Standardised True Fixed Effect Coefficients (Horizontal Axis) Against Standardised Estimates (Vertical Axis) } 
    \includegraphics{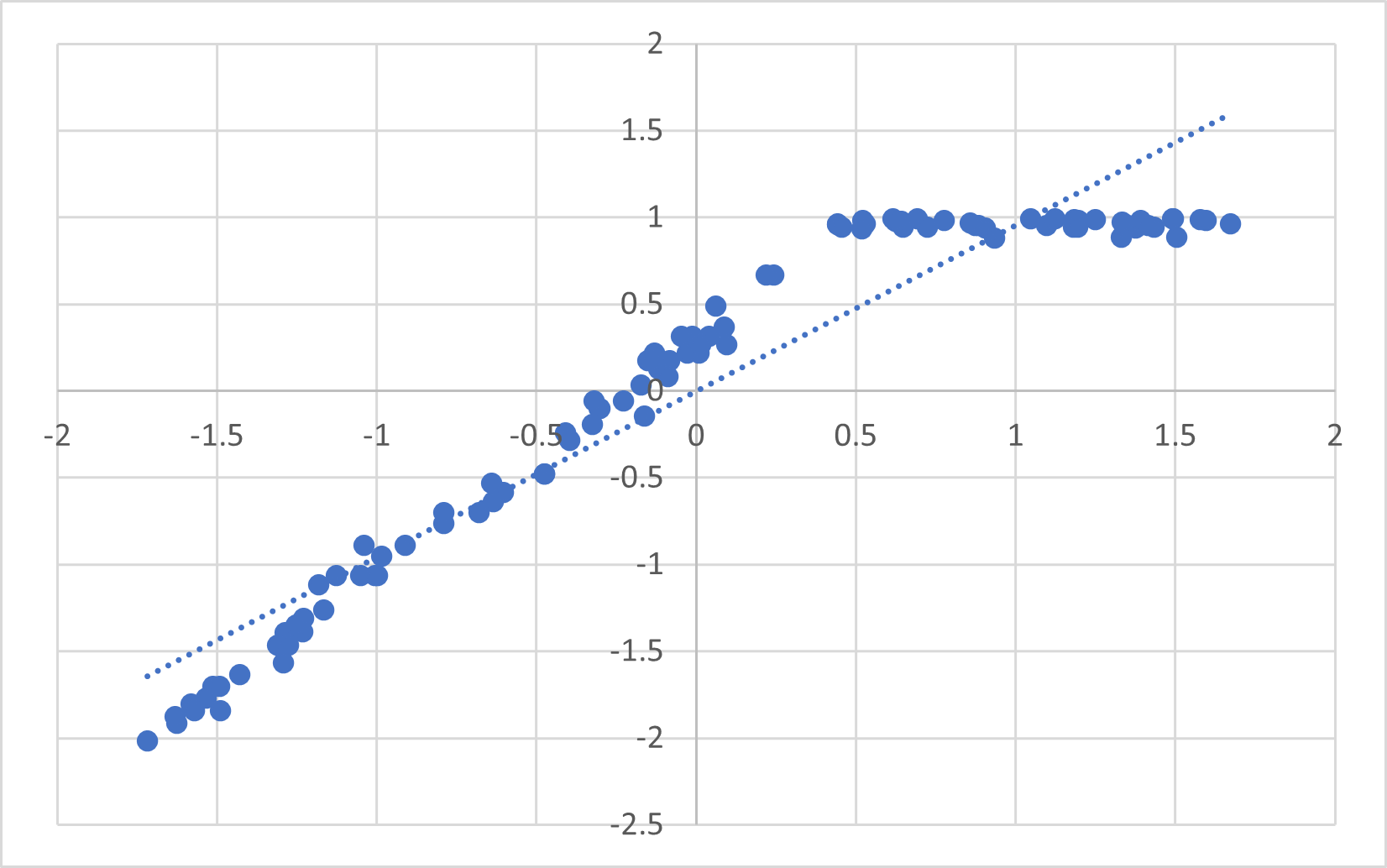}
\end{figure}

\begin{figure}
    \centering
    \caption{Normalised True Fixed Effect Coefficients (Horizontal Axis) Against Normalised Estimates (Vertical Axis), Degree-Based Minimax Normalisation }
    \includegraphics{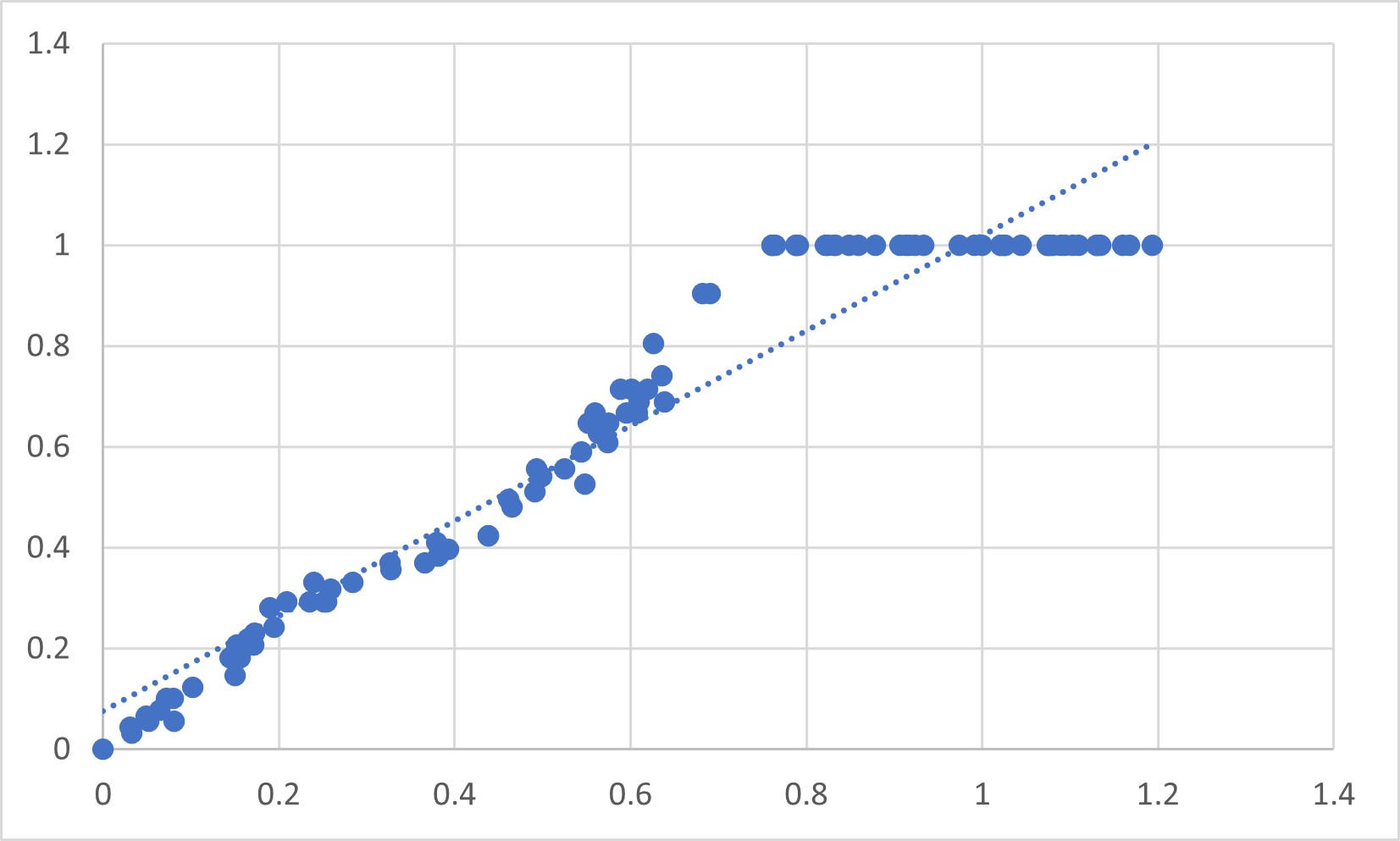}
\end{figure}

\noindent We see that applying the logit estimator essentially leads no information about the true fixed effect coefficients. The semiparametric estimator on the other hand recovers all but very large normalised coefficients with surprising accuracy.

\subsection{Exponential Distribution With Parameter $\lambda=1.5$ (Strongly Concave)}

\begin{figure}[H]
    \centering
      \caption{Degree Distribution}
      \includegraphics{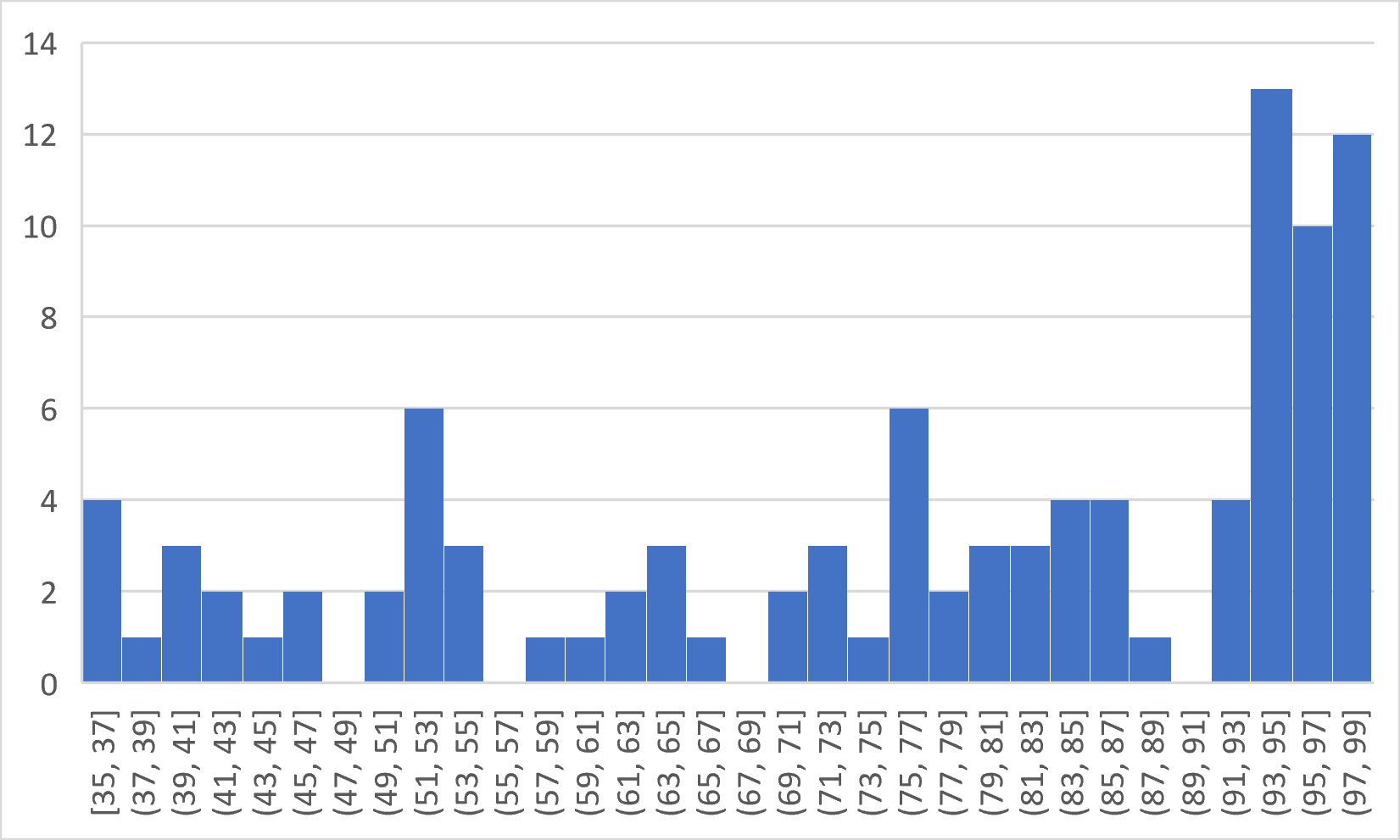}
  
    \label{fig:enter-label}
\end{figure}

\begin{figure}[H]
    \centering
      \caption{True Fixed Effects Coefficients (Horizontal Axis) Against Estimates Assuming Standard Logistic Errors }  \includegraphics{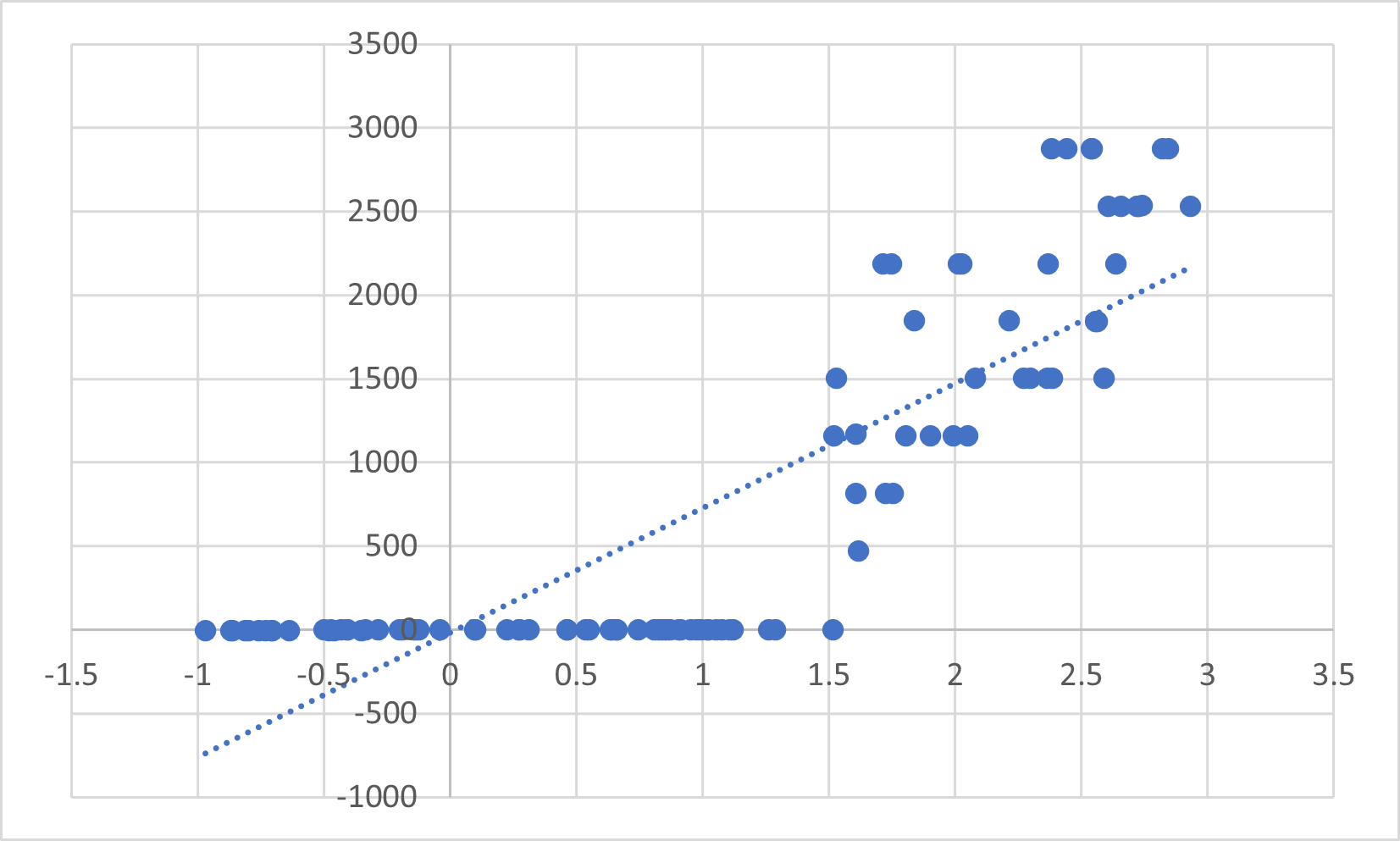}

    \label{fig:enter-label}
\end{figure}

\begin{figure}[H]
    \centering
     \caption{Standardised True Fixed Effect Coefficients (Horizontal Axis) Against Standardised Estimates (Vertical Axis) } \includegraphics{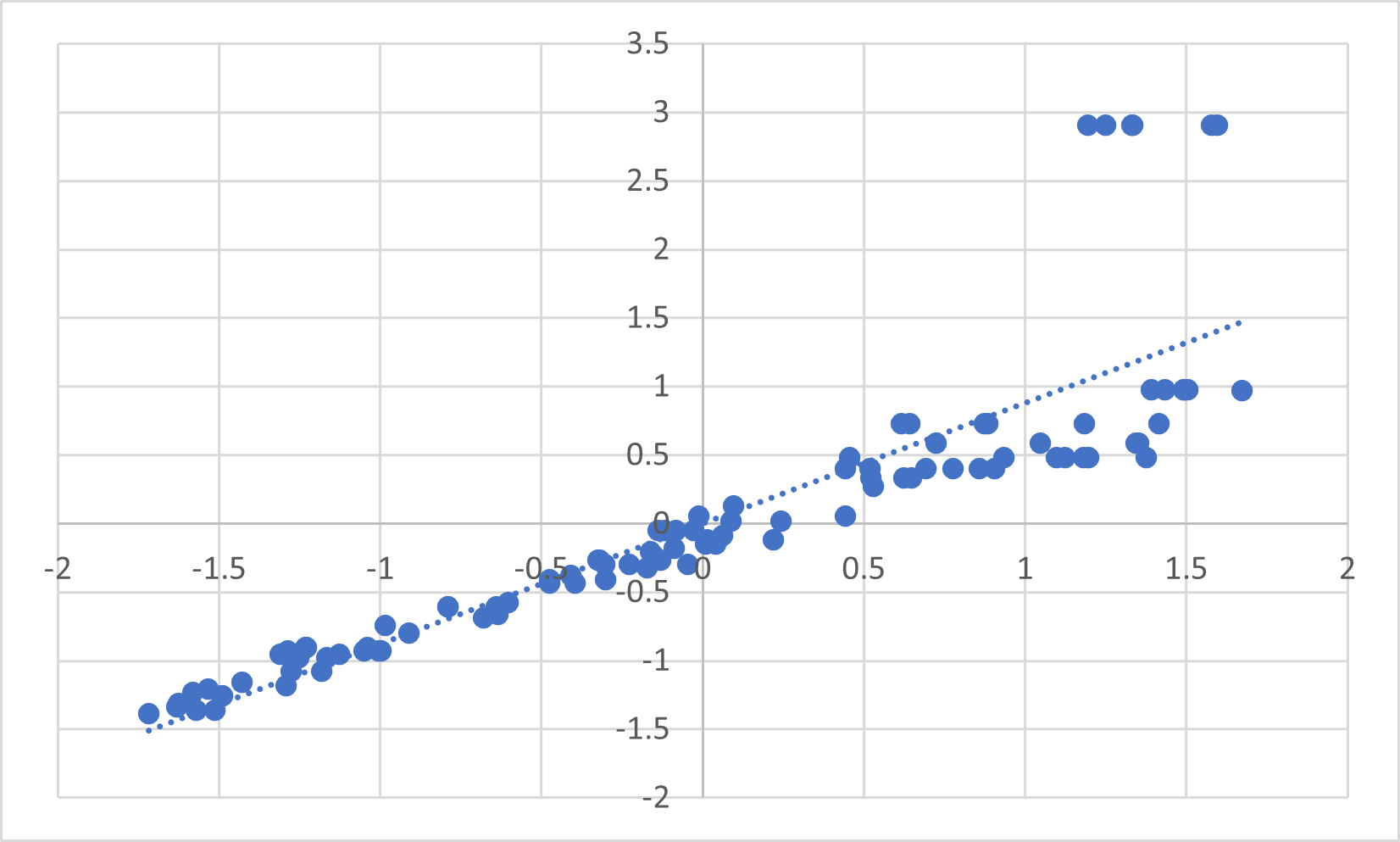}
    \label{fig:enter-label}
\end{figure}

\begin{figure}[H]
    \centering
     \caption{Normalised True Fixed Effect Coefficients (Horizontal Axis) Against Normalised Estimates (Vertical Axis), Degree-Based Minimax Normalisation } \includegraphics{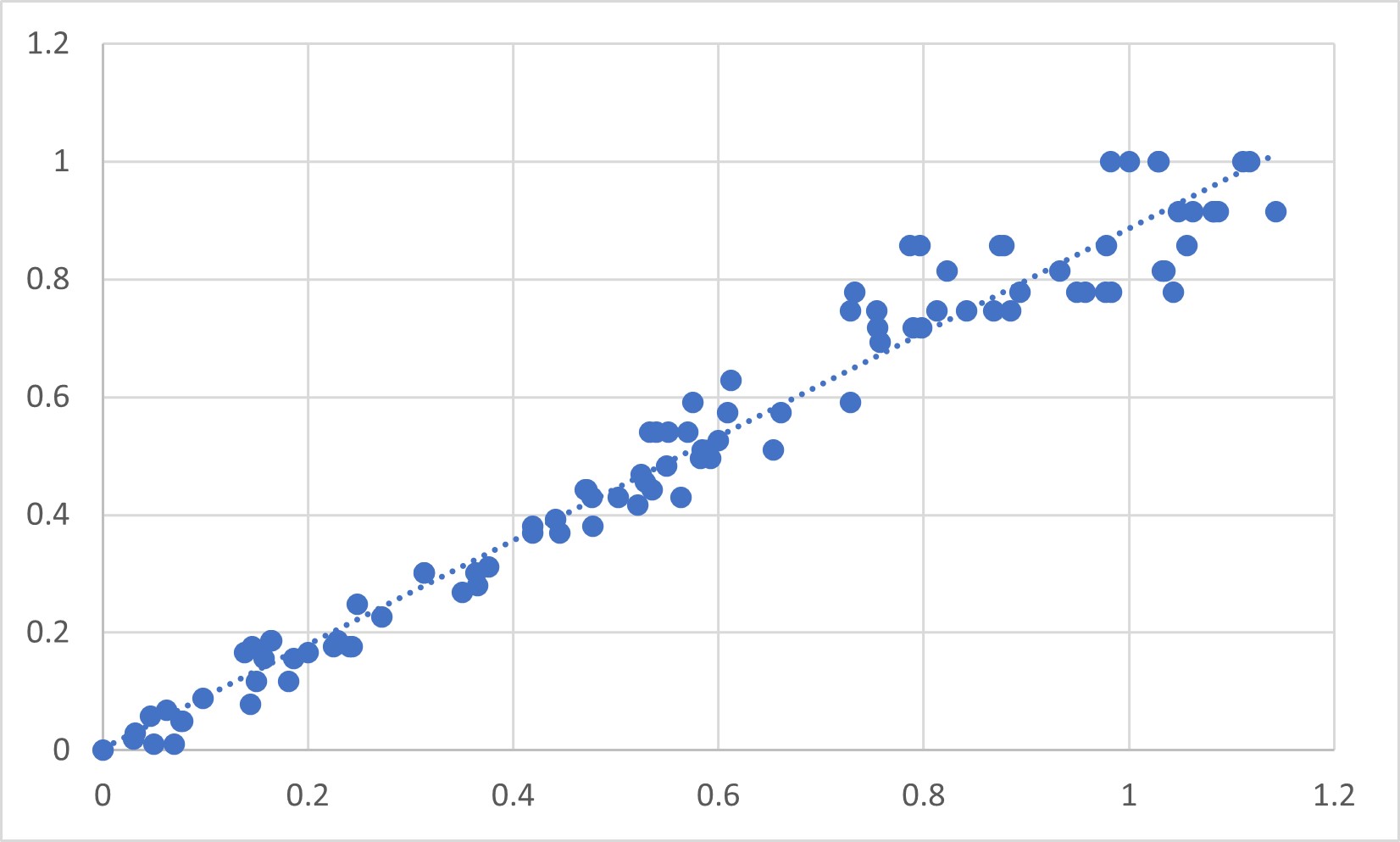}

    \label{fig:enter-label}
\end{figure}
\noindent

\section{Conclusion}

The present paper investigates semiparametric estimation of a parsimonious link formation model, lacking observable covariates. A novel normalisation is introduced and its distinct advantages are highlighted. The asymptotic properties of the normalised estimates are derived and verified by simulation. This shows that the (fortunate) asymptotic properties of (the impossible) parametric estimator can be achieved. 
A comparison with an erroneously applied logit estimator, underpinned by various references to practical applications, highlights that semiparametric estimators can be highly beneficial and should be employed more frequently in applied economic research.



\section{Appendix}

\subsection{Convergence of the Estimated Probabilities}

{\bf Lemma 1: }
 $\forall \boldsymbol{\eta} \in \mathbb{N}  \hspace{.25cm} \widehat{F}_{u}(\boldsymbol{\eta},v_{ij})-
F_{u_0}(v_{ij})=O(N^{-1/2}h^{-1} \underline{p}^{-1})$.
\newline

{\bf Proof:}
With $v_{ij}=w_{ij}'\boldsymbol{\eta}$,
\[  
\hat{p}_{1,ij} \big(\boldsymbol{\eta}\big)=
\frac{1}{h(L-1)}
\sum_{
\substack{
\{k,m \} \in \\
\mathbb{P} \backslash \{i,j\}
}
}
I(g_{km}=1) K \left(\frac{v_{ij}-v_{km}}{h} \right)\]
\[=
\frac{1}{(L-1)}
\sum_{\substack{
\{k,m \} \in \\
\mathbb{P} \backslash \{i,j\}
}}
I(g_{km}=1) K \left(v_{ij}-v_{km} \right)
\]
is a sample average of terms 
\[ I(g_{km}=1) K \left(v_{ij}-v_{km} \right)\]
where $\{g_{km},v_{km}\}$ are i.i.d.. With $h_N \rightarrow 0$ as $N \rightarrow \infty$, by Assumption 6
\[
h_N^{r+1}
\Bigg|
I(g_{km}=1) K \left(v_{ij}-v_{km} \right)\Bigg|<c
\hspace{.75cm}r+1>0
\]
and 
\[
h_N^{s}
\Bigg|
\frac{\partial
I(g_{km}=1) K \left(v_{ij}-v_{km} \right)}{
\partial \boldsymbol{\eta}}
\Bigg|<c
\hspace{.75cm}s>0.
\]
Let $E(\hat{p}_{1,ij}(\boldsymbol{\eta}))$ be the expectation of $\hat{p}_{1,ij}(\boldsymbol{\eta})$ taken over the distribution of ${g_{km},v_{km}}$. Then, for $
\boldsymbol{\eta}
$ in a compact and bounded set, for any $\alpha>0$
\[
N^{
\frac{1-\alpha}{2}} h_N^{r+1}
\underset{\boldsymbol{\eta}}{sup}
|\hat{p}_{1,ij}(\boldsymbol{\eta})-E(\hat{p}_{1,ij}(\boldsymbol{\eta}))|
\rightarrow
\hspace{.2cm}
0 \hspace{.5cm}
a.s..
\] 
Proof: see \citet{KS} Lemma 1.  
As a consequence
\[
N^{
\frac{1}{2}} h_N^{1}
N^{-\frac{\alpha}{2}}h_N^r
|\hat{p}_{1,ij}(\boldsymbol{\eta})-E(\hat{p}_{1,ij}(\boldsymbol{\eta}))|
\rightarrow
\hspace{.2cm}
0 \hspace{.5cm}
a.s..
\]
Then with $h_N=N^{-\frac{1}{7}}$, as long as $-\frac{\alpha}{2}-\frac{r}{7}>0$, i.e.\ $-3.5 \alpha >r>-1$, then the terms $N^{-\frac{\alpha}{2}}h_N^r$ is increasing in $N$ and as such
\[
N^{
\frac{1}{2}} h_N^{1}
|\hat{p}_{1,ij}
(\boldsymbol{\eta})
-E(\hat{p}_{1,ij}(\boldsymbol{\eta}))|=O_p(1)
\hspace{.25cm} \forall \boldsymbol{\eta} \in \mathbb{N}
\]
or equivalently 
\[
|\hat{p}_{1,ij}(\boldsymbol{\eta})-E(\hat{p}_{1,ij}(\boldsymbol{\eta}))|=O_p(N^{-
\frac{1}{2}} h_N^{-1})
\hspace{.25cm} \forall \boldsymbol{\eta} \in \mathbb{N}.
\]
By the same argument 
\[
|\hat{p}_{0,ij}(\boldsymbol{\eta})-E(\hat{p}_{0,ij}(\boldsymbol{\eta})
)|=O_p(N^{-
\frac{1}{2}} h_N^{-1})
\hspace{.25cm} \forall
\boldsymbol{\eta}\in \mathbb{N}
.
\]
Then 
\[
E(\hat{p}_{1,ij} (\boldsymbol{\eta}))=
E(I(g=1) K \left(v_{ij}-v \right))=
\int \int 
I(g=1) K \left(v_{ij}-v \right)f_{gv}(gv)dgdv
\]
\[
= \int  \int I(g=1) K \left(v_{ij}-v\right)f_{v|g}(v)f(g)dvdg=\]
\[P(g=1) \int K \left(v_{ij}-v\right)f_{v|g=1}(v)dv.
\]
Let $z=\frac{v-v_{ij}}{h_N}$ for fixed $v_l$, such that $v=v_{ij}+h_Nz$, then 
\[
E(\hat{p}_{1,ij} (\boldsymbol{\eta}))=
P(g=1) \int K (v_{ij}-v)f_{v|g=1}(v_{ij}+h_Nz)
\frac{dv}{dz}dz=
\]
\[
P(g=1) \int h_N^{-1}K (z)f_{v|g=1}(v_{ij}+h_Nz)
h_Ndz=
P(g=1) \int K (z)f_{v|g=1}(v_{ij}+h_Nz)dz.
\]
If $h_N=0$ then the integrand is $p_{1,ij}$, so we expand around zero
\[
E(\hat{p}_{1,ij} (\boldsymbol{\eta}))=p_{1,ij}(\boldsymbol{\eta})+\]
\[P(g=1) \int K (z)
\frac{\partial f_{v|g=1}(v)}{\partial v}|_{v=v_{ij}}z
dz h_N+\]
\[P(g=1) \int K (z)
\frac{\partial^2 f_{v|g=1}(v)}{\partial^2 v}|_{v=v_{ij}}z^2
dz h_N^2+
\]
\[
P(g=1) \int K (z)
\frac{\partial^3 f_{v|g=1}(v)}{\partial^3 v}|_{v=v_{ij}}z^3
dz h_N^3+
\]
\[
P(g=1) \int K (z)
\frac{\partial^4 f_{v|g=1}(v)}{\partial^4 v}|_{v=v_{ij}+\Bar{h}z}z^4
dz h_N^4
\]
\[=
p_{1,ij}(\boldsymbol{\eta})+\]
\[P(g=1)
\underbrace{ \int  z K (z) dz}_{=0}
\frac{\partial f_{v|g=1}(v)}{\partial v}|_{v=v_{ij}} h_N+\]
\[P(g=1) \underbrace{\int z^2 K (z) dz}_{=0}
\frac{\partial^2 f_{v|g=1}(v)}{\partial^2 v}|_{v=v_{ij}} h_N^2+
\]
\[
P(g=1) \underbrace{\int z^3 K (z) dz}_{=0}
\frac{\partial^3 f_{v|g=1}(v)}{\partial^3 v}|_{v=v_{ij}} h_N^3+
\]
\[
P(g=1) \underbrace{\int z^4 K (z) dz}_{<\infty}
\frac{\partial^4 f_{v|g=1}(v)}{\partial^4 v}|_{v=v_{ij}+\Bar{h}z} h_N^4.\]
Such that 
\[
E(\hat{p}_{1,ij} (\boldsymbol{\eta}))-p_{1,ij}
=
O(h_N^4).
\]
By the same argument 
\[
E(\hat{p}_{0,ij} (\boldsymbol{\eta}))-p_{0,ij}
=
O(h_N^4).
\]
We conclude that $\hat{p}_{1,ij}(\boldsymbol{\eta})$ (respectively $\hat{p}_{0,ij}(\boldsymbol{\eta})$) converges to $p_{1,ij}$ (respectively  $p_{0,ij}$)  uniformly in $\boldsymbol{\eta}$.

Next, let $\widehat{F_u}(\boldsymbol{\eta}, 
 \eta_i+ \eta_j)=\widehat{F_u}(\boldsymbol{\eta},v_{ij})$ be defined as the estimated CDF of the model error (which depends on all parameters), evaluated at $\eta_i+\eta_j$
\[
\widehat{F_u}(\boldsymbol{\eta},v_{ij})=
\frac{\hat{p}_{1,ij}\big(\boldsymbol{\eta}\big)}{\hat{p}_{1,ij}\big(\boldsymbol{\eta}\big)+\hat{p}_{0,ij}\big(\boldsymbol{\eta}\big)}.
\]
Define 
\[
p_{ij}=
p_{1,ij}+p_{0,ij}
\]
and 
\[
\hat{p}_{ij}(\boldsymbol{\eta})=
\hat{p}_{1,ij}(\boldsymbol{\eta})+
\hat{p}_{0,ij}(\boldsymbol{\eta})
\]
and 
\[
\Delta(\boldsymbol{\eta})=
\frac{
p_{ij}-
\hat{p}_{ij}(\boldsymbol{\eta})
}{p_{ij}}.
\]
Then 
\[
\hat{p}_{ij}(\boldsymbol{\eta})=
p_{ij}-
\left(p_{ij}-
\hat{p}_{ij}(\boldsymbol{\eta})\right)=
p_{ij}-
p_{ij}
\frac{p_{ij}-
\hat{p}_{ij}(\boldsymbol{\eta})}{p_{ij}}=
p_{ij}(1-\Delta(\boldsymbol{\eta})).
\]
Consequently
\[
\frac{1}{\hat{p}_{ij}(\boldsymbol{\eta})}=
\frac{1}{p_{ij}}((1-\Delta(\boldsymbol{\eta}))^{-1}.
\]
From the geometric expansion of an inverse
\[
\frac{1}{\hat{p}_{ij}}=
\frac{1}{p_{ij}}
\left(1+\Delta(\boldsymbol{\eta})+
\Delta(\boldsymbol{\eta})^2(1-\Delta(\boldsymbol{\eta}))^{-1}\right).
\]
Thus
\[
\widehat{F_u}(\boldsymbol{\eta},v_{ij})
-
F_{u_0}(v_{ij})=
\hat{p}_{1,ij}(\boldsymbol{\eta})
\frac{1}{p_{ij}}
\left(1+\Delta(\boldsymbol{\eta})+
\Delta(\boldsymbol{\eta})^2(1-\Delta(\boldsymbol{\eta}))^{-1}\right)
-
\frac{p_{1,ij}}{p_{ij}}=
\]
\[
\frac{\hat{p}_{1,ij}(\boldsymbol{\eta})-p_{1,ij}}{p_{ij}}+
\frac{\hat{p}_{1,ij}(\boldsymbol{\eta})}{p_{ij}}
\Delta(\boldsymbol{\eta})+
\frac{\hat{p}_{1,ij}(\boldsymbol{\eta})}{p_{ij}}\Delta(\boldsymbol{\eta})^2(1-\Delta(\boldsymbol{\eta}))^{-1}.
\]
Under the assumption
\[
p_{1,ij}>\underline{p},p_{ij}>\underline{p}
\]
and since 
\[ \Delta(\boldsymbol{\eta})=O(N^{-1/2}h^{-1})\]
\[\widehat{F_u}(\boldsymbol{\eta},v_{ij})
-
F_{u_0}(v_{ij})=
\frac{\hat{p}_{1,ij}(\boldsymbol{\eta})-p_{1,ij}}{p_{ij}}+
O(N^{-1/2}h^{-1} \underline{p}^{-1})+
o(N^{-1/2}h^{-1} \underline{p}^{-1})
\hspace{.25cm} \forall \boldsymbol{\eta} \in \mathbb{N}.
\]
I conclude that $\widehat{F}_{u}(\boldsymbol{\eta},v_{ij})$ converges to $F_{u_0}(v_{ij})$ uniformly in $\boldsymbol{\eta}$ and $v_{ij}$. Since $\hat{p}_{1,ij}(\boldsymbol{\eta})-p_{1,ij}$ is $O(L^{-1/2}h^{-1})$, thus 
\[
\widehat{F}_{u}(\boldsymbol{\eta},v_{ij})-
F_{u_0}(v_{ij})=O(N^{-1/2}h^{-1} \underline{p}^{-1}).
\]

\subsection{Convergence of the Estimated Derivatives}

{\bf Lemma 2: } 
 $\forall \boldsymbol{\eta} \in \mathbb{H} 
 \hspace{.25cm} 
 \widehat{f}_{u}(\boldsymbol{\eta},v_{ij})-
f_{u_0}(v_{ij})=O(N^{-1/2}h^{-2} \underline{p}^{-1})$.
\newline

{\bf Proof:} \\
\[\hat{p}d_{1,ij}(\boldsymbol{\eta})=
\frac{\partial 
\hat{p}_{1,ij}
}{\partial v_{ij} }=
\frac{1}{h(L-1)}
\sum_{
\substack{
\{k,m \} \in \\
\mathbb{P} \backslash \{i,j\}
}
}
I(g_{km}=1)
\frac{\partial K \left(\frac{v_{ij}-v_{km}}{h} \right) }{
\partial 
\left(\frac{v_{ij}-v_{km}}{h} \right)
}
\frac{1}{h}.\]
Since 
\[
K \left(\frac{v_{ij}-v_{km}}{h} \right) \frac{1}{h}=
K \left(v_{ij}-v_{km} \right),
\]
thus
\[
\frac{\partial K \left(\frac{v_{ij}-v_{km}}{h} \right) }{
\partial 
\left(\frac{v_{ij}-v_{km}}{h} \right)
} \frac{\partial \left(\frac{v_{ij}-v_{km}}{h} \right)}
{\partial v_{ij}}\frac{1}{h}=
\frac{\partial K \left(\frac{v_{ij}-v_{km}}{h} \right) }{
\partial 
\left(\frac{v_{ij}-v_{km}}{h} \right)
} 
\frac{1}
{ h^2}=
\frac{\partial K \left(v_{ij}-v_{km} \right)}{
\partial v_{ij}
},
\] so 
\[\hat{p}d_{1,ij}(\boldsymbol{\eta})=
\frac{\partial 
\hat{p}_{1,ij}
}{\partial v_{ij} }=
\frac{1}{(L-1)}
\sum_{\substack{
\{k,m \} \in \\
\mathbb{P} \backslash \{i,j\}
}}
I(g_{km}=1)
\frac{\partial K \left(v_{ij}-v_{km} \right)}{
\partial v_{ij}
}.\]
For any $F_u$, I use $f_u$ to denote the first derivative. 
I let
\[
\widehat{f_u}(\boldsymbol{\eta},v_{ij})=
\frac{
\hat{p}_{1,ij}\big(\boldsymbol{\eta}\big)
\Big(\hat{p}d_{1,ij}\big(\boldsymbol{\eta})+
\hat{p}d_{0,ij}\big(\boldsymbol{\eta})\Big)-
\hat{p}d_{1,ij}\big(\boldsymbol{\eta}\big)
\Big(
\hat{p}_{1,ij}\big(\boldsymbol{\eta}\big)
+
\hat{p}_{0,ij}\big(\boldsymbol{\eta}\big)\Big)
}{\big(\hat{p}_{1,ij}\big(\boldsymbol{\eta}\big)+\hat{p}_{0,ij}\big(\boldsymbol{\eta}\big)\big)^2}.
\]
By the definition of the derivative
\[
\widehat{f_u}(\boldsymbol{\eta},v_{ij})=
\lim_{x\rightarrow 0}
\frac{
\widehat{F_u}(\boldsymbol{\eta},v_{ij}+x)-\widehat{F_u}(\boldsymbol{\eta},v_{ij})
}{x}
\]
and 
\[
f_{u_0}(v_{ij})=
\lim_{x\rightarrow 0}
\frac{F_{u_0}(v_{ij}+x)-F_{u_0}(v_{ij})}{x}.
\]
Since $h \rightarrow 0$ as $N \rightarrow \infty$, therefore, I can replace $x$ by $h$, subtract both equations and rearrange into
\[
\widehat{f_u}(\boldsymbol{\eta},v_{ij})-
f_{u_0}(v_{ij})=
\frac{
\widehat{F_u}(\boldsymbol{\eta},v_{ij}+h)-F_{u_0}(v_{ij}+h)
}{h}-
\frac{
\widehat{F_u}(\boldsymbol{\eta},v_{ij})-F_{u_0}(v_{ij})
}{h}.
\]
Then from the result of the previous section, it follows that $\widehat{f}_{u}(\boldsymbol{\eta},v_{ij})$ converges to $f_{u_0}(v_{ij})$ uniformly in $\boldsymbol{\eta}$ and $v_{ij}$, that is 
\[
\widehat{f_{u}}(\boldsymbol{\eta},v_{ij})-
f_{u_0}(v_{ij})=O(N^{-1/2}h^{-2} \underline{p}^{-1}).
\]
\subsection{Convergence of the Objective Function}

There are two sources of randomness: randomness in estimating the error distribution and randomness in estimating the individual effects.
I first abstract from the second source of randomness, that is, I treat $\boldsymbol{\eta}$ as fixed and $F_u()$ as a random variable. I carry out a Taylor expansion of the approximate objective function around the true error distribution (evaluated at the constant parameter vector $\boldsymbol{\eta}$).  

To ease notation, let $D=
\operatorname{diag}((N-1)^{-1})$ denote the $N \times N$ diagonal matrix with entry $(N-1)^{-1}$ on the diagonal. 
Let $F_u(\boldsymbol{\eta})$ denote the $L \times 1$ vector collecting the scalars  $F_u(w'_{ij} \boldsymbol{\eta}) \hspace{.25cm} i \in N, j \in N, i\neq j$ (that is, the error CDF evaluated at $v_{ij}$) and let $m \left(F_u(\boldsymbol{\eta}) \right)$ denote the $N \times 1$ vector of moment conditions (as a function of the $L$ pair-specific values $F_u(w_{ij}'\boldsymbol{\eta})$) evaluated at $F_u(\boldsymbol{\eta})$. 
Holding the coefficient vector $\boldsymbol{\eta}$ constant, I expand the objective function around $F_{u_0}(w_{ij}'\boldsymbol{\eta})$ for each pair $\{i,j\} \in \mathbb{P}$ and sum over pairs. 
For any particular choice of $F_u, \boldsymbol{\eta}$, because each pair appears in two of the moments, taking derivatives of the moment conditions with respect to the (scalar) $F_u(w_{ij}'\boldsymbol{\eta})$ generates a vector with elements $-(N-1)^{-1}$ at  entries $i$ and $j$ and zeros everywhere else: 
$F_u(w_{ij}'\boldsymbol{\eta})$ is
\[
\frac{\partial 
m(F_u(\boldsymbol{\eta}))
}{
\partial 
F_u(w_{ij}'\boldsymbol{\eta})
}
=
- D w_{ij}.
\]
As a consequence, the derivative of the objective function with respect to the (scalar) $F_u(w_{ij}'\boldsymbol{\eta})$ evaluated at $F_u()=F_{u_0}()$ is
\[ 
\frac{\partial 
Q(F_u(\boldsymbol{\eta}))
}{
\partial 
F_u(w_{ij}'\boldsymbol{\eta})
}\Bigg|_{F_u(w_{ij}'\boldsymbol{\eta})=
F_{u_0}(w_{ij}'\boldsymbol{\eta})}
=-
2  w_{ij}' D m\left(F_{u_0}(\boldsymbol{\eta})\right)
\] 
and therefore, for any $\boldsymbol{\eta}$, 
\[Q(\hat{F}_u
(\boldsymbol{\eta}))=
Q(F_{u_0}
(\boldsymbol{\eta}))+\]
\[
\sum_{
\substack{\{i,j\} \\ \in \mathbb{P}}
} \Bigg(
-
2
w_{ij}'D
m(F_{u_0}(\boldsymbol{\eta}))
\left(\widehat{F_u}(\boldsymbol{\eta},w_{ij}'\boldsymbol{\eta})-F_{u_0}(w_{ij}'\boldsymbol{\eta})\right)+\]
\[ \underbrace{ w_{ij}'DDw_{ij}}_{
\frac{2}{(N-1)^2}
}
\left(\widehat{F_u}(
\boldsymbol{\eta},
w_{ij}'\boldsymbol{\eta})+F_{u_0}(w_{ij}'\boldsymbol{\eta})\right)^2\Bigg).
\] 
Because of Lemma 1, with $h=L^{-1/7}=(N(N-1)/2)^{-1/7}$, we have \[\widehat{F_u}(
\boldsymbol{\eta},w_{ij}' \boldsymbol{\eta})-F_{u_0}(w_{ij}'\boldsymbol{\eta})
=O((N(N-1)/2)^{-5/14}).\] 
Therefore 
\[Q(\hat{F}_u
(\boldsymbol{\eta}))=
Q(F_{u_0}
(\boldsymbol{\eta}))+\]
\[\frac{2}{N-1}
\sum_{
\substack{\{i,j\} \\ \in \mathbb{P}}
} \Bigg(
-
\underbrace{\left(m_i(F_{u_0}(\boldsymbol{\eta}))+m_j(F_{u_0}(\boldsymbol{\eta}))\right)}_{
\substack{O(1)\\
sample \hspace{.1cm} averages \hspace{.1cm}  of \hspace{.1cm}  probabilities
}
}
\underbrace{\left(\widehat{F_u}(\boldsymbol{\eta},w_{ij}'\boldsymbol{\eta})-F_{u_0}(w_{ij}'\boldsymbol{\eta})\right)}_{
\substack{O(N(N-1)/2)^{-5/14})\\
by \hspace{.1cm} Lemma \hspace{.1cm}  1 
}}+\]
\[ 
\frac{1}{(N-1)}\Big(
\underbrace{
\widehat{F_u}(
\boldsymbol{\eta},
w_{ij}'\boldsymbol{\eta})+F_{u_0}(w_{ij}'\boldsymbol{\eta})
}_{
\substack{O(N(N-1)/2)^{-5/14})\\
by \hspace{.1cm} Lemma \hspace{.1cm}  1 
}
}\Big)^2\Bigg).
\]
Because $m(F_{u_0}(\boldsymbol{\eta}))$ is a sample average of probabilities, hence it is stochastically bounded, such that both terms in brackets converge to zero (are $o((N-1)^{-1}$).
We conclude that, at any $\boldsymbol{\eta}$, the objective function using the kernel density estimators is asymptotically equivalent to the objective function using the correctly specified error distributions. Since each $\widehat{F_u}(\boldsymbol{\eta},w_{ij}' \boldsymbol{\eta})$ converges uniformly in $\boldsymbol{\eta}$, convergence of the objective function holds true at $\widetilde{
\hat{
\boldsymbol{\eta}} }(\hat{\boldsymbol{\eta}})$  at a given sample size. 

\subsection{Asymptotic Distribution}

I use 
\[ m(\boldsymbol{\eta})=
m(F_u(\boldsymbol{\eta}))
\] and 
\[M(\boldsymbol{\eta})=
\frac{\partial m(\boldsymbol{\eta})}{\partial \boldsymbol{\eta}}
\]
to denote the vector of moment conditions and the matrix of derivatives of the moment conditions with respect to the parameters for any given choice of $F_u$.  Therefore, I use
\[ \hat{m}(\boldsymbol{\eta})=
m(\hat{F}_u(\boldsymbol{\eta}))
\] and 
\[\hat{M}(\boldsymbol{\eta})=
\frac{\partial \hat{m}(\boldsymbol{\eta})}{\partial \boldsymbol{\eta}}
\] 
when using the kernel estimates of $F_u$ and $f_u$ 
and I use 
\[ m_0(\boldsymbol{\eta})=
m(F_{u_0}(\boldsymbol{\eta}))
\] and 
\[M_0(\boldsymbol{\eta})=
\frac{\partial m_0(\boldsymbol{\eta})}{\partial \boldsymbol{\eta}}
\] when using the error distribution of the true DGP, 
where $\hat{M}(\boldsymbol{\eta})$ and $M_0(\boldsymbol{\eta})$ are matrices of dimension $N\times N$.


In the following, I work with the normalised estimates $\widetilde{
\hat{
\boldsymbol{\eta}} }(\hat{\boldsymbol{\eta}})$ and the normalised coefficients $\Tilde{\boldsymbol{\eta}}_0(
\boldsymbol{\eta}_0)$. 
Since
\[
\hat{m}(
\widetilde{
\hat{
\boldsymbol{\eta}} }(\hat{\boldsymbol{\eta}})
)
=
\hat{m}(
\Tilde{\boldsymbol{\eta}}_0(
\boldsymbol{\eta}_0))
+ \hat{M}(\Bar{\boldsymbol{\eta}})
\left(\widetilde{
\hat{
\boldsymbol{\eta}} }(\hat{\boldsymbol{\eta}})-
\Tilde{\boldsymbol{\eta}}_0(
\boldsymbol{\eta}_0)\right) 
\]\[
\hat{M}(
\widetilde{
\hat{
\boldsymbol{\eta}} }(\hat{\boldsymbol{\eta}})
)'
\hat{m}(\widetilde{
\hat{
\boldsymbol{\eta}} }(\hat{\boldsymbol{\eta}}))=
\hat{M}(\widetilde{
\hat{
\boldsymbol{\eta}} }(\hat{\boldsymbol{\eta}}))'\hat{m}(
\Tilde{\boldsymbol{\eta}}_0(
\boldsymbol{\eta}_0))
+ \hat{M}(\widetilde{
\hat{
\boldsymbol{\eta}} }(\hat{\boldsymbol{\eta}}))'\hat{M}(\Bar{\boldsymbol{\eta}})
\left(\widetilde{
\hat{
\boldsymbol{\eta}} }(\hat{\boldsymbol{\eta}})-
\Tilde{\boldsymbol{\eta}}_0(
\boldsymbol{\eta}_0)\right) =0
\]
\[
\Rightarrow
\widetilde{
\hat{
\boldsymbol{\eta}} }(\hat{\boldsymbol{\eta}})-
\Tilde{\boldsymbol{\eta}}_0(\boldsymbol{\eta}_0)
=
\left(\hat{M}(\widetilde{
\hat{
\boldsymbol{\eta}} }(\hat{\boldsymbol{\eta}}))'
\hat{M}(\Bar{\boldsymbol{\eta}})\right)^{-1} 
\hat{M}(\widetilde{
\hat{
\boldsymbol{\eta}} }(\hat{\boldsymbol{\eta}}))'
\hat{m}(
\Tilde{\boldsymbol{\eta}}_0(
\boldsymbol{\eta}_0)).\]

Recalling that $w_{ij}$ is an $N \times 1$ vector indicating taking the value one at the position of the respective individuals involved in the link, let 
$W$ be a matrix stacking the $L$ vectors $w_{ij} \hspace{.2cm} \{i,j\} \in \mathbb{P}$, resulting in a $L \times N $ matrix. When stacking the vectors, I first arbitrarily order the individuals and then stack them according to this order, that is 
\[W=[w_{12};w_{13};...;w_{1N};...;w_{N(N-1)}].\]


Define the pair specific deviations between true and estimated CDFs at given parameters as
\[ \Delta F_{ij}=
\underbrace{ \left(
\widehat{F_u}(\boldsymbol{\eta},w_{ij}'\boldsymbol{\eta})-F_{u_0}(w_{ij}'\boldsymbol{\eta})
\right)
}_{O_p((\sqrt{L}h\underline{p})^{-1})}.
\]
I use $\Delta F$ to denote the $L \times 1$ vector stacking the deviations 
\[ \Delta F=\left( \Delta F_{12};...;\Delta F_{N(N-1)} \right)
\]
where I stick to the order as introduced when constructing $W$. 
Similarly, let the $L \times N$ matrix $\Delta f$ be defined by concatenating the vectors of derivatives of the $\Delta F$ vector with respect to each coefficient
\[
\Delta f= 
\left[
\frac{\partial \Delta F}{ \eta_1},...,\frac{\partial \Delta F}{ \eta_N}
\right]
\] with
\[
 \frac{\partial \Delta F_{ij}}{ \partial \eta_i}=\underbrace{ \left(
\widehat{f_u}(\boldsymbol{\eta},w_{ij}'\boldsymbol{\eta})-f_{u_0}(w_{ij}'\boldsymbol{\eta})
\right)
}_{O_p(\sqrt{L}^{-1}\underline{p}^{-1}h^{-2})}
\]
and 
\[ \frac{\partial \Delta F_{ij}}{\partial \eta_k}=0 \hspace{.25cm} \forall k \neq i,j.\]

\begin{adjustbox}{angle=90}

$\footnotesize \Delta f=
\begin{bmatrix}
  \hat{f}_u(v_{12})-f_{u_0}(v_{12}) &    
  \hat{f}_u(v_{12})-f_{u_0}(v_{12})&0&&...&&&0\\
    \hat{f}_u(v_{13})-f_{u_0}(v_{13}) & 0 &
    \hat{f}_u(v_{13})-f_{u_0}(v_{13})&0&...&&&0\\
   ... &...&...&...&...&...&...&... \\
    \hat{f}_u(v_{1N})-f_{u_0}(v_{1N}) &0&...&&&&0&    
    \hat{f}_0(v_{1N})-f_{u_0}(v_{1N})\\
  0&
  \hat{f}_u(v_{23})-f_{u_0}(v_{23}) & 
  \hat{f}_u(v_{23})-f_{u_0}(v_{23})&0&...&&0\\
    0&
  \hat{f}_u(v_{24})-f_{u_0}(v_{24}) &  0&  \hat{f}_u(v_{24})-f_{u_0}(v_{24})&0&...&0\\
     ... &...&...&...&...&...&...&... \\
         0& \hat{f}_u(v_{2N})-f_{u_0}(v_{2N}) &0&...&&&0&    
         \hat{f}_u(v_{2N})-f_{u_0}(v_{2N})\\
          ... &...&...&...&...&...&...&... \\
           ... &...&...&...&...&...&...&... \\
            ... &...&...&...&...&...&...&... \\
             ... &...&...&...&...&...&  
             \hat{f}_u(v_{N(N-1)})-f_{u_0}(v_{N(N-1)})&
             \hat{f}_u(v_{N(N-1)})-f_{u_0}(v_{N(N-1)}) \\
  \end{bmatrix}
$ 
\end{adjustbox}

\newpage

\noindent
{\bf Lemma 3 (Convergence of the Moment Conditions): } \\$ \forall \boldsymbol{\eta} \in \mathbb{H }
\hspace{.2cm}
\sqrt{N-1}
\hat{m}(\boldsymbol{\eta}) =
\sqrt{N-1}
m_0(\boldsymbol{\eta})+o(1).$
\newline

Because 
\[
\frac{\partial 
m(F(\boldsymbol{\eta}))
}{
\partial 
F(w_{ij}'\boldsymbol{\eta})
}
=
- D w_{ij}\]
thus 
\begin{equation}
    \hat{m}(\boldsymbol{\eta})=
m_0(\boldsymbol{\eta})-
D W'\Delta F.
\label{mc}
\end{equation}

Recalling that $D=
\operatorname{diag}((N-1)^{-1})$, hence $\sqrt{D}=
\operatorname{diag}((N-1)^{-1/2})$, if
\[
\sqrt{D} W'\Delta F=o(1)
\hspace{1cm}
\Rightarrow 
\sqrt{N-1}
\hat{m}(\boldsymbol{\eta}) \overset{}{\rightarrow}
\sqrt{N-1}
m_0(\boldsymbol{\eta})
\]
pointwise in $\boldsymbol{\eta}$. 
Each element of the $N \times 1$ vector $\sqrt{D} W'\Delta F$ converges to zero, pointwise in the coefficient vector. I exemplify this with the first element.
\[
\left(\sqrt{D} W'\Delta F \right)_{1}=
\frac{1}{\sqrt{N-1}}
\sum_{j=2}^N
\left(\widehat{F_u}(\boldsymbol{\eta},\eta_1+\eta_j)-F_{u_0}(\eta_1+\eta_j)\right).
\]
Given the choice of $h=L^{-1/7}=(N(N-1)/2)^{-1/7}$, we have \[\left(\widehat{F_u}(\boldsymbol{\eta}, \eta_1+\eta_j)-F_{u_0}(\eta_1+\eta_j)\right)=O((N(N-1)/2)^{-5/14}\underline{p}).\] 
 Since a sum of asymptotically bounded terms is also asymptotically bounded (at the highest order at which a summand is bounded), thus (for an arbitrary $\alpha>0$) 
 \[
 \frac{1}{\sqrt{N-1}}
\sum_{j=2}^N
\left(\widehat{F_u}(\boldsymbol{\eta},\eta_1+\eta_j)-F_{u_0}(\eta_1+\eta_j)\right)=
\frac{1}{\sqrt{N-1}}
\sum_{j=2}^N
O((N(N-1)/2)^{-5/14}\underline{p})\]
\[=No((N-1)^{-1/2+\alpha})=o((N-1)^{-1/2+\alpha})=o(1).
 \]
\newline

 {\bf Lemma 4 (Convergence of the Derivative Matrix of the Moment Conditions): } 
$ \forall \boldsymbol{\eta} \in \mathbb{H }
\hspace{.2cm}
    \hat{M}(\boldsymbol{\eta})=
M_0(\boldsymbol{\eta})+o(1).
$\\

If I take the derivative of \eqref{mc} with respect to $\boldsymbol{\eta}$, I obtain 
\begin{equation}
    \hat{M}(\boldsymbol{\eta})=
M_0(\boldsymbol{\eta})-
DW' \Delta f.
\label{MC}
\end{equation}

Recalling that $D=
\operatorname{diag}((N-1)^{-1})$, if
\[
D W'\Delta f=o(1)
\hspace{1cm}
\Rightarrow \hat{M}(\boldsymbol{\eta}) \overset{}{\rightarrow}M(\boldsymbol{\eta})
\]
pointwise in $\boldsymbol{\eta}$. 
Each element of $D W'\Delta f$ converges to zero, pointwise in the coefficient vector. I exemplify this with the first element.
\[
\left(D W'\Delta f \right)_{11}=
\frac{1}{N-1}
\sum_{j=2}^N
\left(\widehat{f_u}(\boldsymbol{\eta},
\eta_1+\eta_j)-f_{u_0}(\eta_1+\eta_j)\right).
\]
Given the choice of $h=L^{-1/7}=(N(N-1)/2)^{-1/7}$, we have \[\left(\widehat{f_u}(\boldsymbol{\eta}, \eta_i+\eta_j)-f_{u_0}(\eta_i+\eta_j)\right)=O((N(N-1)/2)^{-3/14}).\] 
Since a sum of asymptotically bounded terms is also asymptotically bounded (at the highest order at which a summand is bounded), thus (for an arbitrary $\alpha>0$)  
\[
\frac{1}{N-1}
\sum_{j=2}^N
\left(\widehat{f_u}(\boldsymbol{\eta},
\eta_1+\eta_j)-f_{u_0}(\eta_1+\eta_j)\right)=
\frac{1}{N-1}
\sum_{j=2}^N
O((N(N-1)/2)^{-3/14})=\]
\[N o((N-1)^{-1+\alpha})=
o((N-1)^{-1+\alpha})=o(1).
\frac{1}{N-1}
\sum_{j=2}^N
\left(\widehat{f_u}(\boldsymbol{\eta},
\eta_1+\eta_j)-f_{u_0}(\eta_1+\eta_j)\right).
\]


As a consequence
\[
\widetilde{
\hat{
\boldsymbol{\eta}} }(\hat{\boldsymbol{\eta}})-
\Tilde{\boldsymbol{\eta}}_0(\boldsymbol{\eta}_0)
=
\left(\hat{M}(\widetilde{
\hat{
\boldsymbol{\eta}} }(\hat{\boldsymbol{\eta}}))'
\hat{M}(\Bar{\boldsymbol{\eta}})\right)^{-1} 
\hat{M}(\widetilde{
\hat{
\boldsymbol{\eta}} }(\hat{\boldsymbol{\eta}}))'
\hat{m}(
\Tilde{\boldsymbol{\eta}}_0(
\boldsymbol{\eta}_0))\]
implies that 
\[
\widetilde{
\hat{
\boldsymbol{\eta}} }(\hat{\boldsymbol{\eta}})-
\Tilde{\boldsymbol{\eta}}_0(\boldsymbol{\eta}_0)
\overset{p}{\rightarrow}
\left(M(\widetilde{
\hat{
\boldsymbol{\eta}} }(\hat{\boldsymbol{\eta}}))'
M(\Bar{\boldsymbol{\eta}})\right)^{-1} 
M(\widetilde{
\hat{
\boldsymbol{\eta}} }(\hat{\boldsymbol{\eta}}))'
m(
\Tilde{\boldsymbol{\eta}}_0(
\boldsymbol{\eta}_0)),\]
such that the normalised estimates converge to the normalised true coefficients at the parametric rate and are asymptotically normally distributed by standard GMM asymptotic theory.

\renewcommand\refname{Bibliography}
\nocite{*}
\bibliographystyle{chicago} 

\bibliography{my} 

\end{document}